\documentclass[accepted]{uai2025} 
                        
\usepackage[american]{babel}

\usepackage{natbib} 
\usepackage{mathtools} 
\usepackage{booktabs} 
\usepackage{tikz} 

\usepackage[inline]{enumitem}

\usepackage{amsmath,amsfonts,bm}

\def\eqref#1{equation~\ref{#1}}

\def\1{\bm{1}}

\DeclareMathAlphabet{\mathsfit}{\encodingdefault}{\sfdefault}{m}{sl}
\SetMathAlphabet{\mathsfit}{bold}{\encodingdefault}{\sfdefault}{bx}{n}

\DeclareMathOperator*{\argmax}{arg\,max}
\DeclareMathOperator*{\argmin}{arg\,min}

\def\RR{{\mathbb R}}    

\def\11{{\mathbf 1}}    

\def\cA{{\mathcal A}}  \def\cG{{\mathcal G}}      \def\cN{{\mathcal N}}          \def\cP{{\mathcal P}}           \def\cX{{\mathcal X}}

\usepackage{amsmath}
\usepackage{amssymb}
\usepackage{amsthm}
\usepackage{adjustbox}
\newtheorem{theorem}{Theorem}[section]
\newtheorem{proposition}[theorem]{Proposition}
\newtheorem{lemma}[theorem]{Lemma}
\newtheorem{corollary}[theorem]{Corollary}
\newtheorem{definition}[theorem]{Definition}
\newtheorem{assump}[theorem]{Assumption}

\theoremstyle{remark}
\newtheorem{remark}[theorem]{Remark}

\usepackage{comment}             
\usepackage{paralist}            
\usepackage{wrapfig}             
\usepackage{caption}             
\usepackage{pifont}              
\usepackage{minitoc}             
\usepackage{algorithm}           
\usepackage[noend]{algpseudocode}
\usepackage{nicefrac}            
\usepackage{booktabs}            
\usepackage{tikz}
\usetikzlibrary{tikzmark}        
\usepackage{multirow}
\usepackage{stfloats}

\usepackage{todonotes}

\title{Bayesian Optimization for Building Social-Influence-Free Consensus}

\begin{document}
\author[1,2]{Masaki Adachi}
\author[3]{Siu Lun Chau}
\author[4]{Wenjie Xu}
\author[3]{Anurag Singh}
\author[1]{Michael A. Osborne}
\author[3]{Krikamol Muandet}
\affil[1]{%
    Machine Learning Research Group\\
    University of Oxford, UK
}
\affil[2]{%
    Toyota Motor Corporation\\
    Japan
}
\affil[3]{%
    CISPA Helmholtz Center for Information Security\\
    Germany
}
\affil[4]{%
    Automatic Control Laboratory\\
    EPFL, Switzerland
}
\maketitle

\begin{abstract}
  We introduce \emph{Social Bayesian Optimization}~(SBO), a vote-efficient algorithm for consensus-building in collective decision-making. In contrast to single-agent scenarios, collective decision-making encompasses group dynamics that may distort agents' preference feedback, thereby impeding their capacity to achieve a social-influence-free consensus—the most preferable decision based on the aggregated agent utilities. We demonstrate that under mild rationality assumptions, reaching social-influence-free consensus using noisy feedback alone is impossible. To address this, SBO employs a dual voting system: cheap but noisy public votes (e.g., show of hands in a meeting), and more accurate, though expensive, private votes (e.g., one-to-one interview). We model social influence using an unknown social graph and leverage the dual voting system to efficiently learn this graph. Our theoretical findings show that social graph estimation converges faster than the black-box estimation of agents’ utilities, allowing us to reduce reliance on costly private votes early in the process. This enables efficient consensus-building primarily through noisy public votes, which are debiased based on the estimated social graph to infer social-influence-free feedback. We validate the efficacy of SBO across multiple real-world applications, including thermal comfort, team building, travel negotiation, and energy trading collaboration.
\end{abstract}

\doparttoc 
\faketableofcontents 

\section{Introduction}\label{sec:intro}
Building consensus is of paramount importance in collaborative work and is ubiquitous in real life, economics, and human-in-the-loop machine learning—such as preference maximization \citep{eric2007active, gonzalez2017preferential}, human-AI collaboration \citep{carroll2019utility, vodrahalli2022uncalibrated, sarkar2024diverse}, and learning-to-defer \citep{madras2018predict, mao2024two, singh2024domain}. However, reaching a consensus is notoriously challenging. As our illustrative example, imagine you are responsible for selecting the next venue for an international conference. You have a sufficient budget and a wide range of candidate locations worldwide, and you wish to make a decision that reflects the collective preferences of decision-makers. This task presents the following three main challenges.

\textbf{Challenge 1: Aggregation Function.} 
A common approach to collective decision-making is through majority voting, which prioritizes the option with the most votes. This method, following \emph{utilitarianism}, assumes that maximizing the number of satisfied voters leads to the most desirable outcome. While intuitive, majority voting is not always fair nor satisfactory. In the absence of a trivial consensus—where everyone agrees on the best venue—this system inherently disregards the preferences of minority groups. 
For instance, some participants may face visa restrictions that make attending certain locations difficult. \citet{rawls2017theory} argued that consensus should instead prioritize the worst-off individuals. His perspective, known as \emph{egalitarianism}, suggests that decision-making should be robust against uncontrollable factors that could place anyone in the minority. 
Importantly, there is no absolute answer as to which policy is better; reality often lies somewhere in between. Therefore, achieving consensus requires not only aggregating individual preferences but also first selecting the most appropriate aggregation method for a given situation. In economics, this function is known as the \emph{aggregation function}, social welfare function, or choice function. \citet{arrow1950difficulty} proved that no aggregation function can satisfy all fundamental rationality axioms simultaneously (famously known as \emph{Arrow's impossibility theorem}). Consequently, we must determine which aspect of rationality to compromise, effectively choosing between utilitarian and egalitarian principles.

\textbf{Challenge 2: Efficient Utility Function Estimation.} 
Even after choosing an aggregation function, achieving consensus requires maximizing the aggregated preference---a task that presents another significant challenge. Classical economics often assumes that utility functions \citep{fishburn1968utility} are given analytically with nice mathematical properties such as continuity, making optimization straightforward.
In practice, however, people often struggle to introspect and articulate their preferences \citep{kahneman1979interpretation}. For example, asking voters to rank Rio de Janeiro among $100$ candidate cities is difficult. Instead, people are more comfortable making pairwise comparisons, such as ``Do you prefer Rio de Janeiro over Tokyo?'' This aligns with prospect theory \citep{kahneman2013prospect}, which suggests that human decision-making is more reliable when framed in comparative rather than absolute terms \citep{furnkranz2010preference,chau2022spectral,chau2022learning}. 
Yet, this approach introduces another challenge: the need for exhaustive voting. As the number of candidates grows, obtaining all pairwise comparisons from every participant quickly becomes impractical. To address this, we need an adaptive voting scheme that dynamically selects the most informative pairwise comparisons in an online manner and stops voting once a reliable consensus estimate is reached.

\textbf{Challenge 3: Social Influence.} 
Finally, the most critical challenge in collective decision-making is social influence, which can distort voting results. In online voting systems, where quick and flexible feedback loops are essential, the simplest form of collective decision-making might involve a public show of hands: voters see a pair of candidates and raise their hands to indicate their preference. 
However, this approach has a major drawback---susceptibility to social influence. The presence of influential agents, such as conference sponsors or the session chair, can sway the votes of others, leading to biased outcomes. Ideally, a robust decision-making process should be free from such social influence, ensuring that votes reflect individuals' true and genuine preferences rather than the pressures of group dynamics. 
When social influence dominates, it can result in \emph{groupthink}, a failure mode where the group reaches a corrupted consensus. 

\textbf{Our contributions} are summarized as follows.
\begin{figure}
  \centering
  \includegraphics[width=0.93\linewidth]{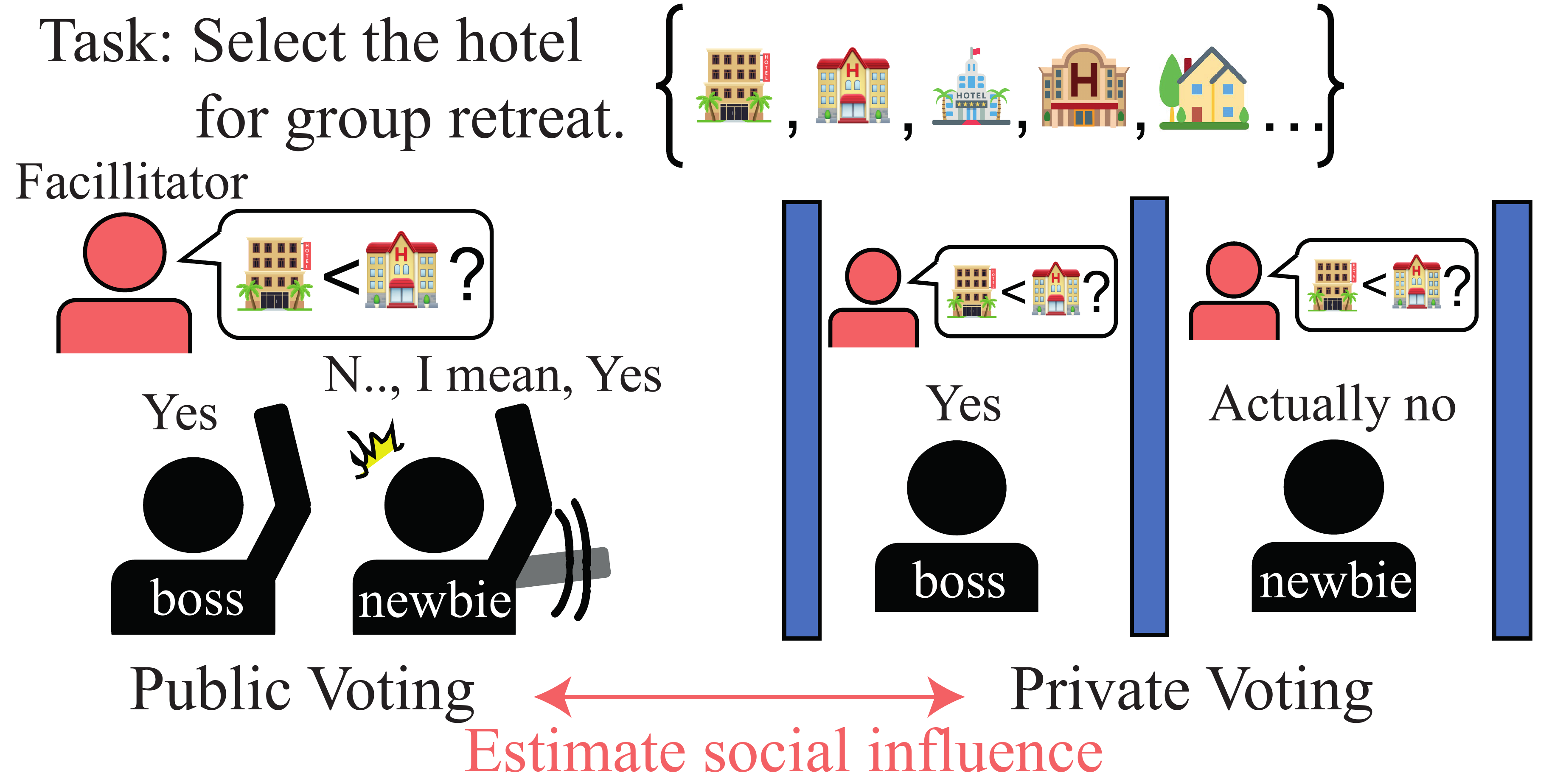}
  \caption{\small{\textbf{A dual voting system:} the difference in public and private votes can identify the social influence. Once identified, social-influence-free consensus can be estimated only from noisy public votes, thereby reducing the total cost.}}
  \label{fig:concept}
  \vspace{-1em}
\end{figure}
\textbf{(1) Impossibility theorem of groupthink-proof consensus.} In the absence of a trivial consensus—where everyone agrees on the best option—we prove that no aggregation function can consistently produce a consensus unchanged under any social influence. 
\textbf{(2) Algorithm for social-influence-free consensus.} 
To circumvent this impossibility result, we propose a dual voting system consisting of a cheap but noisy public vote (e.g., a show of hands in a meeting) and a private vote (e.g., a one-on-one interview), which is free from social influence but costly to obtain (see Figure~\ref{fig:concept} for an illustration of our procedure). 
If we rely solely on private votes, we can achieve a consensus free from social influence, but at a significantly high cost, which is not desirable. To improve cost-effectiveness, we leverage cheap public votes to reduce reliance on costly private votes. We formalize social influence as an unknown social graph convolution operator and estimate the underlying social graph using pairs of public and private votes (\emph{dual votes}). Once we can accurately infer the social graph, private vote outcomes can be recovered from public votes, allowing us to stop querying expensive private votes at a certain point while continuing to estimate the consensus using only public votes. 
We refer to the algorithm that infers the unknown social graph and utilities based on dual votes, while strategically selects the most informative votes to efficiently reach a consensus in an online manner, as \emph{Social Bayesian Optimization (SBO)}.
\textbf{(3) Theoretical guarantees.} 
We prove that our algorithm achieves \emph{no regret}, meaning that the error between the estimated and true consensus asymptotically converges to zero, as does the number of required private votes. Consequently, SBO provably reaches a social-influence-free consensus at a lower cost. 
\textbf{(4) Real-world contributions.} 
We demonstrate SBO’s fast convergence and cost-effectiveness compared to baselines in 10 synthetic and real-world tasks.

\vspace{-0.5em}
\section{Social Bayesian Optimization}
\vspace{-0.5em}
Our goal is to find the consensus $x^\star$,
\begin{equation}
\label{objective: SBO}
x^\star \in \argmax_{x \in \mathcal{X}} \mathcal{A}[u(x, :)], \quad u(x, :) = \{ u(x, i) \}_{i \in V},
\vspace{-0.5em}
\end{equation}
where $V$ represents a social group of $n$ agents, i.e., $|V|=n$, $\mathcal{X}$ is a set of options that is a bounded subset of $\mathbb{R}^d$, and $\cA$ is an aggregation function that produces the \emph{social utility}. 
For each agent $i\in V$ and option $x\in\mathcal{X}$, $u(x, i)$ represents the utility of option $x$ for agent $i$, and $u(x, :) = \{ u(x, i) \}_{i \in V}$ represents the set of utilities for all agents given option $x$. We assume throughout that $u: \mathcal{X} \times V \to \mathbb{R}$ is a black-box utility function that rationalizes the collective preferences of the social group $V$. We iteratively collect votes from agents to minimize regret to reach $x^\star$ within the given budget.

To solve (\ref{objective: SBO}), we rely on preference feedback, or \emph{votes}, where each agent expresses preferences between pairs of options \( [x, x'] \). Following standard preference modeling notations, agent \( i \) prefers option \( x_1 \) over \( x_2 \), denoted \( x_1 \succ_{u(\cdot,i)} x_2 \), if and only if \( u(x_1, i) > u(x_2, i) \), where \( \succ_{u(\cdot,i)} \) represents agent \( i \)'s preference relation. Our optimization procedure thus involves estimating \(n\) utility functions \(\{u(\cdot, i)\}_{i \in V}\) from their \emph{votes}. When the context is clear, this set also denotes the vector of utilities queried at \(x\). 
In our setting, we make two key assumptions.
\begin{assump}[\textbf{Facilitator}]\label{assump:facilitator}
    There exists a single facilitator (or social planner) who facilitates the decision-making process, and decides the aggregation rule $\mathcal{A}$.
\end{assump}
\begin{assump}[\textbf{Pairwise feedback}]\label{assump:pairwise}
    Given a pair of options $(x_t, x^\prime_t)$ at time step $t$, there exists an oracle that returns a preference signal $\textbf{1}_{x \succ x\prime}^{(i)}$ from the $i$-th agent where $\textbf{1}_{x \succ x\prime}^{(i)} :=1$ if $x$ is preferred and zero if $x^\prime$ is preferred.
    The feedback $\textbf{1}_{x \succ x\prime}^{(i)}$ from the oracle follows the Bernoulli distribution with
    $\mathbb{P}(\textbf{1}_{x \succ x\prime}^{(i)} = 1) = p_{x \succ x\prime}^{(i)} = \sigma(u(x,i) - u(x^\prime,i))$, 
    where $\sigma(z) = (1+\exp(-z))^{-1}$. 
\end{assump}
\vspace{-0.25em}
Assumption~\ref{assump:pairwise} is the widely accepted Bradly-Terry model~\citep{bradley1952rank}.

\vspace{-0.25em}
\subsection{Aggregation Functions}
\label{subsec: aggregation}
\vspace{-0.25em}
The crucial part of problem (\ref{objective: SBO}) is choosing a desirable $\mathcal{A}$ that aggregates agents' utilities into a social utility.

\begin{definition}[\textbf{Aggregation function}]\label{def:aggregate}
    The aggregation function \(\mathcal{A}\) combines individual utilities via a positive linear combination \(\mathcal{S}(x) := \mathcal{A}[u(x,:)] = \textbf{w}^\top u(x, :)\) where \(\mathcal{S}(x)\) represents the social utility and $\textbf{w} \in \RR^n_{\geq 0}$ depends on $u(x,:)$. \(\mathcal{A}\) is provided \emph{a priori} by the facilitator and is independent of both the option \(x\) and the time step $t$, meaning it is homogeneous and stationary.
    \vspace{-0.5em}
\end{definition}
\citet{harsanyi1955cardinal} demonstrated that a positive linear combination of individual utilities is the only aggregation rule that satisfies both the von Neumann-Morgenstern (VNM) axioms~\citep{von2007theory} and Bayes optimality~\citep{brown1981complete}. Popular aggregation function such as the utilitarian rule $\mathcal{A}[u(x,:)] = \nicefrac{1}{n}\sum_{i\in V}u(x, i)$ and the egalitarian rule 
$\mathcal{A}[u(x, :)] = \min_{i\in V}u(x, i)$ are positive linear combinations~(c.f. Appendix~\ref{app:aggregate}). Given the range of candidate methods, we adopt the generalized Gini social-evaluation welfare function (GSF; \citet{weymark1981generalized, sim2021collaborative}) as it can interpolate between utilitarian and egalitarian approaches:
\begin{equation}
\begin{aligned}\label{eq:gini}
    &\mathcal{A}[u_t(x,:)] := \textbf{w}^\top \phi(u_t(x,:)),\\
    \text{s.t.} \quad &w_i := \nicefrac{\rho^{i-1}}{\textbf{w}^\top \textbf{1}}, \,\, w_i > 0, \,\,  0 < \rho \leq 1,
\end{aligned}
\end{equation}
where $\textbf{w} := (w_i)_{i \in V}$ is a weight vector, $\textbf{1}$ is the one-vector, and $\phi$ is a sorting function that arranges the elements of the input vector in ascending order and returns the sorted vector.
\begin{proposition}[Proposition 1 in \citet{sim2021collaborative}]\label{proposition:gsf}
    GSF in Eq.~(\ref{eq:gini}) satisfies monotonicity and the Pigou-Dalton principle (PDP; \citet{pigou1912wealth,dalton1920measurement}) on fairness. Moreover, when
    \begin{compactenum}
        \item[(a)] $\rho=1$, $\mathcal{A}$ is utilitarian such that it satisfies the PDP in the weak sense;
        \item[(b)] $0 <\rho<1$, $\mathcal{A}$ satisfies the PDP in the strong sense;
        \item[(c)] $\rho \rightarrow 0$, then $w_i/w_1 \rightarrow 0$ for $i = 2,\ldots,n$, $\mathcal{A}$ converges to egalitarian.
        \vspace{-0.5em}
    \end{compactenum}
\end{proposition}
See Appendix~\ref{proof:gsf} for the proof and further details. The key takeaway is that the GSF interpolates between two popular aggregation rules through a single real parameter, $\rho$, while adhering to the fairness principle of PDP and maintaining monotonicity for regret analysis. Although $\rho$ must be defined a priori, we assume this is defined by the facilitator (Assumption~\ref{assump:facilitator}). We argue that selecting a single parameter is far simpler than choosing an arbitrary aggregation function. Moreover, GSF is intuitive: $\rho$ controls the balance between prioritizing the group average and the worst-off agent. Nonetheless, our algorithm---described later---applies to any aggregation function that is a positive linear combination of utilities and monotonic, encompassing a wide range of popular aggregation rules.

\vspace{-0.25em}
\subsection{Modelling Social Influence}\label{sec:graph}
\vspace{-0.25em}

While various methods exist to model social interactions among agents, we chose a graph convolutional approach. With a slight abuse of notation, let  $V$  represent the set of nodes (agents) and  $E$  the set of weighted directed edges, forming the social influence graph  $G = (V, E)$ , with  $A$  as the corresponding adjacency matrix.
\begin{definition}[\textbf{Social influence}]\label{def:bandwagon}
    Given influence graph $G$, social-influence-free utility $u$, and agent $i\in V$, the corrupted utilities of agent $i$, $v(\cdot, i)$ can be expressed as $h\left(u(\cdot,i), \{u(\cdot,j) \mid j \in N_G(i)\}\right)$, for some generic function $h$ that specifies how the signals interact, and $N_G(i)$ is the (in)-neighbour of $i$ in $G$, defined as $\{j: A_{ij} \neq 0\}$.
    \vspace{-0.5em}
\end{definition}

Under Definition~\ref{def:bandwagon}, it is natural to model the function $h$ as a graph convolution operation
\begin{equation}
    \begin{aligned}
    \label{eq:graph_conv}
    v(\cdot, :)^\top &= Au(\cdot, :)^\top = \sum_{j \in N_G(i)\cup \{i\}} A_{ij} u(\cdot, j)^\top
\end{aligned}
\end{equation}
where adjacency matrix $A$ such that $\sum_{j=1}^n A_{ij} = 1$ and $A_{ij} > 0$ for all $i,j$.
This definition provides useful bounds and conditions (see Lemma~\ref{lemma:graph} for details). While it can be extended to other graph structures (see Appendix~\ref{app:other_graph}), we focus on the simplest one as our main example.

\vspace{-0.5em}
\section{Mitigating Undue Social Influence}
\vspace{-0.5em}

As discussed, the main challenge in achieving consensus lies in social influences that prevent the facilitator from accessing the agents' true utilities. Consequently, the facilitator can only observe the feedback based on distorted utilities, leading to a consensus that misrepresents the agents' actual preferences---an outcome known as groupthink. To overcome this, we propose a dual voting mechanism in Section \ref{sec:dual-voting}, which combines a cost-effective but noisy public vote with a costly private vote that is free from social influence. 

\vspace{-0.25em}
\subsection{Impossibility Theorem}\label{sec:impossibility}
\vspace{-0.25em}

Before introducing our dual voting mechanism, we first illustrate the difficulty of consensus building based solely on distorted utilities through the impossibility theorem. To this end, we define groupthink-proofness as desirable property of any aggregation function.
\begin{definition}[\textbf{Groupthink-proof}]\label{def:bandwagon_proof}
    An aggregation function $\mathcal{A}$ is groupthink-proof if, for any social-influence graph $G$, social-influence-free utility $u$, and corrupted utility $v$, $\argmax_{x \in \mathcal{X}} \mathcal{A}[u(x,:)] = \argmax_{x \in \mathcal{X}} \mathcal{A}[v(x,:)]$.
    \vspace{-0.5em}
\end{definition}
Intuitively, the aggregation function $\mathcal{A}$ is groupthink-proof if its consensus is preserved under any social influence graph. For example, any aggregation function is groupthink-proof if $v(\cdot,:) = u(\cdot,:)$, i.e., no social influence. 
In addition, we define the triviality of the social consensus $x^*$ as
\begin{definition}[\textbf{Trivial social consensus}]\label{def:dictatorship}
    $x^*$ is the trivial social consensus if for all $i\in|V|$, $x^*=\argmax_{x}u(x,i)$
    \vspace{-0.5em}
\end{definition}
Non-triviality of social consensus excludes the situations where all agents unanimously agree on the best option.
The following theorem states that, in the absence of a trivial consensus, no aggregation function when applied on the distorted utilities satisfies groupthink-proofness.
\begin{theorem}[\textbf{Impossibility of groupthink-proof aggregation}]\label{thm:impossibility}
    Under Definitions~\ref{def:aggregate}, \ref{def:bandwagon_proof}, \ref{def:dictatorship}, there exists no aggregation rule $\mathcal{A}$ satisfying groupthink-proof in the absence of a trivial consensus.
    \vspace{-0.5em}
\end{theorem}
Theorem \ref{thm:impossibility} implies that if the agents are not unanimously in consensus on the best option with respect to their true utilities, then the facilitator cannot identify an aggregation function that is groupthink-proof. Appendix~\ref{proof:impossibility} provides a detailed proof, which follows a proof-by-contradiction approach. The core idea is that any groupthink-proof aggregation rule $\mathcal{A}$ must lead to a trivial social consensus.

\begin{figure}[t]
  \centering
  \includegraphics[width=0.35\textwidth]{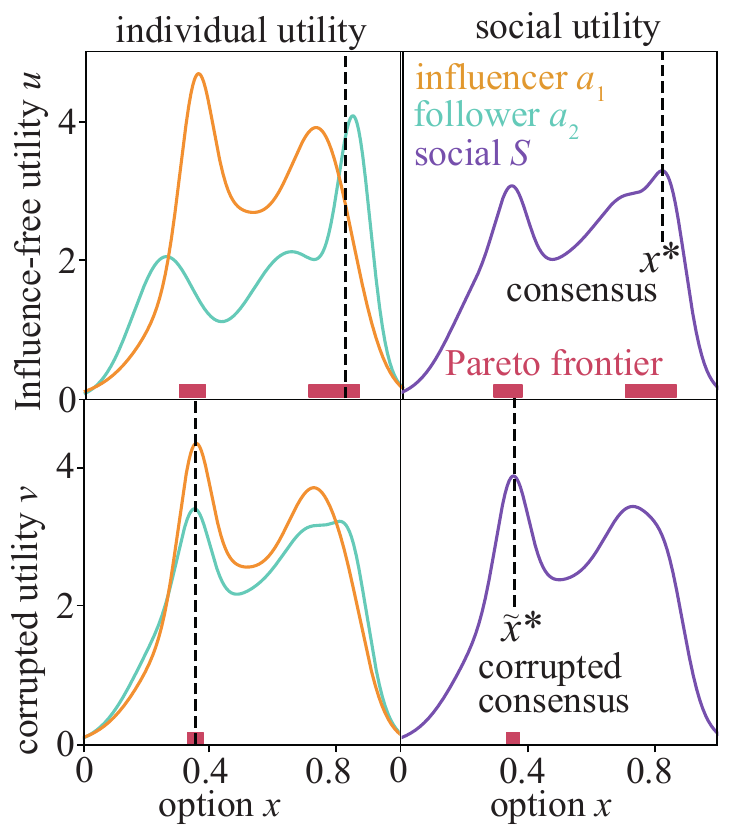}
  \caption{Pareto frontier corrupted by the social influence can exclude the true consensus $x^\star$.}
  \label{fig:example}
  \vspace{-1em}
\end{figure} 
To illustrate the direct implications of our impossibility results, we show that the Pareto front of distorted utilities is not groupthink-proof. The Pareto frontier is a widely used objective in multi-objective optimization, as it is expected to contain the consensus point $x^\star$. While this holds for the true utility $u$, it does not necessarily hold for the corrupted utility $v$. Fig.~\ref{fig:example} provides a counterexample. Consider two agents: an influencer ($a_1$) and a follower ($a_2$).\footnote{Note that the decision-maker is not the influencer but the facilitator. The facilitator seeks to elicit the social-influence-free consensus, while the influencer distorts the entire voting process.} Here, we assume that $v$ is given and that the aggregation rule $\mathcal{A}$ is utilitarian (see Section~\ref{subsec: aggregation}), but we do not have access to $u$ or $\mathcal{A}$. We set the ground truth as $A = \big( \begin{smallmatrix} 0.9 & 0.1 \\ 0.6 & 0.4\\ \end{smallmatrix} \big)$. This indicates that $a_1$ prioritizes their own utility nine times more than $a_2$'s, while $a_2$ values $a_1$’s utility 1.5 times more than their own. 
As a result, the corrupted utilities $v$ become nearly identical. While the true consensus is at $x^\star = 0.82$, the corrupted one shifts to $\tilde{x}^\star = 0.38$. Moreover, the corrupted Pareto frontier does not contain the true consensus $x^\star$. 

\vspace{-0.25em}
\subsection{Dual Voting Mechanism with Approximated Social Graph}\label{sec:dual-voting}
\vspace{-0.25em}
\begin{figure*}[t]
  \centering
  \includegraphics[width=1\hsize]{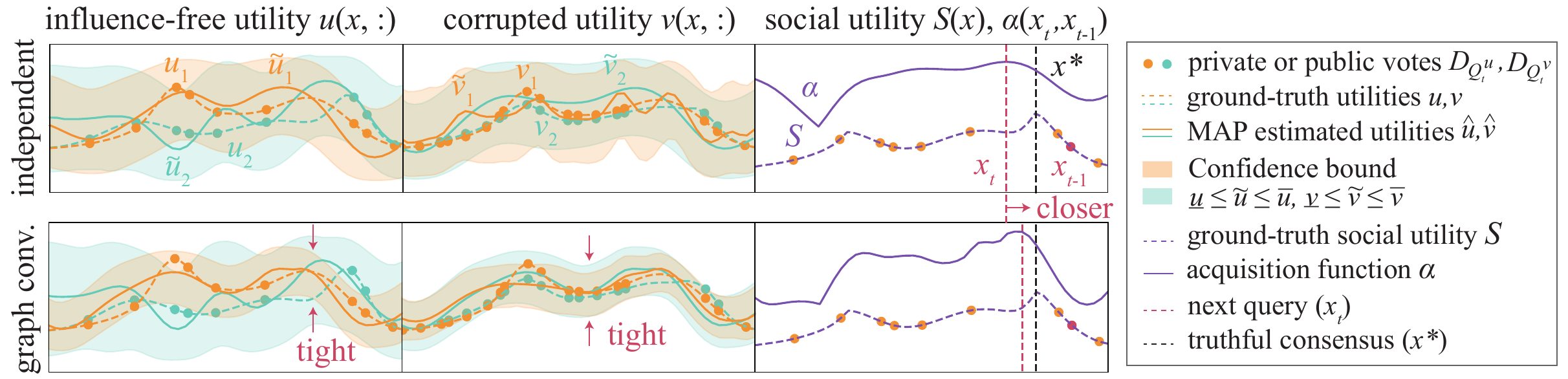}
  \caption{\small{
  The dotted and solid lines represent the ground truth \(u, v\) and estimated utilities \(\tilde{u}, \tilde{v}\), respectively, with the shaded area indicating the confidence interval. Dots mark the queried points, where \(|\mathcal{Q}^u_t|=10\) and \(|\mathcal{Q}^v_t|=20\). Since utility values are not directly observable, the dots are for visual guidance. The acquisition function \(\alpha(\cdot, x_{t-1})\) represents the upper confidence bound of improvement from the previous query \(x_{t-1}\), with its argmax determining the next query \(x_t\) (red vertical line). While the independent model assumes no relationship between $u, v$, the graph convolution model enforces a linear constraint $v = A u$. This yields a tighter predictive interval, resulting in a next query closer to the true consensus \(x^\star\) (black vertical line).
  }}
  \label{fig:visual_exp}
  \vspace{-1em}
\end{figure*}
\textbf{Dual voting.} 
We begin by formally defining the dual voting system. Let $D_{\mathcal{Q}^u_t} := (x_\tau, x_\tau^\prime, \textbf{1}_\tau^{(i)} )_{\tau\in \mathcal{Q}^u_t, i \in V}$ be the \textbf{private} votes, $D_{\mathcal{Q}^v_t} := (x_\tau, x_\tau^\prime, \textbf{1}_\tau^{(i)} )_{\tau\in \mathcal{Q}^v_t, i \in V}$ be the \textbf{public} votes, $\textbf{1}_\tau^{(i)} \in \{0,1\}$ be the realization of the Bernoulli random variable $\textbf{1}_{x \succ x^\prime}^{(i)}$, $[t] := \{1, \ldots, t\}$, $\mathcal{Q}_t^v := \{\tau \in [t-1] \mid \text{if } v \text{ is queried in step }\tau \}$ be public queries, $\mathcal{Q}_t^u$ be private queries $(t \geq |\mathcal{Q}_t^u|, t \geq |\mathcal{Q}_t^v|)$. 
See Appendix~\ref{sec:notation} for the summarised table of notations.
\begin{assump}[\textbf{Public and private votes}]\label{assump:votes}
    While private votes reflect the social-influence-free utility $u$, public votes reflect the (possibly) influenced utility $v$. The query costs satisfy the relationship $\lambda_u \gg \lambda_v$.
    \vspace{-0.5em}
\end{assump}
The cost suggests $|\mathcal{Q}_t^u| \ll |\mathcal{Q}_t^v|$ is more cost-effective. 

\textbf{Bayesian modelling.} 
We now introduce the procedure for learning the social graph $A$ and the surrogate models for $u$ and $v$. To maintain focus on the core ideas, we omit detailed modelling explanations in the main text. 
We assume that both $u$ and $v$ are functions lying in reproducing kernel Hilbert space (RKHS; see Assumption~\ref{assump:bounded_norm} for the formal statement), similar to standard BO methods. RKHS functions are a useful and common assumption in the Bayesian Optimization (BO) literature, as they provide regularity by ensuring that the functions have bounded norms in the corresponding RKHS. 
In a Bayesian sense, this corresponds to placing a prior over the function space, $u \in \mathcal{B}^u_0$ and $v \in \mathcal{B}^v_0$, where $\mathcal{B}^u_0$ and $\mathcal{B}^v_0$ represent the respective function spaces of $u$ and $v$ within the confidence interval (also known as the \emph{confidence set}). The choice of kernel (e.g., RBF or Matérn) encodes prior knowledge about the smoothness of the true utility functions. Additionally, we impose a prior on the graph $A$ using a modified Dirichlet distribution, while the likelihood follows the Bradley-Terry model (Assumption~\ref{assump:pairwise}). After the $t$-th observations, $D^u_t, D^v_t$, the confidence sets are updated to $\mathcal{B}^u_t, \mathcal{B}^v_t$, contracting over time (see \citep{fukumizu2013kernel} for the connection between kernels and Bayes' rule).

A key challenge is that standard Gaussian process (GP) approaches (e.g., \citet{chu2005preference}) do not yield conjugate posteriors, making both computation and theoretical analysis difficult. To address this, we adopt a likelihood-ratio approach \citep{liu2023optimistic,emmenegger2024likelihood, xu2024principled}, which provides predictive confidence intervals for a given test input $x$ rather than a full predictive distribution. This is well-suited for BO, as widely used algorithms such as GP-UCB \citep{srinivas2010gaussian} rely only on the upper confidence bound rather than the full distribution. We extend this approach to incorporate a graph prior and leverage the convolutional relationship described in Eq.~(\ref{eq:graph_conv}). 
The resulting method, termed \emph{optimistic MAP} (maximum a posteriori), provides upper confidence bounds for the individual utilities $u$ and $v$, the aggregated social utility $\mathcal{S}$, and the social graph $\mathcal{A}$, conditioned on the observed votes $(D_{\mathcal{Q}^u_t}, D_{\mathcal{Q}^v_t})$. A key advantage of our approach is that it offers a principled algorithm with theoretical guarantees. A detailed explanation is provided in Appendix~\ref{app:bayesian}.

Fig.~\ref{fig:visual_exp} provides a visual example of our modeling, comparing two approaches: the independent model and the graph convolutional model. We consider a scenario with data imbalance, where $\lvert \mathcal{Q}^v_t \rvert > \lvert \mathcal{Q}^u_t \rvert$. This imbalance is desirable, as we aim to leverage the noisy yet cheap public votes $\mathcal{Q}^v_t$ to reduce reliance on the clean but costly private votes $\mathcal{Q}^u_t$. While the independent model, which assumes no relationship between $u$ and $v$, produces large confidence bounds, the graph convolutional model captures correlations, leading to tighter confidence bounds. A more precise utility estimation yields a sharper upper confidence bound for the social utility $\mathcal{S}$, i.e., $\alpha$, improving the true consensus estimation $x^\star$ as $x_t$.

\vspace{-0.25em}
\subsection{Proposed Algorithm}
\vspace{-0.25em}
\begin{figure}[t]
\begin{algorithm}[H]
\caption{Social Bayesian Optimisation (SBO)}
\label{alg:SBO}
\begin{algorithmic}[1]
\normalsize
\State \textbf{Input}: decay rate $0 < q < 1$
\State Set $\mathcal{Q}^u_0=\emptyset$, $\mathcal{Q}^v_0=\emptyset$, $\mathcal{B}^{u,A,v}_1 = \mathcal{B}^{u,A,v}$, and draw the initial point $x_0 \in \mathcal{X}$.

\For{$t\in[T]$}
\State Solve $x_t = \argmax_{x \in \mathcal{X}} \alpha(x, x_{t-1})$
\label{alg_line:query} 

\State Query public vote $\textbf{1}_t^v$ on $(x_t, x_{t-1})$.
\State Update $D_{\mathcal{Q}_t^v} = D_{\mathcal{Q}_{t-1}^v} \cup (x_t, x_{t-1}, \textbf{1}_t^v)$ and models. 

\If{$w^u_t(x_t, x_{t-1})\geq \max \left\{ \frac{1}{t^{q}}, \,\, w^v_t(x_t, x_{t-1}) \right\}$} 
\label{alg_line:handover}
\State Query private vote $\textbf{1}_t^u$ on $(x_t, x_{t-1})$.
\State Update $D_{\mathcal{Q}_t^u} = D_{\mathcal{Q}_{t-1}^u} \cup (x_t, x_{t-1}, \textbf{1}_t^u)$. 
\EndIf
\EndFor
\end{algorithmic}
\end{algorithm}
\vspace{-2em}
\end{figure}
\begin{table*}[!hb]
    \centering
    \caption{Theoretical bound for cumulative regret $R_T$ and sample complexity $\lvert \mathcal{Q}^u_t \rvert$.}
    \label{tab:bounds}
    \begin{tabular}{llcc}
    \toprule
    \multicolumn{2}{l}{Voting scheme and Assumptions} & cumulative regret $R_T$ & sample complexity of private votes $|\mathcal{Q}^u_t|$ \\ 
    \midrule
    Private only & - & $\mathcal{O} \left(L_\mathcal{A} \sqrt{\beta_T^u \gamma_{\mathcal{Q}_T}^{uu^\prime} T}
    \right)$ & $T$ \\
    Oracle (Public only) & (a)(b)(c) & $\mathcal{O} \left(n L_\mathcal{A} \sqrt{\beta_T^v \gamma_{\mathcal{Q}_T}^{vv^\prime} T}
    \right)$ & 0 \\
    SBO (dual voting) &(c) & $\mathcal{O}\left(
     L_\mathcal{A} T^{1-\frac{q}{4}} +
     L_\mathcal{A} \sqrt{(\beta_T^u \gamma^{uu^\prime}_T + \beta_T^v \gamma^{vv^\prime}_T) T} 
    \right)$ & $\mathcal{O}\left(
      T^{q} \left( \gamma^{vv^\prime}_T \right)^{2} \log\frac{T \mathcal{N}(\mathcal{B}^u, \nicefrac{1}{T},\lVert \cdot \rVert_\infty)}{\delta}
      \right)$ \\
    \bottomrule
    \end{tabular}
\end{table*}
We first provide an overview of our SBO Algorithm~\ref{alg:SBO} before detailing each component. Recall our goal is to reach consensus $x^\star$ with as few voting iterations $t$ as possible while minimizing the number of expensive private votes $\mathcal{Q}^u_t$, using them only when necessary. Line~\ref{alg_line:query} selects the next vote $(x_t, x_{t-1})$ by maximizing the acquisition function $\alpha$. Line~\ref{alg_line:handover} defines the stopping criterion, which terminates expensive private queries once the graph $A$ is accurately estimated.

\textbf{Acquisition function.} 
We propose an optimistic algorithm, akin to GP-UCB \citep{srinivas2010gaussian}:
\begin{align}\label{prob:ucb_acquisition}
    \alpha(x_t, x_{t-1}) = \max_{\tilde{u} \in \mathcal{B}^{u}_t}  \mathcal{A} [\tilde{u}(x_t,:)] - \mathcal{A}[\tilde{u}(x_{t-1},:)].
\end{align}
Intuitively, this acquisition function operates similarly to the expected improvement approach \citep{mockus1974bayesian, osborne2008gaussian}, aiming to identify the point with the highest potential improvement over previous evaluations at $t-1$. The term $\max_{\tilde{u} \in \mathcal{B}^u_t}$ represents the maximum utility within the confidence set, corresponding to the upper confidence bound of this improvement.

Iteratively maximizing $\alpha$ as in Line~\ref{alg_line:query} ensures that the algorithm converges to the true consensus, i.e., $\text{lim}_{t \to \infty} x_t = x^\star$, with probability at least $1 - \delta$, similar to GP-UCB. Intuitively, at each time step $t$, the estimated utility function $\tilde{u} \in \mathcal{B}^u_t$ contains the ground truth function $u$ with probability at least $1 - \delta$ (see Lemma~\ref{lemma:conf_set} for the proof). Hence, we can define a confidence interval for the consensus estimate: $\max_{x \in \mathcal{X}} \min_{\tilde{u} \in \mathcal{B}^u_t} \mathcal{A}[\tilde{u}(x,:)] \leq \mathcal{A}[u(x^\star,:)] \leq \max_{x \in \mathcal{X}} \max_{\tilde{u} \in \mathcal{B}^u_t} \mathcal{A}[\tilde{u}(x,:)]$. As more data $D^u_t$ and $D^v_t$ are observed, this confidence bound tightens over time, and its upper bound asymptotically approaches the true consensus $x^\star$. While our setting is more complex due to the pairwise comparison structure $(x, x^\prime)$, setting $x^\prime = x_{t-1}$ simplifies the extension.

\textbf{Stopping criterion.} 
As a stopping criterion for determining whether additional expensive private votes are necessary, we use uncertainty, specifically the confidence interval of pairwise comparisons:
\begin{equation}
\begin{aligned}
   w^u_t(x_t, x_t^\prime)
   =\sup_{\tilde{u}, \tilde{u}^\prime \in\mathcal{B}_t^{v}} \lVert &\tilde{u}(x_t,:)- \tilde{u}(x_{t-1},:)\\
   &- \left( \tilde{u}^\prime(x_t,:)-\tilde{u}^\prime(x_{t-1},:) \right) \rVert,
\end{aligned}\label{prob:projection}
\end{equation}
\normalsize
where graph $A$ is more confidently estimated at the point $(x_t, x_{t-1})$, indicating that additional private votes are no longer required. The decay rate $q$ controls the threshold for meeting this stopping condition.

\vspace{-0.25em}
\subsection{Theoretical Analyses}\label{sec:theory}
\vspace{-0.25em}
We now analyze the theoretical bounds of our SBO algorithm to understand its guarantees and limitations. 

\textbf{Cumulative regret and cost.} We define two performance metrics: cumulative regret, $R_T := \sum_{t=1}^T (\mathcal{A}[u(x^*,:)] - \mathcal{A}[u(x_t,:)])$, and the sample complexity of private votes, $|\mathcal{Q}_T^u|$. The average of cumulative regret, $\nicefrac{R_T}{T}$, represents the mean error of the consensus estimate. \emph{No-regret} is defined as $\lim_{T \to \infty} \nicefrac{R_T}{T} = 0$, ensuring that our algorithm asymptotically converges to the true consensus $x^\star$ (i.e., the estimation error vanishes over time). The sample complexity $|\mathcal{Q}_T^u|$ quantifies the cost-effectiveness of the algorithm, measuring the number of costly private votes required to achieve no-regret.

\textbf{Graph properties.} The regret bound is affected by the following properties of social influence graph $A$:
\begin{compactenum}
    \item[(a)] \textbf{Given}: We know $A$, otherwise $A$ is unknown a priori and we need to estimate it.
    \item[(b)] \textbf{Invertible}: There exists $A^{-1}$,  otherwise $A$ is singular.
    \item[(c)] \textbf{Identifiable}: $A$ is identifiable from the votes, otherwise dual voting cannot identify true $A$.
\end{compactenum}

\begin{theorem}[\textbf{Regret bound}]\label{thm:regret}
    Under Assumptions~\ref{assump:pairwise} to \ref{assump:lipchitz}, Algorithm~\ref{alg:SBO} satisfies the bound described in Table~\ref{tab:bounds} with probability at least $1-\delta$, where 
    $L_\mathcal{A} := \sqrt{n} (\lVert \textbf{w} \rVert + 2 L_u C_L)$, 
    $\gamma_{\mathcal{Q}_T}^{vv^\prime}, \gamma_{\mathcal{Q}_T}^{uu^\prime}$ are maximum information gain for the corresponding kernel, $0 < q < 1$ is a user-defined parameter. 
\end{theorem}
\begin{table*}[!hb]
   \centering
   \caption{Kernel-specific bounds for main algorithm where $\nu>\nicefrac{d}{4}(3+d+\sqrt{d^2+14d+17})=\Theta(d^2)$.}
    \label{tab:kern_spec_bounds}
    \small
    \setlength\tabcolsep{1.0pt} 
    \resizebox{1\textwidth}{!}{
    \begin{tabular}{lccc}
    \toprule
         Metric & Linear &  RBF & Mat\'ern  \\ 
    \midrule
    $R_{T}$ & $\mathcal{O}\left(
    T^{1-\frac{q}{4}} + T^{\frac{3}{4}} (\log T)^{\frac{3}{4}}
    \right)$ & $\mathcal{O}\left(
    T^{1-\frac{q}{4}} + T^{\frac{3}{4}} (\log T)^{\frac{3}{4} (d+1)}
    \right)$ & $\mathcal{O}\left(
    T^{1-\frac{q}{4}} + T^{\frac{d}{4\nu} + \frac{d(d+1)}{4\nu + 2d(d+1)}} (\log T)^{\frac{3}{4}}
    \right)$ \\ 
    $|\mathcal{Q}^u_T|$  & $\mathcal{O}\left( T^{q}(\log T)^{3}\right)$ & $\mathcal{O}\left(T^{q}(\log T)^{3(d+1)}\right)$ & $\mathcal{O}\left( T^{q + \frac{d}{\nu} \frac{2d(d+1)}{2\nu + d(d+1)}} (\log T)^{3}\right)$ \\
    \bottomrule
    \end{tabular}
    }
    \normalsize
    \vspace{-1.5em}
\end{table*}
Appendix~\ref{proof:regret} provides the proof and more details. 

\textbf{Interpretation.} 
We compare three representative cases in Table~\ref{tab:bounds}. The first row represents the scenario where only clean private votes are used, without incorporating noisy public votes. Since all votes are clean, social influence is no longer a concern, resulting in the best possible regret bound. However, the sample complexity $\lvert \mathcal{Q}_T^u \rvert$ is the worst, exhibiting linear dependence on $T$. This implies that as $T \to \infty$, an infinite number of private votes would be required to achieve no-regret. 
The second row describes the best-case scenario for our algorithm, where only public votes are queried, leading to $\lvert \mathcal{Q}_T^u \rvert = 0$, making it the most cost-effective approach. This case assumes that the exact graph $A$ is known and invertible, allowing for a perfect estimation of the social-influence-free utility: $\hat{u}(\cdot,:)^\top = A^{-1} \hat{v}(\cdot, :)^\top$. Here, the only source of estimation error is the noise in the utility estimator $\hat{v}$. Even in this ideal case, an additional factor of $n$ appears compared to the private-votes-only scenario, due to the amplification of estimation noise by $\lVert A^{-1} \rVert \leq n$ in the worst case (see Lemma~\ref{lemma:matrix_norm_bound}). 
The third row presents a more practical setting, where the graph $A$ is unknown and may be non-invertible, but is at least identifiable from votes. While the regret bound includes an additional term, the algorithm remains no-regret. More importantly, the sample complexity $\lvert \mathcal{Q}_T^u \rvert$ is also no-regret, i.e., $\lim_{t \to \infty} \nicefrac{\lvert \mathcal{Q}^u_t \rvert}{t} = 0$. This ensures that, asymptotically, SBO stops querying private votes. 

For all cases, the aggregation function \(\mathcal{A}\) affects \(R_T\), with egalitarian being the slowest (\(L_\mathcal{A} \rightarrow \sqrt{n}\)) and utilitarian the fastest (\(L_\mathcal{A} = 1\)). \(R_T\) is sublinear to the number of agents. i.e., $\sqrt{n}$. The dimension \(d\) scales similarly to standard BO, techniques like additive kernels~\citep{kandasamy2015high} can improve dimensional scalability.

\textbf{Kernel specific bounds.}
To further analyze our approach, we derive kernel-specific bounds for SBO in Table~\ref{tab:kern_spec_bounds}, based on maximum information gain bounds~\citep{srinivas2012information, vakili2021information} and covering number bounds~\citep{wu2017lecture, xu2024lower, bull2011convergence, zhou2002covering}. Within the range $0 < q < 1$, we confirm the no-regret property for both $R_T$ and $\lvert \mathcal{Q}_T^u \rvert$ across these popular kernels, as all $T$-dependent terms remain sublinear. This implies that our SBO is \emph{groupthink-proof}, ensuring no-regret performance regardless of $A$ (as long as it is identifiable). We also observe a trade-off in the decay parameter $q$: a larger $q$ accelerates the convergence of $R_T$ but increases the required number of private queries, $\lvert \mathcal{Q}_T^u \rvert$. This trade-off is reasonable, as later iterations must infer consensus from corrupted public votes. Thus, 
$q$ can be interpreted as a necessary compromise for handling graph estimation error, aligning with the impossibility theorem in Section~\ref{sec:impossibility}. 
Importantly, under typical conditions ($T \leq 1000, d \geq 2$, RBF kernel), the $q$-dependent term in $R_T$ is always smaller than the non-$q$-dependent terms, suggesting that in practice, the convergence rate remains comparable.

\textbf{Graph identification bound.} 
We further analyzed the asymptotic convergence rate of graph $A$ and utility function estimation errors using the minimum excess risk (MER) approach~\citep{xu2022minimum}.
\begin{theorem}[\textbf{Graph identification}]\label{thm:asymptotic_conv}
    Asymptotic convergence rates of the estimation errors are
    \small
    \begin{align*}
        &\lVert \tilde{A}_t - A \rVert \leq \mathcal{O}\left( 2^{-1/2} |\mathcal{Q}^u_t|^{-1/2} \right),\\
        &\lvert \tilde{u}_t(x_t) - \tilde{u}_t(x_{t-1}) -(u(x_t) - u(x_{t-1})) \rvert \leq \mathcal{O}\left( L_{k, \mathcal{Q}^u_t}^{1/4} |\mathcal{Q}^u_t|^{-1/4} \right),
    \end{align*}
    \normalsize
    for $\max_{i \in V} \kappa_i = 1 + \nicefrac{\delta_A^2}{n^2} - 2 \xi \delta_A^2$, $\kappa_i > 1$, $\xi < \nicefrac{1}{2n^2}$ and $\tilde{A}_t, \tilde{u}_t \in \mathcal{B}_t^{u, A, v}$ defined in Lemma~\ref{lemma:conf_set}.
\end{theorem}
Appendix~\ref{proof:asymptotic_conv} provides the proof and more details. Theorem~\ref{thm:asymptotic_conv} shows the graph identification error converges faster than pointwise utility estimation error in the worst-case scenarios. Intuitively, the estimation error for the linear model (\(A\)) converges faster than for the black-box non-linear functions (\(u, v\)). Thus, we can stop querying private votes \(|\mathcal{Q}^u_t|\) once we reach sufficient confidence. This results also suggests $q = \nicefrac{1}{2}$ can be optimal as the cumulative error of $\sum_{i \in [T]} t^{-1/2} \leq \mathcal{O}(T^{1/2})$ to match the rate of $|\mathcal{Q}^u_t|$. Based on this analysis, we set $q = 1/2$ throughout the experiments.

\vspace{-0.5em}
\section{Related Works}
\vspace{-0.5em}

\textbf{BO for black-box games.} 
In the BO community, the game-theoretic approach has been researched as the \emph{application} to multi-objective BO (MOBO; \citet{hernandez2016predictive, daulton2020differentiable}). Direct consideration of the multi-agentic scenario is limited, which typically assumes the specific aggregation rule (Nash equilibrium; \citet{al2018approximating, picheny2019bayesian, han2024noregret}, Kalai-Smorodinsky solution; \citet{binois2020kalai}) or Chebyshev scalarisation function \citep{astudillo2024preferential}, which requires discrete domain to ensure the existence of solution. Ours is the \emph{first-of-its-kind} principled work that addresses the social-influence issues on continuous domain.

\textbf{Preferential BO.} 
Preferential BO is a single-agentic preference maximisation algorithms \citep{gonzalez2017preferential, astudillo2023qeubo, xu2024principled}, extended to diverse scenarios; choice data \citep{benavoli2023bayesian}, top-$k$ ranking \citep{nguyen2021top}, preference over objectives on MOBO \citep{abdolshah2019multi, ozaki2024multi}, human-AI collaboration \citep{adachi2024looping, xu2025principled}. Our work is the first to study the multi-agentic social influence, and is orthogonal to these works.

\textbf{Other BO.} 
Multitask BO \citep{kandasamy2016gaussian} addresses scenarios where cheap but lower-fidelity information is available. However, \citet{mikkola2023multi} showed that this approach can converge significantly more slowly than standard BO if the low-fidelity information is unreliable, as is the case in our setting. While cost-aware BO \citep{lee2020cost} accounts for location-dependent cost functions, our public votes do not exhibit such dependency.
\vspace{-0.5em}
\section{Experiments}\label{sec:experiment}
\vspace{-0.5em} 
\begin{figure*}
    \centering
    \includegraphics[width=0.8\linewidth]{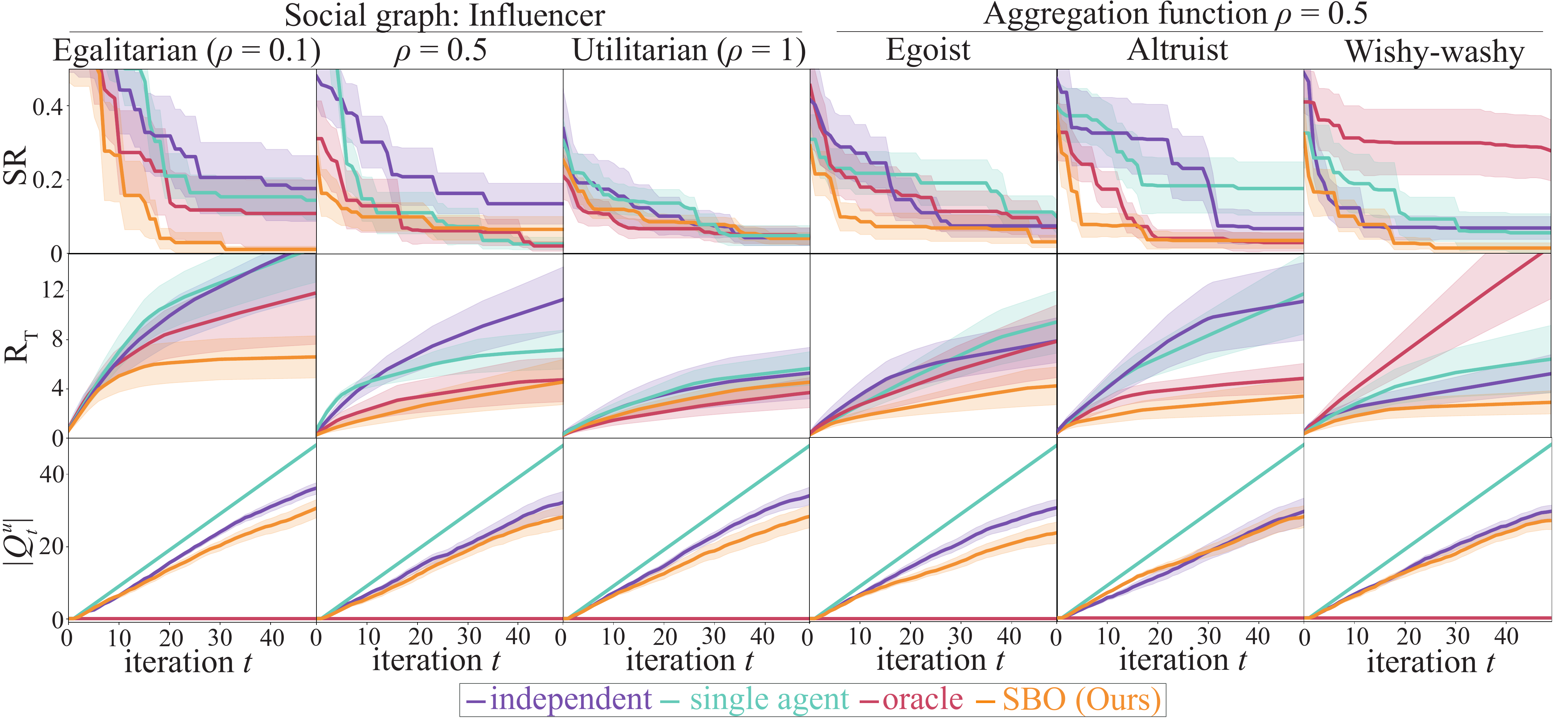}
    \caption{\small{Robustness analysis was conducted using the function shown in Fig.~\ref{fig:visual_exp}. The lines and shaded areas represent the mean $\pm$ 1 standard error. The cumulative regret $R_T$ reaches a plateau, confirming the no-regret property, while the cumulative queries $|\mathcal{Q}^u_t|$ demonstrate sublinear convergence.}}
    \label{fig:social_influence}
  \vspace{-1.5em}
\end{figure*}
\begin{figure}
    \centering
    \includegraphics[width=1\linewidth]{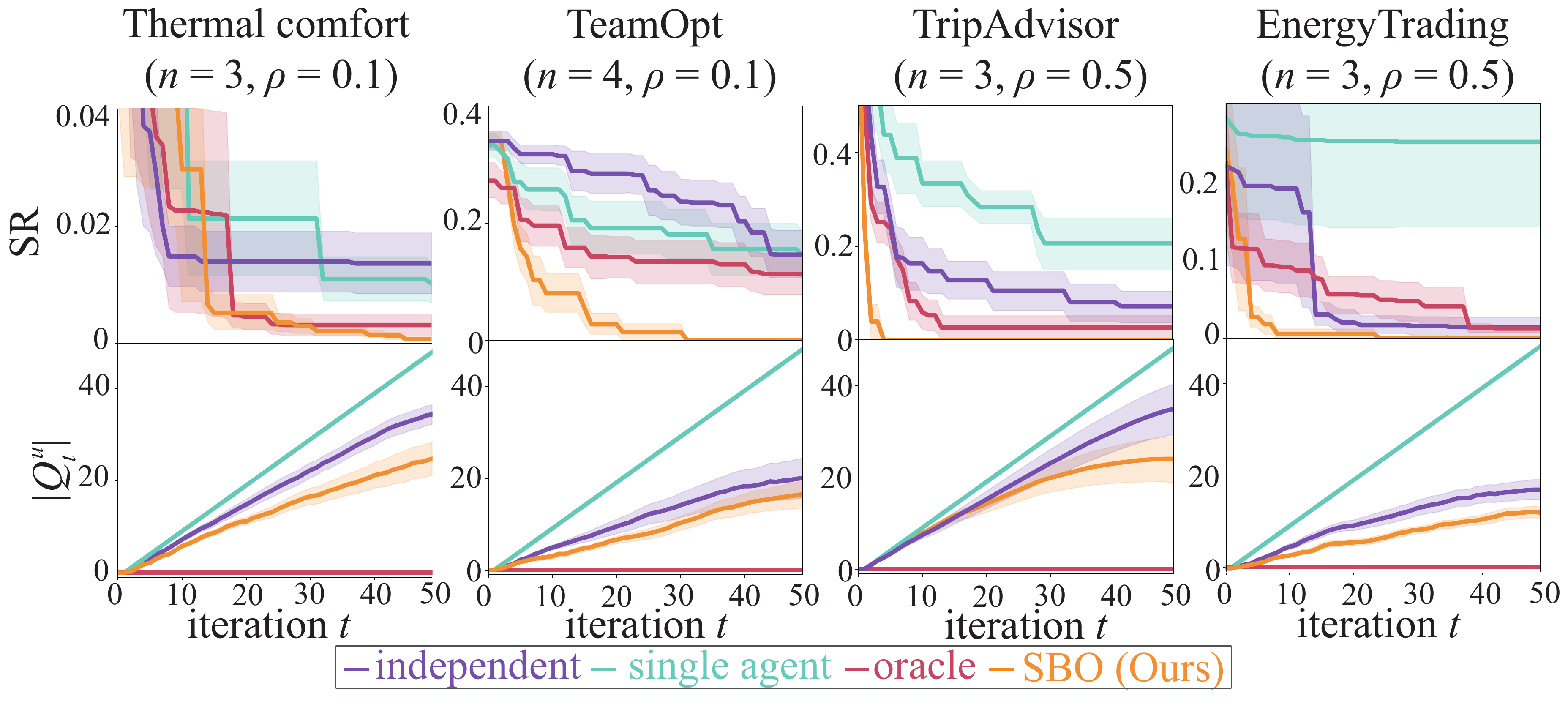}
    \caption{\small{Real-world experiments with varying number of agents $n$ and aggregation rule $\rho$}}
    \label{fig:real-world}
    \vspace{-1.5em}
\end{figure}
Since our problem setting—optimization with preference feedback under social influence—is novel, we benchmark our proposed algorithm against simpler versions of our method by systematically removing one design choice at a time: \textbf{(a) Oracle}: $A$ is known, thus only $v$ is queried; \textbf{(b) Single Agent}: Ignores agent heterogeneity, performing single-agent preference maximization (reduced to standard preferential BO); \textbf{(c) Independent}: Assumes $u$ and $v$ are completely independent; \textbf{(d) SBO (ours)}: Assumes $v = Au$ with the graph prior $p(A)$. See Appendix~\ref{app:exp} for experimental setup details for reproducibility. Along with cumulative regret and query counts, we also report simple regret, defined as $\text{SR}_{t} := \min_{\tau \in [T]}(\mathcal{S}(x^\star) - \mathcal{S}(x_\tau))$.

\textbf{Robustness to aggregate function and social graph.} 
We first evaluate the robustness of our algorithm against variations in the aggregation rule $\rho$ and social graph $A$. The results support Theorem~\ref{thm:regret}, showing that the egalitarian rule leads to the slowest convergence, while the utilitarian rule achieves the fastest. Interestingly, in the egalitarian case, our SBO model outperforms the oracle model. This is because, unlike the oracle model—which relies solely on public votes—our approach leverages both public and private votes, incorporating more data and improving early-stage predictions. 
Similarly, the oracle model performs the worst in the wishy-washy case due to the non-invertible $A$, making it impossible to identify $u = A^{-1} v$. In the altruist case, where a selfless voter causes public votes to be uniformly corrupted, the problem ironically becomes one of the most challenging for the single-agent baseline, which assumes homogeneity. In contrast, our SBO model remains robust regardless of the structure of graph $A$, as its minimal reliance on strong assumptions enhances its adaptivity.

\textbf{Real-world tasks.} 
We introduce four new real-world collective decision-making tasks to evaluate our algorithm. See Appendix~\ref{app:exp} for details on the experimental setup and the newly introduced conditions. Fig.~\ref{fig:real-world} shows that our SBO consistently outperforms other baselines, while the sample complexity $|\mathcal{Q}^u_t|$ remains the lowest among all baselines (except for the oracle model). 
Additional experiments on computational efficiency and Gaussian process applications are provided in Appendices~\ref{app:gp_exp} and~\ref{app:compute_time}.

\vspace{-1em}
\section{Conclusion and Limitations}
\vspace{-1em}
We investigated the impact of social influence, a prevalent yet underexplored cognitive bias in collective decision-making. To overcome the impossibility of groupthink-proof aggregation, we proposed a dual voting mechanism and developed a learning algorithm for social influence graphs using optimistic MAP. This approach accelerated social-influence-free consensus-building across 10 tasks. Our approach generalizes to any positive linear combination of aggregation functions and apply to both provable likelihood-ratio models and popular GP-based approaches.

While our algorithm is the \emph{first-of-its-kind} to provide a general theoretical guarantee in the social-influence setting, it shares limitations common to optimistic algorithms like GP-UCB~\citep{srinivas2010gaussian}, particularly in high-dimensional problems. Additionally, our current framework requires the aggregation function to be specified in advance, assuming stationarity and homogeneity. Extending to heterogeneous and dynamic settings is an important direction for future research, where the probabilistic choice function approach~\citep{benavoli2023learning} offers a promising pathway. While we framed social influence as a bias to eliminate, positive social influence—such as debiasing confusion or leveraging nudge theory~\citep{thaler2008nudge} to guide behavior—can also be beneficial. Since our algorithm is symmetric with respect to $u$ and $v$, this inverse approach is feasible and represents another promising direction for future research, as discussed further in Appendix~\ref{app:positive}.

\bibliography{uai2025-template}

\newpage
\onecolumn
\addcontentsline{toc}{section}{Appendix} 
\part{Appendix} 
\parttoc 

\appendix
\section{Preliminary}
\subsection{Table of notations}\label{sec:notation}
\begin{table}[H]
    \centering
    \caption{Notations and Descriptions (Part I)}
    \begin{tabular}{llll}
    \toprule
 Category & Symbol & Description & Reference\\ 
 \midrule
 \multirow{5}{*}{Domain} & $x$& Option & Eq.~(\ref{objective: SBO})\\ 
 &$x^*$& Consensus (global optimum) & Eq.~(\ref{objective: SBO})\\
 &$x_t$& Queried option at $t$-th step & Eq.~(\ref{objective: SBO})\\
 &$\mathcal{X} \in \mathbb{R}^d$& (continuous) domain & Eq.~(\ref{objective: SBO})\\ 
 &$d$& Number of dimensions & Eq.~(\ref{objective: SBO})\\ 
 \midrule
 \multirow{16}{*}{Utility} &$n$& Number of agents & Eq.~(\ref{objective: SBO})\\
 & $V$ & Set of $n$ agents & Eq.~(\ref{objective: SBO})\\
 & $i$ & Index of agents & Eq.~(\ref{objective: SBO})\\
 &$u(x,i)$& Truthful utility for the $i$-th agent. & Eq.~(\ref{objective: SBO})\\ 
 &$v(x,i)$& Non-truthful utility for the $i$-th agent. & Eq.~(\ref{objective: SBO})\\
 &$u(x,:)$ & Utilities of all agents & Eq.~(\ref{objective: SBO})\\
 &$\succ_{u(\cdot,i)}$ & Preference of agent $i$ induced by utility $u(\cdot,i)$ & Assumption~\ref{assump:pairwise}\\
 &$\succ_{\mathcal{S}}$ & Preference of group induced by social utility $\mathcal{S}$ & Definition~\ref{def:dictatorship}\\
 &$p(u), p(v)$ & Prior over utilities (uniform prior) & Section~\ref{sec:dual-voting}\\
 &$\hat{u}, \hat{v}$& MAP estimate of utilities. & Lemma~\ref{lemma:conf_set}\\
 &$v(x,i)$& Non-truthful utility for the $i$-th agent. & Eq.~(\ref{objective: SBO})\\
 &$u(x,:)$ & Utilities of all agents & Eq.~(\ref{objective: SBO})\\
 &$\tilde{u}, \tilde{v}$& Utility function sample from confidence set. & Assumption~\ref{assump:bounded_norm}\\ 
 &$\hat{u}_t, \hat{v}_t$& MAP estimated utility function at $t$-th step. & Assumption~\ref{assump:bounded_norm}\\ 
 &$\underline{u}_t, \underline{v}_t$& The lower confidence bound of utility. & Remark~\ref{remark:confidence_bound}\\ 
 &$\overline{u}_t, \overline{v}_t$& The upper confidence bound of utility. & Remark~\ref{remark:confidence_bound}\\ 
 \midrule
 \multirow{4}{*}{Likelihood}
 &$\sigma$& Sigmoid function & Assumption~\ref{assump:pairwise}\\
 &$\ell_t(u), \ell_t(v)$& Log Likelihood (LL) function & Corollary~\ref{corollary:pref_likelihood}\\
 &$\hat{\ell}^u_t, \hat{\ell}^v_t$& MLE estimate of LL values& Lemma~\ref{lemma:conf_set}\\
 &$\mathcal{L}_t(u,A,v)$& Unnormalized negative log posterior & Appendix~\ref{app:bayesian}\\
 \midrule
 \multirow{5}{*}{\shortstack[l]{Aggregation\\function}}
 &$\mathcal{A}$& Aggregation function & Definition~\ref{def:aggregate}\\
 &$\mathcal{S}$& Social utility& Definition~\ref{def:aggregate}\\
 &$\textbf{w}$ & Weight function of $\mathcal{A}$ & Proposition~\ref{proof:gsf}\\
 &$\rho$ & GSF interpolation parameter & Proposition~\ref{proof:gsf}\\
 &$\phi$ & Sorting function & Proposition~\ref{proof:gsf}\\
 \midrule
 \multirow{11}{*}{RKHS}
 &$k_i(x, x^\prime)$&Kernel of $i$-th agent's utility & Assumption~\ref{assump:bounded_norm}\\
 &$\mathcal{H}_{k_i}$&RKHS corresponding to the kernel $k_i$ & Assumption~\ref{assump:bounded_norm}\\
 &$||\cdot||_{k_i}$ & Norm induced by inner product in RKHS $\mathcal{H}_{k_i}$ & Assumption~\ref{assump:bounded_norm}\\ 
 &$L^v$&Isotropic norm bound of $\mathcal{H}_{k_i}$ & Assumption~\ref{assump:bounded_norm}\\
 &$\mathcal{B}^{v_i}, \mathcal{B}^{u_i}$ & Set of utility functions of $i$-th agent & Assumption~\ref{assump:bounded_norm}\\
 &$\mathcal{B}^{v}, \mathcal{B}^{u}$ & Superset of utility functions of all agent & Assumption~\ref{assump:bounded_norm}\\
 &$\mathcal{B}^{v}_t, \mathcal{B}^{u}_t$ & Superset of MLE-estimated utility functions at $t$-th step & Lemma~\ref{lemma:conf_set}\\
 &$\mathcal{B}^{u,A,v}_t$ & Superset of MAP-estimated utility functions at $t$-th step & Lemma~\ref{lemma:conf_set}\\
 &$\gamma^u_t, \gamma^v_t, \gamma^{uu}_t, \gamma^{vv}_t$ & Maximum information gain for corresponding kernels & Lemma~\ref{lemma:conf_set}\\
 &$L_{k, t}$ & Kernel specific term & Theorem~\ref{thm:asymptotic_conv}\\
 &$\nu$ & Matérn kernel smoothness paramter & Table~\ref{tab:kern_spec_bounds}\\
 \midrule
 \multirow{6}{*}{\shortstack[l]{Confidence\\set}}
 &$\mathcal{F}_t$&Filtration at the step $t$ & Lemma~\ref{lemma:conf_set}\\
 &$\delta$ & Probability that $\mathcal{B}^{u,A,v}_t$ does not contain $u,A,v$ & Lemma~\ref{lemma:conf_set}\\
 &$\epsilon$ &Radius of the function space ball $L_{k_i}$ & Lemma~\ref{lemma:conf_set}\\
 &$\mathcal{N}(\mathcal{B}^u, \epsilon, \lVert \cdot \rVert_\infty)$ & Covering number of the set $\mathcal{B}^u$ & Lemma~\ref{lemma:conf_set}\\
 &$\beta^u_t, \beta^v_t$ & MLE-based confidence set bound parameter & Lemma~\ref{lemma:conf_set}\\
 &$\beta_t$ & MAP-based confidence set bound parameter  & Lemma~\ref{lemma:conf_set}\\
 \bottomrule
\end{tabular}
\label{tab:notation1}
\end{table}

\begin{table}[t]
    \centering
    \caption{Notations and Descriptions Part II}
    \begin{tabular}{llll}
    \toprule
 Category & Symbol & Description & Reference\\ 
 \midrule
 \multirow{11}{*}{Graph}
 &$E$&The set of weighted directed edges. & Definition~\ref{def:bandwagon}\\
 &$G=(V,E)$&The social-influence graph & Definition~\ref{def:bandwagon}\\
 &$A$&The adjacency matrix of $G$. & Definition~\ref{def:bandwagon}\\
 &$N_G(i)$&The (in)-neighbour of $i$ in $G$. & Definition~\ref{def:bandwagon}\\
 &$p(A)$& Graph prior & Eq.~(\ref{eq:graph_prior})\\
 &$\xi$& Tiknohov parameter (invertibility regularizer) & Eq.~(\ref{eq:graph_prior})\\
 &$\kappa_i$& Dirichlet concentration parameter of $i$-th row & Eq.~(\ref{eq:graph_prior})\\
 &$\delta_A$& The smallest element of $A$ & Eq.~(\ref{eq:graph_prior})\\
 &$Z$& Normalising constant of prior $p(A)$ & Eq.~(\ref{eq:graph_prior})\\
 &$\tilde{A}$& graph sample from confidence set. & Assumption~\ref{assump:bounded_norm}\\ 
 &$\hat{A}_t$& MAP estimated graph. & Assumption~\ref{assump:bounded_norm}\\ 
 \midrule
 \multirow{8}{*}{Queries}
 &$t$&The step of iteration & Assumption~\ref{assump:votes}\\
 &$T$&The running horizon & Assumption~\ref{assump:votes}\\
 &$\mathcal{Q}^v_t$& public queries & Assumption~\ref{assump:votes}\\
 &$\mathcal{Q}^u_t$& private queries & Assumption~\ref{assump:votes}\\
 &$\mathcal{Q}^{uv}_t$& combined set of private and public queries & Lemma~\ref{lemma:conf_set}\\
 &$D_{\mathcal{Q}^v_t}$ & Public vote & Assumption~\ref{assump:votes}\\
 &$D_{\mathcal{Q}^u_t}$ & Private vote & Assumption~\ref{assump:votes}\\
 &$\lambda_u$, $\lambda_v$ & Query costs of private and public votes & Assumption~\ref{assump:votes}\\
 \midrule
 \multirow{6}{*}{Algorithm}
 &$\alpha$ & Acquisition function & Eq.~(\ref{prob:ucb_acquisition})\\
 &$w^u_t, w^v_t$ & Projection weight function & Eq.~(\ref{prob:projection})\\
 &$q$ & Decay rate & Algorithm~\ref{alg:SBO}\\
 &$R_T$ & Cumulative regret & Theorem~\ref{thm:regret}\\
 & SR & Simple regret & Section~\ref{sec:experiment}\\
 &$L_\mathcal{A}$ & Regret constant from aggregate function & Theorem~\ref{thm:regret}\\
 \bottomrule
\end{tabular}
\label{tab:notation2}
\end{table}

\subsection{Hyperparmater list}
\begin{table}[hbt!]
    \centering
    \caption{The complete list of hyperparameters and their settings.}
    \label{tab:hypers}
    \resizebox{1\textwidth}{!}{
    \begin{tabular}{lccccccc}
    \toprule
         hyperparameters &
         initial value&
         data-driven optimisation?&
         tuning method
         \\
    \midrule
         kernel hyperparamters & GPyTorch default & \ding{51} & the method in Appendix~\ref{app:hypers_update}\\
         $\gamma_{T}^u, \gamma_{T}^v$ in Theorem~\ref{thm:regret} & -- & \ding{51} & algorithm using \cite{hong2023optimization}\\
         $\delta_A$ in Eq.~\ref{eq:graph_prior} & $0.01$ & fixed & --\\
         $\xi$ in Eq.~\ref{eq:graph_prior} & $\nicefrac{1}{2 \delta_A^2 n^2}$ & fixed & --\\
         $\kappa_i$ in Eq.~\ref{eq:graph_prior} & $1 + \nicefrac{1}{n^2} (2 \delta^2 - 1)$ & fixed & --\\
         $L_v$ in Assumption~\ref{assump:bounded_norm} & 1.5 & \ding{51} & the method in Appendix~\ref{app:hypers_update}\\
         $\beta_t, \beta_t^u, \beta_t^v$ in Lemma~\ref{lemma:conf_set} & 0.5 & \ding{51} & the method in Appendix~\ref{app:hypers_update}\\
         $q$ in line.~\ref{alg_line:handover} in Alg.~\ref{alg:SBO} & 0.5 & fixed & --\\
    \bottomrule
    \end{tabular}
    }
\end{table}

\subsection{Definitions}
At first, we formally define the definitions we introduced:
\begin{definition}[\textbf{Invertibility}]
    The graph matrix $A$ is invertible (also known as nonsingular or nondegenerate) if there exists a square matrix $B$ such that
    \begin{equation}
        A B = B A = I,
    \end{equation}
    and we denote such $B$ as $A^{-1}$.
\end{definition}
\begin{definition}[\textbf{Identifiability}]
    Let $\mathcal{M} = \{\mathcal{M}_A: v(x,:) = A u(x,:) \mid A \in \mathbb{R}^{n \times n}\}$ be a statistical model of social influence. $\mathcal{M}$ is identifiable if the mapping $A \mapsto \mathcal{M}_A$ is one-to-one:
    \begin{equation}
        \mathcal{M}_{A_1} = \mathcal{M}_{A_2} \Rightarrow A_1 = A_2, \quad \forall A_1, A_2 \in \mathbb{R}^{n \times n}.
    \end{equation}
\end{definition}
In an unidentifiable case, the above relationship is not one-to-one. Such cases can be found. For instance, if $v_i=u_j=u,\forall i, j$, then matrix $A$ can be any matrix with $\sum_jA_{ij}=1$. This assumption excludes such cases.
\begin{definition}[\textbf{Strong convexity}]\label{defn:convex}
    A function $h(x)$ is said to be strongly convex with parameter $m > 0$ if, for all matrices $A, B \in \mathbb{R}^{n \times n}$, the following inequality holds:
    \begin{equation}
        h(B) \geq  h(A) + \nabla h(A)^\top (B - A) + \frac{m}{2} \lVert B - A \rVert^2,
    \end{equation}
    where $\nabla h(A)$ denotes the gradient of $h$ at $A$.
\end{definition}
\begin{definition}[\textbf{Monotonicity}]\label{defn:monotonic}
    For any $t \in [T], i \in V, x, x^\prime \in \mathcal{X}$, a social utility $\mathcal{S}$ is monotonic, $\mathcal{S}(x) > \mathcal{S}(x^\prime)$ if:
    \begin{align*}
        (u(x, i) > u(x^\prime, i)) \wedge (\forall j \in V \backslash \{i\} \,\, u(x, j) = u(x^\prime, j)).
    \end{align*}
\end{definition}
Intuitively, a social utility $\mathcal{S}(x)$ should increase if any individual utility $u_t(x, i)$ improves. Monotonicity guarantees the maximising the social utility leads to the improvement of individual utilities.
\begin{definition}[\textbf{Pigou-Dalton principle (PDP)}]\label{defn:pdp}
     An social utility $\mathcal{S}$ satisfies $\mathcal{S}(x) > \mathcal{S}(x^\prime)$ if:
     
     1. \textbf{In a strong sense}, $u(x,:) \succ u(x^\prime,:)$ for $x \in \mathcal{X}$ whenever there exist $i,j \in V$ such that (a) $\forall k \in V\backslash\{i,j\} \,\, u(x,k) = u(x^\prime,k)$, (b) $u(x,i) + u(x,j) = u(x^\prime,i) + u(x^\prime,j)$, and (c) $|u(x^\prime,i) - u(x^\prime,j)| > |u(x,i) - u(x,j)|$.\\
     2. \textbf{In a weak sense}, we only require that $u(x,:) \succeq u(x^\prime,:)$.
\end{definition}
Intuitively, given a utility vector $u(x,:)$, if an agent with a higher utility transfers $\leq \nicefrac{1}{2}$ of its excess utility to another worse-off agent, the aggregate function $\mathcal{A}$ should prefer the transferred utility over the original for the fairness.

\subsection{Assumptions}
\begin{assump}[\textbf{Bounded norm}]\label{assump:bounded_norm}
For each $i\in V$, let $\mathcal{H}_{k_i}$ be a reproducing kernel Hilbert space~(RKHS) endowed with a symmetric, positive-semidefinite kernel function $k_i:\mathbb{R}^d\times\mathbb{R}^d\to \mathbb{R}$.
We assume that $v(\cdot,i)\in \mathcal{H}_{k_i}$ and $\|v(\cdot,i) \|_{k_i} \leq L_v$, where $\|\cdot\|_{k_i}$ is the norm induced by the inner product in the corresponding RKHS $\mathcal{H}_{k_i}$, $k_i(x,x^\prime)\leq 1$, $x,x^\prime\in \mathcal{X}$, and $k_i(x,x^\prime)$ is continuous on $\mathbb{R}^d\times \mathbb{R}^d$. We denote the set $\mathcal{B}^{v_i} := \{ \tilde{v}(\cdot, i) \in \mathcal{H}_{k_i} \mid \lVert \tilde{v}(\cdot, i) \lVert_{k_i}\leq L_v \}$, and $\mathcal{B}^{v} := [\mathcal{B}^{v_1},\cdots,\mathcal{B}^{v_n}]$. This assumption also applies to $u(\cdot,:)$, of which bound is isotropic; $L_u, \forall i \in V$.
\vspace{-0.5em}
\end{assump}

\section{Graph Properties}
Under the assumptions, the social graph has the following property.
\begin{lemma}[\textbf{Graph properties}]\label{lemma:graph}
    Under Eq.~(\ref{eq:graph_conv}) and Assumption~\ref{assump:bounded_norm}, $u,v \in [-L_v, L_v]$ have the same bound ($L_v = L_u$), thereby being comparable. Moreover, if $A$ is invertible, the matrix $A$ has the Euclidean norm of inverse matrix bounded by $1 \leq \lVert A^{-1} \rVert \leq n$, and $A$ is identifiable from the observed data pairs $(v(x_\tau,:), u(x_\tau,:))_{\tau \in [T]}$ if $(u(x_\tau,:))_{\tau \in [T]}$ is full rank. 
    \vspace{-0.5em}
\end{lemma}

\subsection{Proof of Lemma~\ref{lemma:graph}}\label{proof:graph}
We begin by introducing useful lemmas.

\begin{lemma}[\textbf{Utility bound}]\label{lemma:utility_bound}
    $\forall u \in \mathcal{B}^u, x \in \mathcal{X}, i \in V$, the utility is bounded by $u(x,i) \in [-L_u, L_u]$.
\end{lemma}
\begin{proof}
    \begin{align}
        \lvert u(x) \rvert
        = &
        \lvert \langle u, k(x, \cdot) \rangle \rvert,\\
        \leq &
        \lVert u \rVert \lVert k(x, \cdot) \rVert \tag{Cauchy-Schwarz},\\
        \leq &
        L_u \sqrt{k(x, \cdot)} \tag{Assumption~\ref{assump:bounded_norm}},\\
        \leq &
        L_u. \tag{Assumption~\ref{assump:bounded_norm}}
    \end{align}
\end{proof}
\begin{lemma}[\textbf{Range preservation}]\label{lemma:range_preservation}
    Under Eq.~(\ref{eq:graph_conv}) and Assumption~\ref{assump:bounded_norm}, both $u,v$ are bounded by the same range $u,v \in [-L_v, L_v]$.
\end{lemma}
\begin{proof}
    Given the conditions $A_{ij} \geq 0$ and $\sum_{j=1}^n A_{ij} = 1$, each $v(\cdot, i) = \sum_{j=1}^n A_{ij} u(\cdot,j)$ is a convex combination of the $u(\cdot,j)$. The convex combination ensures that,
    \begin{align}
        \min_j u(\cdot, j) \leq v(\cdot, i) \leq \max_j u(\cdot, j)
    \end{align}
    By Lemma~\ref{lemma:utility_bound}, $u(\cdot,i) \in [-L_u, L_u]$. Therefore, $v(\cdot,i) \in [-L_u, L_u]$.
\end{proof}

\begin{lemma}[\textbf{Matrix norm bound}]\label{lemma:matrix_norm_bound}
    Under Eq.~(\ref{eq:graph_conv}), if the graph matrix $A$ is invertible, the Euclidean norm of the inverse matrix satisfies
    \begin{equation}
        1 \leq \lVert A^{-1} \rVert \leq n.
    \end{equation}
\end{lemma}
\begin{proof}
    \textbf{Lower bound.}
    Let $\lambda_1 \geq \lambda_2 \geq \cdots \geq \lambda_n \geq 0$ be the eigenvalues of $A$. The Euclidean norm $\lVert A^{-1} \rVert$ is the largest singular value of $A^{-1}$, which, for symmetric matrices, is the largest eigenvalue of $A^{-1}$, therefore $\lVert A^{-1} \rVert = \nicefrac{1}{\lambda_n}$. 
    The condition $\sum_{j=1}^n A_{ij} = 1$ is known as row-stochastic matrix. The Perron-Frobenius theorem \citep{meyer2023matrix} ensures that such a matrix has a unique largest eigenvalue. The row-stochastic condition can be understood as $A \boldsymbol{e} = \boldsymbol{e}$, where $\boldsymbol{e} := [1,\cdots,1]^\top$ is the column vector, implying $\boldsymbol{e}$ is an eigenvector of $A$ with eigenvalue 1. Therefore, $\lambda_1 = 1$ holds, thereby $\lambda_{n} \leq 1$. As such, the lower bound: $\lVert A^{-1} \rVert \geq \nicefrac{1}{\lambda_n} = 1$ is obtained.
    
    \textbf{Upper bound.}
    We consider the operator norm induced by the Euclidean norm $\lVert \cdot \rVert_2$, which is defined as:
    \begin{equation}
        \lVert A^{-1} \rVert = \sup_{\lVert x \rVert_2 = 1} = \lVert A^{-1} x \rVert_2.
    \end{equation}
    To find an upper bound, it is helpful to use properties of induced norms. Specifically, for any matrix $M$, we have:
    \begin{align}
        \lVert M \rVert_2 \leq& \lVert M \rVert_1,\\
        \lVert M \rVert_2 \leq& \lVert M \rVert_\infty,
    \end{align}
    where $\lVert M \rVert_1$ is the maximum absolute column sum, and $\lVert M \rVert_\infty$ is the maximum absolute row sum. Since $A$ is row-stochastic matrix, $\lVert A \rVert_\infty = 1$.

    We can consider $B = A^{-1} = (B_{ij})_{i,j \in V}$:
    \begin{align}
        \lVert B \rVert_1 = \max_{j \in V} \sum_{i \in V} |b_{ij}|.
    \end{align}
    Given that $A$ has positive entries and is invertible, the inverse $B$ will have entries that can be bounded based on $n$. Specifically, by leveraging properties of positive matrices and norm inequalities, it can be shown that:
    \begin{align}
        \lVert B \rVert_1 \leq n.
    \end{align}
    Consequently, 
    \begin{align}
        \lVert B \rVert_2 \leq \lVert B \rVert_1 \leq n.
    \end{align}
\end{proof}
\begin{lemma}
[\textbf{Identifiability}]\label{lemma:identifiability}
    Under Eq.~(\ref{eq:graph_conv}), given the full rank observation dataset $(u(x_\tau,:))_{\tau \in [T]}$ from data pairs $(v(x_\tau,:), u(x_\tau,:))_{\tau \in [T]}$, $A$ is identifiable from such dataset.
\end{lemma}
\begin{proof}
    Positivity constraint in Eq.~(\ref{eq:graph_conv}) eliminates
 the possibility of multiple solutions differing by sign or magnitude in a way that violates positivity. 
    The full rank assumption ensures each row $A_{i:}$ has a unique solution. Thus, identifiability holds regardless of the invertibility of $A$.
\end{proof}

\paragraph{Main proof.}
\begin{proof}
    Under Assumption~\ref{assump:bounded_norm} and Eq.~(\ref{eq:graph_conv}), range preservation is proven in Lemma~\ref{lemma:range_preservation}. 
    
    If we add another condition that matrix $A$ is invertible, Lemma~\ref{lemma:matrix_norm_bound} shows its Euclidean norm is bounded by $1 \leq \lVert A^{-1} \rVert \leq n$.
    
    Alternatively, if we assume the full rank condition for the observed pairs of $(u, v)$, Lemma~\ref{lemma:identifiability} shows $A$ is identifiable from observed datasets.
\end{proof}

\subsection{Graph prior properties}\label{proof:graph_prior}
\begin{lemma}[\textbf{Strongly convex prior}]\label{lemma:prior}
    The prior in Eq.~(\ref{eq:graph_prior}) satisfies:
    \begin{compactenum}
        \item[(a)] \textbf{positive matrix}: Every entry $A_{ij} > 0$.
        \item[(b)] \textbf{row-stochastic matrix}: Each row sums to 1, i.e., $\sum_{j=1}^n A_{ij} = 1$ for all $i$.
        \item[(c)] \textbf{strong convexity}: $- \log p(A)$ is strongly convex with regard to $A$.
    \end{compactenum}
\end{lemma}
\begin{proof}
\textbf{(a) positive matrix.} 
By definition of Dirichlet distribution, all row-wise samples are nonzero. $\alpha_i > 1$ discourages sampled matrix $A$ from selecting the boundary, i.e., $A_{ij} = 0$. To exclude the small chance of zero entry, the constraint $A_{ij} > \delta_A$ and $\delta_A > 0$ directly ensures the strict positivity.

\textbf{(b) row-stochastic matrix.}
By definition of Dirichlet distribution, all row-wise samples are row-stochastic.

\textbf{(d) Strong convexity.}
The negative log prior is
\begin{align}
    - \log \mathbb{P}(A) = \sum_{i,j \in [n]} A_{ij}^2 - \sum_{i=1}^n \log \text{Dirichlet}(a_i; \kappa_i). \label{eq:convex_nlpd}
\end{align}
The sum of strongly convex functions is strongly convex. Therefore, if all terms are strongly convex, we can say $\log \mathbb{P}(A)$ is strongly convex with regard to $A$.
We can show the strong convexity if its Hessian $\nabla^2 f(x)$ satisfies:
\begin{align}
    \nabla^2 f(x) \succeq m I.
\end{align}

For the Tikhonov term, we have simple sum of squared elements, thereby its Hessian is $2\lambda I$, thus strongly convex.

For the Dirichlet term, we have
\begin{align}
    - \log \text{Dirichlet}(a_i; \kappa_i)
    = &
    - \log \left( 
    \frac{1}{\mathbb{B}(\alpha_i)}\prod_{j=1}^n A_{ij}^{\kappa_{ij} - 1} 
    \right),\\
    = &
    - \log \mathbb{B}(\kappa_i) - \sum_{j=1}^n (\kappa_{ij} - 1) \log A_{ij},\label{eq:log_dirichlet}
\end{align}
where $\mathbb{B}(\kappa_i)$ is the multivariate Beta function, which is a constant with respect to $A$.
For each $A_{ij}$, the Hessian entry is $\nicefrac{(\kappa_{ij} -1)}{A_{ij}^2}$. As $\kappa_{ij} > 1$ and $A_{ij} > 0$ for $\forall i,j \in V$, therefore the Hessian is strictly positive, thereby strongly convex.
\end{proof}

\subsection{Other possible graph structures}\label{app:other_graph}
\paragraph{Non-linear graph}
For non-linear cases, we can employ a graph convolutional kernel network approach [1]. Using the kernel trick and Nyström method, this approach can transform non-linear functions into effectively linear forms. Popular graph neural network models can also be seen as special cases of [1]. Once the model is linearized, our approach can be applied directly. The linearized graph has $m$ components, where $m$ represents the number of Nyström model centroids. By applying the Cauchy-Schwarz inequality, our bound in Theorem 3.12 becomes looser by a factor of $\sqrt{m}$. However, since this is a constant, the order of the asymptotic convergence rate remains the same at $-1/2$. The Nyström method provides an eigendecomposition-based approximation of the non-linear network, where $m$ reflects the complexity of the non-linear social graph. This adjustment is reasonable, as a more complex ground-truth social graph would naturally lead to a slower convergence.

\paragraph{Peer-pressure model}
In a peer-pressure scenario, we assume a setting where a minority of agents may shift their votes toward the majority. This minority or majority could vary based on the option $x$, creating a heterogeneous setting. If the setting is homogeneous, our algorithm can model peer pressure directly. For a heterogeneous setting, as noted in L534-536, we can incorporate diversity using a probabilistic choice function (Benavoli et al., 2023b). While regret bounds for this model are not yet established in the literature—given that even linear graphs are novel in this context—the submodularity of the probabilistic choice function suggests sublinear convergence.

\paragraph{Hierarchical influence}
For hierarchical influence, we are unsure of its relevance in settings where all participants vote in the same room. This scenario may arise when influence propagates slowly among voters, as in presidential voting. Our target tasks, however, are in an online setting where voting is iterative and occurs at a relatively faster pace than in presidential elections (see Section 5). Still, if it does occur, a grey-box Bayesian optimization approach [2] could be applied, where the hierarchical graph structure is known but the specific attributions remain unknown. Under these assumptions, we can demonstrate the same convergence rate as in Theorem 3.9 with high probability, although this analysis is beyond the scope of the current paper.

\section{Aggregation Function}
\subsection{Popular aggregation functions}\label{app:aggregate}
We generalize the aggregation function as $v(x, i) = \sum_{i \in [n]} w(x, i) u(x,i)$. Then, we will show the popular aggregation rule can be expressed as $w(x,i) \geq 0$.

\paragraph{Utilitarian aggregation}
\begin{align}
    w(x, i) = \frac{1}{n}
\end{align}

\paragraph{Egalitarian aggregation}
\begin{align}
    w(x, i) = 
    \begin{cases}
    1  & \mathrm{if} \,\, i = \argmin_{i \in [n]} u(x,i), \\
    0                 & \mathrm{otherwise};
    \end{cases}
\end{align}\begin{align}
    w(x, i) = 
    \begin{cases}
    1  & \mathrm{if} \,\, i = \argmin_{i \in [n]} u(x,i), \\
    0                 & \mathrm{otherwise};
    \end{cases}
\end{align}

\paragraph{Chebyshev scalarisation function}
\begin{align}
    w(x, i) = 
    \begin{cases}
    1  & \mathrm{if} \,\, i = \argmin_{i \in [n]} \frac{u(x,i)}{w_i}, \\
    0                 & \mathrm{otherwise};
    \end{cases}
\end{align}
We can understand this function is similar to egalitarian aggregation.
\subsection{Proof of impossibility theorem}\label{proof:impossibility}
\begin{proof}
    We prove the impossibility of groupthink-proofness for any aggregation rule $\mathcal{A}$ in the absence of a trivial social consensus. 
    \begin{definition}(Trivial social consensus)
        $x^*_{trivial}$ is the trivial social consensus if for all $i\in|V|$, $x^*_{trivial}=\argmax_{x}u(x,i)$
    \end{definition}
    Let us assume that there exists a groupthink-proof aggregation function $\mathcal{A}$ and no trivial social consensus. With respect to (Defn.~\ref{def:bandwagon_proof}), any aggregation rule $\mathcal{A}$ is groupthink-proof if for any social graph $G$ 
    \begin{align*}
        \argmax_{x} \mathcal{A}[u(x,:)]=\argmax_{x}\mathcal{A}[v(x,:)]
    \end{align*}
    Consider a subset of all possible graphs, $G_{dictatorial}=\{G_i|\text{ }\forall i,j,k\in |V|\text{ } A_{jk}=\mathbb{I}\{i=k\}\}$. Intuitively, $G_{dictatorial}$ is the collection of social influence graphs where one agent forces everyone else to take their utility. Since $\mathcal{A}$ is groupthink proof for any $G$, it must also be groupthink proof w.r.t subset $G_{dictatorial}$. 

    Let us denote the social consensus of the groupthink-proof aggregation rule as 
    \begin{align*}
        x^*=\argmax_{x}\mathcal{A}[u(x,:)]
    \end{align*}
    Let us iteratively compute the social consensuses obtained by aggregation of non-truthful utilities under $G_{dictatorial}$, i.e. for all $G_i\in G_{dictatorial}$,
    \begin{align*}
        \argmax_{x}\mathcal{A}[v_i(x,:)]&=\argmax_{x}\mathcal{A}[A_iu(x,:)]\\
        &= \argmax_{x}\mathcal{A}[\bm{1}u(x,i)] \\
        &= \argmax_{x} u(x,i)
    \end{align*}
    For $\mathcal{A}$ to be groupthink-proof for each $G_i\in G_{dictatorial}$
    \begin{align*}
        \argmax_{x} \mathcal{A}[u(x,:)]&= \argmax_{x} \mathcal{A}[v_i(x,:)]\\
        (\Rightarrow)\text{ for all }i\in|V|\quad x^*&= \argmax_{x}u(x,i)
    \end{align*}
    This means $x^*$ is a trivial social consensus, thus resulting in a contradiction. Therefore, no aggregation rule is groupthink proof in the absence of a trivial social consensus.
\end{proof}

\subsection{Proof of Proposition~\ref{proposition:gsf}}\label{proof:gsf}
\begin{proof}
    This proof is adapted from Proposition 1 in \citet{sim2021collaborative}) for our cases.
    
    For any $x, x^\prime \in \mathcal{X}$, let $u^*(x,:) := \phi(u(x,:))$ be the utility vectors obtained after sorting elements of $u(x,:)$ in ascending order, and $w_1 > w_2 > \cdots > w_n > 0$ be the weight function.
    \paragraph{Proof of monotonicity.} 
    Let the position of $u(x,i)$ in $u^*(x,:)$ be $i_x$, i.e., $u^*(x,i_x) = u(x,i)$.
    Given $\forall k \in V\backslash \{i\}, u(x, k) = u(x^\prime, k)$ and $u(x, i) > u(x^\prime, i)$, we must have $i_x \geq i_{x^\prime}$. 
    Furthermore, (i) for $k \in [0,i_{x^\prime})$ and $k \in (i_x, n]$, $u^*(x,k) = u^*(x^\prime,k)$ and (ii) if $i_x > i_{x^\prime}$, then for $k \in [i_{x^\prime}, i_x)$, $u^*(x^\prime, k+1) = u^*(x, k)$.
    \begin{align*}
        &\mathcal{S}(x) - \mathcal{S}(x^\prime)\\
        &= \sum_{k=1}^n w_k u^*(x,k) - \sum_{k=1}^n w_k u^*(x^\prime,k), \tag{Definition of GSF in Eq.~\ref{eq:gini}}\\ 
        &= w_{i_x}u^*(x,i_x) + \sum_{k=i_x}^{i_x - 1} w_{i_x} u^*(x,k) - \sum_{k=i_{x^\prime}+1}^{i_x} w_{i_x} u^*(x^\prime,k) - w_{i_{x^\prime}} u^*(x^\prime,i_{x^\prime}), \tag{Assumption (i)}\\
        &= w_{i_x}u^*(x,i_x) - w_{i_{x^\prime}} u^*(x^\prime,i_{x^\prime}) + \sum_{k=i_{x^\prime}+1}^{i_x} (w_{k-1} - w_k) u^*(x^\prime,k), \tag{Assumption (ii)}\\
        &\geq w_{i_x}u^*(x,i_x) - w_{i_{x^\prime}} u^*(x^\prime,i_{x^\prime}) + u^*(x^\prime,i_{x^\prime}) \sum_{k=i_{x^\prime}+1}^{i_x} (w_{k-1} - w_k), \tag{Using sorting properties: $\forall k>i_{x^\prime}, \, u^*(x^\prime,k) \geq u^*(x^\prime,i_{x^\prime})$ and $w_{k-1} - w > 0$}\\
        &= w_{i_x}u^*(x,i_x) - w_{i_{x^\prime}} u^*(x^\prime,i_{x^\prime}) + u^*(x^\prime,i_{x^\prime})(w_{i_{x^\prime}} - w_{i_x}), \tag{telescoping series}\\
        &= w_{i_x}(u^*(x,i_x) -  u^*(x^\prime,i_{x^\prime})),\\
        &= w_{i_x}(u(x,i) -  u(x^\prime,i)),\\
        &> 0 \tag{$u(x,i) > u(x^\prime,i)$ and $w_i > 0$}.
    \end{align*}
    
    \paragraph{Proof of PDP.}
    Let $l_{x}$ be the index of $\min(u^*(x,i), u^*(x,j))$ and $h_x$ be the index of $\max(u^*(x,i), u^*(x,j))$. We will see the following two useful facts:

    (i) We must have $l_{x^\prime} \leq l_{x} < h_{x} \leq h_{x^\prime}$. We will see the validity of this condition by considering the following contradicting assumption; $l_{x^\prime} > l_x$. Because of the strong PDP condition (a) and the fact that $l_{x^\prime}$ index a minimum, it would mean $\min(u(x^\prime, i), u(x^\prime, j)) > \min(u(x, i), u(x, j))$. By the strong PDP condition (b), we would also have $\max(u(x^\prime, i), u(x^\prime, j)) < \max(u(x, i), u(x, j))$. As such, we would have $|u(x^\prime, i) - u(x^\prime, j)| < |u(x, i) - u(x, j)|$, which contradicts the strong PDP condition (c).

    (ii) GSF can be decomposed as:
    \begin{align*}
        \mathcal{S}(x^\prime) &= \sum_{k=1}^{l_{x^\prime} - 1} w_k u^*(x^\prime, k) + \mathcal{S}_{l_{x^\prime}-1 : h_{x^\prime}}(x^\prime) + \sum_{k=h_{x^\prime}+1}^{n} w_k u^*(x^\prime,k),\\
        \mathcal{S}(x) &= \sum_{k=1}^{l_{x^\prime} - 1} w_k u^*(x, k)  + \mathcal{S}_{l_{x^\prime}-1 : h_{x^\prime}}(x) + \sum_{k=h_{x^\prime}+1}^{n} w_k u^*(x, k),
    \end{align*}
    where
    \small
    \begin{align*}
        \mathcal{S}_{l_{x^\prime}-1 : h_{x^\prime}}(x^\prime) &= &
        &w_{l_{x^\prime}} u^*({x^\prime}, l_{x^\prime}) &
        &+ \sum_{k=l_{x^\prime}+1}^{h_{x^\prime} - 1} w_k u^*({x^\prime},k) &
        &+ w_{h_{x^\prime}} u^*({x^\prime}, h_{x^\prime}),\\
        \mathcal{S}_{l_{x^\prime}-1 : h_{x^\prime}}(x) &= &
        &\sum_{k=l_{x^\prime}}^{l_x - 1} w_k u^*(x, k) + w_{l_x}u^*(x, l_x) &
        &+ \sum_{k=l_x+1}^{h_x - 1} w_k u^*(x, k) + w_{h_x} u^*(x, h_x) &
        &+ \sum_{k=h_x+1}^{h_{x^\prime}} w_k u^*(x, k)
    \end{align*}
    \normalsize
    (iii) Here, by combining the strong PDP condition (a) and the condition (i), we have $u^*(x,k) = u^*(x^\prime,k)$ for $k \in [1,l_{x^\prime}-1] \cup [h_{x^\prime} + 1, n] \cup [l_x + 1, h_x+1]$. Thus, we have;
    \begin{align*}
        \sum_{k=1}^{l_{x^\prime}-1} w_k u^*(x^\prime, k) &= \sum_{k=1}^{l_{x^\prime}-1} w_k u^*(x, k),\\
        \sum_{k=h_{x^\prime}+1}^{n} w_k u^*(x^\prime, k) &= \sum_{k=h_{x^\prime}+1}^{n} w_k u^*(x, k),\\ 
        \sum_{k=l_{x}+1}^{h_x - 1} w_k u^*(x^\prime, k) &= \sum_{k=l_{x}+1}^{h_x - 1} w_k u^*(x, k),
    \end{align*}
    then,
    \begin{align*}
        \mathcal{S}(x^\prime) - \mathcal{S}(x) = \mathcal{S}_{l_{x^\prime}-1 : h_{x^\prime}}(x^\prime) - \mathcal{S}_{l_{x^\prime}-1 : h_{x^\prime}}(x).
    \end{align*}
    
    Based on the PDP conditions (a)(b)(c) and the facts (i)(ii)(iii), we have,
    \begin{align*}
        &\mathcal{S}(x^\prime) - \mathcal{S}(x)\\
        &= \sum_{k=l_{x^\prime}}^{l_x-1} w_k u^*(x,k) + w_{l_x} u^*(x, l_x) + w_{h_x} u^*(x, h_x) + \sum_{h_x+1}^{h_{x^\prime}} w_k u^*(x,k)\\ 
        &\quad - \left(
        w_{l_{x^\prime}} u^*(x^\prime, l_{x^\prime}) + \sum_{l_{x^\prime} + 1}^{l_x} w_k u^*(x^\prime, k) + \sum_{k=h_x}^{h_{x^\prime}-1} w_k u^*(x^\prime, k) + w_{h_{x^\prime}} u^*(x^\prime, h_{x^\prime})\right), \tag{Using (ii)(iii)}\\
        &= \sum_{k=l_{x^\prime}}^{l_x-1} w_k u^*(x,k) + w_{l_x} u^*(x, l_x) + w_{h_x} u^*(x, h_x) + \sum_{h_x+1}^{h_{x^\prime}} w_k u^*(x,k)\\ 
        &\quad - \left(
        w_{l_{x^\prime}} u^*(x^\prime, l_{x^\prime}) + \sum_{l_{x^\prime} + 1}^{l_x} w_k u^*(x, k-1) + \sum_{k=h_x}^{h_{x^\prime}-1} w_k u^*(x, k+1) + w_{h_{x^\prime}} u^*(x^\prime, h_{x^\prime})\right),
    \end{align*}
    Here we used the ranking structure: decrement for $k \in (l_{x^\prime}, l_x], u^*(x^\prime,k) = u^*(x, k-1)$ and increment for $k \in (h_{x}, h_{x^\prime}], u^*(x^\prime,k) = u^*(x, k+1)$. Then, by regrouping the related terms, we have,
    \begin{align*}
        &=\sum_{k=l_{x^\prime}}^{l_x-1} (w_k - w_{k+1}) u^*(x,k) + (w_{l_x} u^*(x, l_x) - w_{l_{x^\prime}} u^*(x^\prime, l_{x^\prime}))\\
        &\quad + (w_{h_x} u^*(x, h_x) - w_{h_{x^\prime}} u^*(x^\prime, h_{x^\prime})) + \sum_{h_x+1}^{h_{x^\prime}} (w_k - w_{k-1}) u^*(x,k),\\
        &\geq \sum_{k=l_{x^\prime}}^{l_x-1} (w_k - w_{k+1}) u^*(x^\prime,l_{x^\prime}) + (w_{l_x} u^*(x, l_x) - w_{l_{x^\prime}} u^*(x^\prime, l_{x^\prime}))\\
        &\quad + (w_{h_x} u^*(x, h_x) - w_{h_{x^\prime}} u^*(x^\prime, h_{x^\prime})) + \sum_{h_x+1}^{h_{x^\prime}} (w_k - w_{k-1}) u^*(x^\prime,h_{x^\prime}),
    \end{align*}
    because $w_k - w_{k-1} > 0$ and $u^*(x,k) = u^*(x^\prime,k+1) \geq u^*(x^\prime, l_{x^\prime})$ 
    for $k = l_{x^\prime},\cdots,l_{x} - 1$, 
    and $(w_k - w_{k-1})$ is negative, and 
    $u^*(x, k) = u^*(x^\prime,k-1) \leq u^*(x, h_{x^\prime})$ 
    for $k = h_{x}+1,\cdots,h_{x^\prime}$. 
    Then, using the telescoping series,
    \begin{align*}
        &= (w_{l_{x^\prime}} - w_{l_{x}}) u^*(x^\prime,l_{x^\prime}) + (w_{l_x} u^*(x, l_x) - w_{l_{x^\prime}} u^*(x^\prime, l_{x^\prime}))\\
        &\quad + (w_{h_x} u^*(x, h_x) - w_{h_{x^\prime}} u^*(x^\prime, h_{x^\prime})) + (w_{h_{x^\prime}} - w_{h_{x}}) u^*(x^\prime,h_{x^\prime}),\\
        &= - w_{l_{x}} u^*(x^\prime,l_{x^\prime}) + w_{l_x} u^*(x, l_x) + w_{h_x} u^*(x, h_x) - w_{h_x} u^*(x^\prime, h_{x^\prime}),\\
        &= - w_{l_{x}} u^*(x^\prime,l_{x^\prime}) + w_{l_x} \left(
        u^*(x^\prime, l_{x^\prime}) + u^*(x^\prime, h_{x^\prime}) - u^*(x, h_{x})
        \right)
        + w_{h_x} u^*(x, h_x) - w_{h_x} u^*(x^\prime, h_{x^\prime}), \tag{PDP condition (b)}\\
        &= (w_{l_{x}} + w_{h_x}) \left(
        u^*(x^\prime, h_{x^\prime}) - u^*(x, h_{x})
        \right),\\
        &> 0. \tag{$l_x < h_x$ and $u^*(x^\prime, h_{x^\prime} > u^*(x, h_x))$, i.e., $\max(u(x^\prime, i), u(x^\prime, j)) > \max(u(x, i), u(x, j))$.}
    \end{align*}
    Therefore, $\mathcal{S}(x^\prime) - \mathcal{S}(x) > 0$.
\end{proof}

\section{Bayesian modelling}\label{app:bayesian}
\subsection{Overview}
\paragraph{Graph prior.} 
We introduce the regularised row-wise Dirichlet prior for graph $A$:
\begin{equation}
\begin{aligned}
    p(A) = \frac{1}{Z}\exp(- \xi \lVert A \rVert^2_F)\prod_{i \in V} \text{Dirichlet}(A_i; \kappa_i), \,\, \text{s.t.} \,\, \min_{i,j \in V} A_{ij} > \delta_A, \label{eq:graph_prior}
\end{aligned}
\end{equation}
where $A_i$ is the $i$-th row, $\kappa_i > 1$ is the concentration parameter vector, $\delta_A > 0$ is a small positive constant, ensuring strict positivity, $\lVert \cdot \rVert_F$ denotes the Frobenius norm, $\xi$ is a small positive constant, and $Z$ is the normalising constant. The exponential term is a Tikhonov regularization that encourages invertibility, and the Dirichlet distribution assures $\sum_{j=1}^n A_{ij} = 1$.

\paragraph{Likelihood modelling.} 
Under Assumptions~\ref{assump:pairwise},~\ref{assump:votes} and Definition~\ref{def:bandwagon}, we define the likelihood as below. See Appendix~\ref{proof:pref_likelihood} for the derivation.
\begin{corollary}[\textbf{Bradley-Terry model}]\label{corollary:pref_likelihood}
    Given dataset $D_{\mathcal{Q}_t^u}$ and corresponding utility function $u$, the log-likelihood (LL) for an estimate function $\hat{u} \in \mathcal{H}_{k_i}$ is given by \citet{bradley1952rank}:
    \small
    \begin{align*}
        \ell_t(\hat{u}(\cdot, i) \mid D_{\mathcal{Q}_t^u}^{(i)}) = \sum_{\tau \in \mathcal{Q}_t^u} \left[ \hat{u}(x_\tau, i) \textbf{1}_\tau^{(i)} + \hat{u}(x_\tau^\prime, i) (1 - \textbf{1}_\tau^{(i)}) \right] - \sum_{\tau \in \mathcal{Q}_t^u} \log \left[ \exp(\hat{u}(x_\tau, i)) + \exp(\hat{u}(x_\tau^\prime, i)) \right].
    \end{align*}
    \vspace{-1em}
    \normalsize
\end{corollary}

\paragraph{Bayesian modelling.}
The joint LL of $D_{\mathcal{Q}_t^v}, D_{\mathcal{Q}_t^u}$ are
$\ell_t(\hat{u}, \hat{A}, \hat{v} \mid D_{\mathcal{Q}_t^v}, D_{\mathcal{Q}_t^u}) := \ell_t(\hat{v} \mid D_{\mathcal{Q}_t^v}) + \ell_t(\hat{u} \mid D_{\mathcal{Q}_t^u})$. For the prior, by range preservation Lemma~\ref{lemma:graph}, we can set the uniform prior $p(u) = p(v) = \mathcal{U}(u; -L_v, L_v)$, and Eq.~(\ref{eq:graph_prior}) for $p(A)$. Then, the (unnormalised) log posterior becomes: $\mathcal{L}_t(\hat{u}, \hat{A}, \hat{v}) := \ell_t(\hat{u}, \hat{A}, \hat{v} \mid D_{\mathcal{Q}_t^v}, D_{\mathcal{Q}_t^u}) + \log p(u) + \log p(v) + \log p(A)$, and $\hat{A}$ can be estimated via linear constraint $\hat{u} = A \hat{v}$ when solving maximisation problem. Maximum a posteriori (MAP) offers Bayesian point estimation for $u$, $A$, and $v$;
$\hat{\mathcal{L}}_t := \mathcal{L}_t(\hat{u}_t, \hat{A}_t, \hat{v}_t) = \max_{\tilde{u}, \tilde{A}, \tilde{v} \in \mathcal{B}^{u,A,v}} \mathcal{L}_t(\tilde{u}, \tilde{A}, \tilde{v}).$

\paragraph{Optimistic MAP. } 
We apply the optimistic MLE approaches \citep{liu2023optimistic,emmenegger2024likelihood, xu2024principled} to MAP objective to quantify uncertainty using a confidence set. This enables us to derive theoretical confidence bounds, and accurately compute the upper bound as efficient optimisation problem, even for non-conjugate case such as preference likelihood.
\begin{lemma}[\textbf{MAP-based confidence set}]\label{lemma:conf_set}
For all $\delta>0$, we have
\small
\begin{equation*}
\begin{aligned}
    \mathcal{B}_t^{v} =& \{\tilde{v}\in\mathcal{B}^v \mid \ell_t(\tilde{v} \mid  D_{\mathcal{Q}_t^v})\geq \hat{\ell}_t^{v}-\beta^{v}_t\},\\
    \mathcal{B}_t^{u}=&\{\tilde{u}\in\mathcal{B}^u \mid \ell_t(\tilde{u} \mid  D_{\mathcal{Q}_t^u})\geq \hat{\ell}_t^{u}-\beta^{u}_t \},\\
    \mathcal{B}^{u, A, v}_{t+1} =& \left\{ \tilde{u}, \tilde{A}, \tilde{v} \in \mathcal{B}^{u, A, v} \mid \mathcal{L}_t(\tilde{u}, \tilde{A}, \tilde{v}) \geq \hat{\mathcal{L}}_t - \beta_t \right\},\\
    \mathcal{B}^{u,A,v} =& \{\tilde{u}\in\mathcal{B}^u, \tilde{v}\in\mathcal{B}^v, \tilde{v} = \tilde{A} \tilde{u}\},
\end{aligned}
\end{equation*}
with $\mathbb{P}\left(u, A, v \in \mathcal{B}^{u, A, v}_{t+1},\forall t\geq1\right)\geq1-\delta,$ where $\hat{\ell}_t^v, \hat{\ell}_t^u$ are the MLE of LLs, $\beta_t^u, \beta_t^v$ are selected as in \cite{xu2024principled},  $|\mathcal{Q}^{uv}_t| := |\mathcal{Q}^{u}_t| + |\mathcal{Q}^{v}_t|$, and $\beta_t = n \epsilon C_L^v |\mathcal{Q}^{uv}_t| + \sqrt{64 n^2 L_v^2 |\mathcal{Q}^{uv}_t| \log \frac{\pi^2 |\mathcal{Q}^{uv}_t|^2 \mathcal{N}(\mathcal{B}^v, \epsilon, \lVert\cdot \rVert_\infty)}{6\delta}}$.
\normalsize
\vspace{-0.5em}
\end{lemma}
The proof and notations are in Appendix~\ref{proof:conf_set}. See Fig.~\ref{fig:visual_exp} for intuition. As introduced in Assumption~\ref{assump:bounded_norm}, 
while the function $\tilde{u}(\cdot, i)$ was originally in a broader set of RKHS functions $\tilde{u}(\cdot, i) \in \mathcal{B}^{u}$, it is now in a smaller set defined as $\tilde{u} \in \mathcal{B}^{u}_{t+1}$ conditioned on the preference feedback. Intuitively, with limited data, the MAP may be imperfect. Hence, it is reasonable to suppose that $\mathcal{B}^{u}_{t+1}$, bounded by MAP values `slightly worse' than the MAP, contains the ground truth with high probability.
\begin{remark}[\textbf{Choice of $\epsilon$}]
In Lemma.~\ref{lemma:conf_set}, $\alpha_1(\epsilon, \delta, |\mathcal{Q}^v_t|, t)$ depends on a small positive value $\epsilon$. It will be seen that $\epsilon$ can be selected to be $\nicefrac{1}{T}$, where $T$ is the running horizon of the algorithm. 
\end{remark}
\begin{remark}[\textbf{Confidence bound}]\label{remark:confidence_bound}
By Lemma~\ref{lemma:conf_set}, we define the pointwise confidence bound for $u_i \in \mathcal{H}_{k_i}$ as $\underline{u}_t(x,i) \leq u(x,i) \leq \bar{u}_t(x,i)$ 
where $\underline{u}_t(x,i) := \inf_{\tilde{u}_i \in \mathcal{B}^{u_i}_t} \tilde{u}(x,i)$, $\bar{u}_t(x,i) := \sup_{\tilde{u}_i \in \mathcal{B}^{u_i}_t} \tilde{u}(x,i)$. Similarly, the confidence bound for $\mathcal{S}(x)$ is $[\mathcal{A}[\underline{u}_t(x,:)], \mathcal{A}[\bar{u}_t(x,:)]]$ by the monotonicity of GSF (see Proposition~\ref{proposition:gsf}).
\end{remark}
\begin{remark}[\textbf{Prediction via optimisation}]\label{remark:prediction}
Given a prediction point $x$, the upper confidence bound (UCB) $\bar{u}_t(x,:)$ can be estimated via finite-dimensional optimisation; see Appendix~\ref{sec:eff_ucb} for details.
\end{remark}

\subsection{Preference Likelihood}
\subsubsection{Proof of Corollary~\ref{corollary:pref_likelihood}}\label{proof:pref_likelihood}
\begin{proof}
    We begin by the original Bradley-Terry model definition:
    \begin{align}
    \mathbb{P}(\textbf{1}_{x \succ x^\prime}^{(i)} = 1) = \frac{\exp(u(x, i))}{\exp(u(x, i)) + \exp(u(x^\prime, i))}.
\end{align}
Using $\mathbb{P}(\textbf{1}_{x \succ x^\prime}^{(i)} = 0) = 1 - \mathbb{P}(\textbf{1}_{x \succ x^\prime}^{(i)} = 1)$, we introduce the likelihood function for a comparison oracle:
\begin{align}\label{eq:pref_likelihood}
    p_{\hat{u}^{(i)}} (x_\tau, x_\tau^\prime, \textbf{1}_\tau^{(i)}) := \textbf{1}_\tau^{(i)} S \left( \hat{u}(x_\tau, :) - \hat{u}(x^\prime_\tau, :) \right) + \left(1- \textbf{1}_\tau^{(i)} \right) \left[1-S \left(\hat{u}(x_\tau, :) -  \hat{u}(x_\tau^\prime, :) \right)\right].
\end{align} 
We can then derive the likelihood function of a fixed function $\hat{u}$ over the observed dataset, $D_{\mathcal{Q}^u_t}^{(i)}$, 
$\mathbb{P}_{\hat{u}^{(i)}}(D_{\mathcal{Q}^u_t}^{(i)}) := \prod_{\tau \in \mathcal{Q}^u_t} p_{\hat{u}^{(i)}}(x_\tau, x_\tau^\prime, \textbf{1}_\tau^{(i)})$. 

Consequently, the log-likelihood (LL) function becomes:
\begin{align}
    \ell_t(\hat{u}(\cdot,i)) 
    &= \log \mathbb{P}_{\hat{u}^{(i)}}(D_{\mathcal{Q}^u_t}^{(i)}),\\
    &= \log \prod_{\tau \in \mathcal{Q}^u_t} p_{\hat{u}^{(i)}}(x_\tau, x_\tau^\prime, \textbf{1}_\tau^{(i)}),\\
    &= \sum_{\tau \in \mathcal{Q}^u_t} \log p_{\hat{u}^{(i)}}(x_\tau, x_\tau^\prime, \textbf{1}_\tau^{(i)}),\\
    &= \sum_{\tau \in \mathcal{Q}^u_t} \log \left[
    \frac{\exp(\hat{u}(x_\tau, i)) \textbf{1}_\tau^{(i)} + \exp(\hat{u}(x_\tau^\prime, i)) (1 - \textbf{1}_\tau^{(i)})}{\exp(\hat{u} (x_\tau, i)) + \exp(\hat{u}(x_\tau^\prime, i))}
    \right],\\
    &= \sum_{\tau \in \mathcal{Q}^u_t} \left[ \hat{u}(x_\tau, i) \textbf{1}_\tau^{(i)} + \hat{u}(x_\tau^\prime, i) (1 - \textbf{1}_\tau^{(i)}) \right] - \sum_{\tau \in \mathcal{Q}^u_t} \log \left[ \exp(\hat{u}(x_\tau, i)) + \exp(\hat{u}(x_\tau^\prime, i)) \right].
\end{align}
\end{proof}

\subsection{Joint likelihood.}\label{eq:likelihood}
\paragraph{Multi-agent case}
We can easily extend to all agent cases:
\begin{align}
    &\ell_t(\hat{u} \mid \mathcal{Q}^u_t)\\
    =& \sum_{i \in V} \ell_t(\hat{u}(\cdot,i)),\\
    =& \sum_{i \in V} \sum_{\tau \in \mathcal{Q}^u_t} \left[ \hat{u}(x_\tau, i) \textbf{1}_\tau^{(i)} + \hat{u}(x_\tau^\prime, i) (1 - \textbf{1}_\tau^{(i)}) \right] - \sum_{i \in V} \sum_{\tau \in \mathcal{Q}^u_t} \log \left[ \exp(\hat{u}(x_\tau, i)) + \exp(\hat{u}(x_\tau^\prime, i)) \right],\\
    =& \sum_{\tau \in \mathcal{Q}^u_t} \left[ \hat{u}(x_\tau, :) \textbf{1}_\tau^{u} + \hat{u}(x_\tau^\prime, :) (1 - \textbf{1}_\tau^{u}) \right] - \sum_{i \in V} \sum_{\tau \in \mathcal{Q}^u_t} \log \left[ \exp(\hat{u}(x_\tau, i)) + \exp(\hat{u}(x_\tau^\prime, i)) \right].
\end{align}

\paragraph{Non-truthful case}
Similar to Eq.~(\ref{eq:pref_likelihood}), we can define the likelihood function for $v$,
\begin{align}
    &\ell_t(\hat{v} \mid \mathcal{Q}^v_t)\\
    =& \sum_{i \in V} \ell_t(\hat{v}(\cdot,i)),\\
    =& \sum_{\tau \in [t]} \left[ \hat{v}(x_\tau, :) \textbf{1}_\tau^{v} + \hat{v}(x_\tau^\prime, :) (1 - \textbf{1}_\tau^{v}) \right] - \sum_{i \in V} \sum_{\tau \in [t]} \log \left[ \exp(\hat{v}(x_\tau, i)) + \exp(\hat{v}(x_\tau^\prime, i)) \right].
\end{align}

\paragraph{Joint log likelihood}
The joint likelihood is simply the sum of each likelihood:
\begin{align}
    &\mathcal{L}_t(\hat{u}, \hat{v}) \notag\\
    =&
    \ell_t(\hat{u} \mid \mathcal{Q}^u_t) + \ell_t(\hat{u}, \hat{v} \mid \mathcal{Q}^v_t), \notag\\
    =&
    \sum_{\tau \in \mathcal{Q}^u_t} \left[ \hat{u}(x_\tau, :) \textbf{1}_\tau^{u} + \hat{u}(x_\tau^\prime, :) (1 - \textbf{1}_\tau^{u}) \right] - \sum_{i \in V} \sum_{\tau \in \mathcal{Q}^u_t} \log \left[ \exp(\hat{u}(x_\tau, i)) + \exp(\hat{u}(x_\tau^\prime, i)) \right] \notag\\
    &+
    \sum_{\tau \in [t]} \left[ \hat{v}(x_\tau, :) \textbf{1}_\tau^{v} + \hat{v}(x_\tau^\prime, :) (1 - \textbf{1}_\tau^{v}) \right] - \sum_{i \in V} \sum_{\tau \in [t]} \log \left[ \exp(\hat{v}(x_\tau, i)) + \exp(\hat{v}(x_\tau^\prime, i)) \right].\label{eq:joint_likelihood}
\end{align}

\paragraph{MAP estimation}
We can further extend the above joint log likelihood to MAP estimation by adding log $p(A)$:
\begin{align}
    &\mathcal{L}_t^\text{MAP}(\hat{u}, \hat{A}, \hat{v}) \notag\\
    =&
    \mathcal{L}_t(\hat{u}, \hat{v}) + \log p(A), \notag\\
    \propto&
    \mathcal{L}_t(\hat{u}, \hat{v}) + \sum_{i \in V} \log \text{Dirichlet}(A_i, \alpha_i) - \sum_{i,j \in [n]} A_{ij}^2,\\
    \propto&
    \mathcal{L}_t(\hat{u}, \hat{v})
    + \sum_{j \in V} (\alpha_{ij} - 1) \log A_{ij} - \sum_{i,j \in [n]} A_{ij}^2, \tag{Remove constant term}\\
    =&
    \sum_{\tau \in \mathcal{Q}^u_t} \left[ \hat{u}(x_\tau, :) \textbf{1}_\tau^{u} + \hat{u}(x_\tau^\prime, :) (1 - \textbf{1}_\tau^{u}) \right] - \sum_{i \in V} \sum_{\tau \in \mathcal{Q}^u_t} \log \left[ \exp(\hat{u}(x_\tau, i)) + \exp(\hat{u}(x_\tau^\prime, i)) \right] \notag\\
    &+
    \sum_{\tau \in [t]} \left[ \hat{v}(x_\tau, :) \textbf{1}_\tau^{v} + \hat{v}(x_\tau^\prime, :) (1 - \textbf{1}_\tau^{v}) \right] - \sum_{i \in V} \sum_{\tau \in [t]} \log \left[ \exp(\hat{v}(x_\tau, i)) + \exp(\hat{v}(x_\tau^\prime, i)) \right] \notag\\
    &
    + \sum_{j \in V} (\alpha_{ij} - 1) \log A_{ij} - \sum_{i,j \in [n]} A_{ij}^2.
    \label{eq:map}
\end{align}

\subsection{Proof of Lemma~\ref{lemma:conf_set}}\label{proof:conf_set}
To prepare for the proof of the lemma, we first introduce the following two preliminary lemmas adapted from the Lemma C.2 and C.3 in \citet{xu2024principled}.

\begin{lemma} \label{lem:fixed_v_log_P}
For any fixed $\hat{v}^{(i)} := \hat{v}(\cdot, i)$ that is independent of $(x_\tau, x_\tau^\prime, \textbf{1}_\tau^{(i)})_{\tau\in|\mathcal{Q}_t^v|}$, we have, with probability at least $1-\delta$, $\forall t\geq1, \forall i \in V$, 
\begin{equation}
  \log \mathbb{P}_{\hat{v}^{(i)}} \left( (x_\tau, x_\tau^\prime, \textbf{1}_\tau^{(i)})_{\tau\in|\mathcal{Q}_t^v|}
  \right) - 
  \log \mathbb{P}_{v^{(i)}}\left( (x_\tau, x_\tau^\prime, \textbf{1}_\tau^{(i)})_{\tau\in|\mathcal{Q}_t^v|}
  \right) \leq \sqrt{32 |\mathcal{Q}^v_t| L_v^2 \log \frac{\pi^2 |\mathcal{Q}^v_t|^2}{6\delta}},
\end{equation}
where ${v^{(i)}}$ is the ground truth function.
\end{lemma}
\begin{lemma}\label{lem:close_f}
There exists an independent constant $C_L^v > 0$, such that, $\forall \epsilon > 0, \forall v_1^{(i)}, v_2^{(i)} \in \mathcal{B}^u$ that satisfies $\lVert v_1^{(i)} - v_2^{(i)} \rVert_\infty \leq \epsilon$, we have,
\begin{equation}
  \log \mathbb{P}_{v_1^{(i)}}\left( (x_\tau, x_\tau^\prime, \textbf{1}_\tau^{(i)})_{\tau\in|\mathcal{Q}_t^v|}
  \right) - 
  \log \mathbb{P}_{v_2^{(i)}}\left( (x_\tau, x_\tau^\prime, \textbf{1}_\tau^{(i)})_{\tau\in|\mathcal{Q}_t^v|}
  \right) \leq C_L^v \epsilon |\mathcal{Q}_t^v|,
\end{equation}
where $C_L^v := 1 + \frac{2}{1 + e^{-2 L_v}}$.
\end{lemma}

\paragraph{Main proof.}
We use $\mathcal{N}(\mathcal{B}^v, \epsilon, \lVert \cdot \rVert_\infty)$ to denote the covering number of the set $\mathcal{B}^v$, with $(v_j^{(i),\epsilon})_{j=1}^{\mathcal{N}(\mathcal{B}^v, \epsilon, \lVert \cdot \rVert_\infty)}$ be a set of $\epsilon$-covering for the set $\mathcal{B}^v$. Reset the `$\delta$' in Lemma~\ref{lem:fixed_v_log_P} as $\nicefrac{\delta}{\mathcal{N}(\mathcal{B}^v, \epsilon, \lVert \cdot \rVert_\infty)}$ and applying the probability union bound, we have, with probability at least $1 - \delta, \forall v_j^{(i),\epsilon}, t \geq 1, i \in V$,
\begin{equation}
  \log \mathbb{P}_{v_j^{(i),\epsilon}}\left( (x_\tau, x_\tau^\prime, \textbf{1}_\tau^{(i)})_{\tau\in|\mathcal{Q}_t^v|}
  \right) - 
  \log \mathbb{P}_{v^{(i)}}\left( (x_\tau, x_\tau^\prime, \textbf{1}_\tau^{(i)})_{\tau\in|\mathcal{Q}_t^v|}
  \right) \leq
  \sqrt{32 |\mathcal{Q}^v_t| L_v^2 \log \frac{\pi^2 |\mathcal{Q}^v_t|^2}{6\delta}}.
\end{equation}
By the definition of $\epsilon$-covering, there exists $k \in [\mathcal{N}(\mathcal{B}^v, \epsilon, \lVert \cdot \rVert_\infty)]$, such that, 
\begin{equation}
  \lVert \hat{v} - v_k^\epsilon \rVert_\infty \leq \epsilon,
\end{equation}
Hence, with probability at least $1 - \delta$,
\begin{equation}
\begin{aligned}
  &\log \mathbb{P}_{\hat{v}^{(i)}}\left( (x_\tau, x_\tau^\prime, \textbf{1}_\tau^{(i)})_{\tau\in|\mathcal{Q}_t^v|}
  \right) - 
  \log \mathbb{P}_{v^{(i)}}\left( (x_\tau, x_\tau^\prime, \textbf{1}_\tau^{(i)})_{\tau\in|\mathcal{Q}_t^v|}
  \right)\\
  = &\log \mathbb{P}_{\hat{v}^{(i)}}\left( (x_\tau, x_\tau^\prime, \textbf{1}_\tau^{(i)})_{\tau\in|\mathcal{Q}_t^v|}
  \right)
  - \log \mathbb{P}_{v_j^{(i),\epsilon}}\left( (x_\tau, x_\tau^\prime, \textbf{1}_\tau^{(i)})_{\tau\in|\mathcal{Q}_t^v|}
  \right)\\
  &+ \log \mathbb{P}_{v_j^{(i),\epsilon}}\left( (x_\tau, x_\tau^\prime, \textbf{1}_\tau^{(i)})_{\tau\in|\mathcal{Q}_t^v|}
  \right)
  - \log \mathbb{P}_{v^{(i)}}\left( (x_\tau, x_\tau^\prime, \textbf{1}_\tau^{(i)})_{\tau\in|\mathcal{Q}_t^v|}
  \right),\\
  \leq
  &C_L^v \epsilon |\mathcal{Q}^v_t| + \sqrt{32 |\mathcal{Q}^v_t| L_v^2 \log \frac{\pi^2 |\mathcal{Q}^v_t|^2 \mathcal{N}(\mathcal{B}^v, \epsilon, \lVert\cdot \rVert_\infty)}{6\delta}}.
\end{aligned}
\end{equation}
Under the isotropic norm bound assumption~\ref{assump:bounded_norm}, this easily extends to $n$ utilities,
\begin{equation}
\begin{aligned}
  &\log \mathbb{P}_{\hat{v}}\left( (x_\tau, x_\tau^\prime, \textbf{1}_\tau^v)_{\tau\in|\mathcal{Q}_t^v|}
  \right) - 
  \log \mathbb{P}_{v}\left( (x_\tau, x_\tau^\prime, \textbf{1}_\tau^{v})_{\tau\in|\mathcal{Q}_t^v|}
  \right)\\
  \leq
  &nC_L^v \epsilon |\mathcal{Q}^v_t| + \sqrt{32 |\mathcal{Q}^v_t| n^2 L_v^2 \log \frac{\pi^2 |\mathcal{Q}^v_t|^2 \mathcal{N}(\mathcal{B}^v, \epsilon, \lVert\cdot \rVert_\infty)}{6\delta}}.
\end{aligned}
\end{equation}
Similarly, the same applies to $u$ under Assumption~\ref{assump:bounded_norm},
\begin{equation}
\begin{aligned}
  &\log \mathbb{P}_{\hat{u}}\left( (x_\tau, x_\tau^\prime, \textbf{1}_\tau^u)_{\tau\in|\mathcal{Q}_t^u|}
  \right) - 
  \log \mathbb{P}_{u}\left( (x_\tau, x_\tau^\prime, \textbf{1}_\tau^{u})_{\tau\in|\mathcal{Q}_t^u|}
  \right)\\
  \leq
  &nC_L^u \epsilon |\mathcal{Q}^u_t| + \sqrt{32 |\mathcal{Q}^u_t| n^2 L_u^2 \log \frac{\pi^2 |\mathcal{Q}^u_t|^2 \mathcal{N}(\mathcal{B}^v, \epsilon, \lVert\cdot \rVert_\infty)}{6\delta}}.
\end{aligned}
\end{equation}
Therefore, the joint likelihood $\mathcal{L}_t(\tilde{u}, \tilde{A}, \tilde{v})$ is bounded by:
\begin{align}
  &\mathcal{L}_t(\tilde{u}, \tilde{A}, \tilde{v}) \notag\\
  = &\log \mathbb{P}_{\hat{v}}\left( (x_\tau, x_\tau^\prime, \textbf{1}_\tau^v)_{\tau\in|\mathcal{Q}_t^v|}
  \right) - 
  \log \mathbb{P}_{v}\left( (x_\tau, x_\tau^\prime, \textbf{1}_\tau^{v})_{\tau\in|\mathcal{Q}_t^v|}
  \right) \notag\\
  &+ \log \mathbb{P}_{\hat{u}}\left( (x_\tau, x_\tau^\prime, \textbf{1}_\tau^u)_{\tau\in|\mathcal{Q}_t^u|}
  \right) - 
  \log \mathbb{P}_{u}\left( (x_\tau, x_\tau^\prime, \textbf{1}_\tau^{u})_{\tau\in|\mathcal{Q}_t^u|}
  \right), \notag\\
  \leq
  &nC_L^v \epsilon |\mathcal{Q}^v_t| + \sqrt{32 |\mathcal{Q}^v_t| n^2 L_v^2 \log \frac{\pi^2 |\mathcal{Q}^v_t|^2 \mathcal{N}(\mathcal{B}^v, \epsilon, \lVert\cdot \rVert_\infty)}{6\delta}} \notag\\
  &+ nC_L^u \epsilon |\mathcal{Q}^u_t| + \sqrt{32 |\mathcal{Q}^u_t| n^2 L_u^2 \log \frac{\pi^2 |\mathcal{Q}^u_t|^2 \mathcal{N}(\mathcal{B}^v, \epsilon, \lVert\cdot \rVert_\infty)}{6\delta}}. \label{eq:temp_beta}
\end{align}
By range preservation lemma~\ref{lemma:range_preservation}, $L_v = L_u$, thereby $C_L^v = C_L^u$. For brevity, we introduce the notations $C_\epsilon := \frac{\pi^2 \mathcal{N}(\mathcal{B}^v, \epsilon, \lVert\cdot \rVert_\infty)}{6\delta}$, 
$|\mathcal{Q}^{uv}_t| := |\mathcal{Q}^v_t| + |\mathcal{Q}^u_t|$.

Therefore,
\begin{align}
    &\mathcal{L}_t(\tilde{u}, \tilde{A}, \tilde{v}) \notag\\
    \leq&
    n \epsilon C_L^v (|\mathcal{Q}^u_t| + |\mathcal{Q}^v_t|) + \sqrt{32 n^2 L_v^2} \left(\sqrt{|\mathcal{Q}^u_t| \log C_\epsilon |\mathcal{Q}^u_t|^2} + \sqrt{|\mathcal{Q}^v_t| \log C_\epsilon |\mathcal{Q}^v_t|^2)} \right)
    , \tag{Rearranging Eq.~(\ref{eq:temp_beta})}\\
    \leq&
    n \epsilon C_L^v |\mathcal{Q}^{uv}_t|
    + \sqrt{32 n^2 L_v^2} \left(\sqrt{|\mathcal{Q}^u_t| \log C_\epsilon |\mathcal{Q}^{uv}_t|^2} + \sqrt{|\mathcal{Q}^v_t| \log C_\epsilon |\mathcal{Q}^{uv}_t|^2)} \right)
    , \tag{$|\mathcal{Q}^{u}_t|, |\mathcal{Q}^{v}_t| < |\mathcal{Q}^{uv}_t|$}\\
    =&
    n \epsilon C_L^v |\mathcal{Q}^{uv}_t|
    + \sqrt{32 n^2 L_v^2} (\sqrt{|\mathcal{Q}^u_t|} + \sqrt{|\mathcal{Q}^v_t|}) \sqrt{\log C_\epsilon |\mathcal{Q}^{uv}_t|^2}, \tag{factor out}\\
    \leq&
    n \epsilon C_L^v |\mathcal{Q}^{uv}_t|
    + \sqrt{64 n^2 L_v^2 |\mathcal{Q}^{uv}_t| \log C_\epsilon |\mathcal{Q}^{uv}_t|^2}, \tag{Cauchy-Schwarz}\\
    =&
    n \epsilon C_L^v |\mathcal{Q}^{uv}_t|
    + \sqrt{64 n^2 L_v^2 |\mathcal{Q}^{uv}_t| \log \frac{\pi^2 |\mathcal{Q}^{uv}_t|^2 \mathcal{N}(\mathcal{B}^v, \epsilon, \lVert\cdot \rVert_\infty)}{6\delta}}, \tag{unpack $C_\epsilon$}\\
    :=&
    \beta_1(\epsilon, \delta, n, |\mathcal{Q}^{uv}_t|). \tag{define $\beta$}
\end{align}
The last inequality was derived from Cauchy-Schwarz inequality, $\sqrt{a} + \sqrt{b} \leq \sqrt{2(a+b)}$.

\section{Algorithm}
\subsection{Efficient computations}\label{sec:eff_ucb}
So far, we have introduced several optimization problems (MAP, Prob~(\ref{prob:ucb_acquisition}), Remark~\ref{remark:prediction}, Line~\ref{alg_line:query} in Alg.~\ref{alg:SBO}, and Probs.~(\ref{prob:projection})). However, the functions $\tilde{u}$ and $\tilde{v}$ exist in an infinite-dimensional space. Fortunately, by applying the representer theorem \citep{scholkopf2001generalized} and utilizing the RKHS property, we can kernelize these problems into tractable, finite-dimensional optimization problems. For example, MAP and Prob.~(\ref{prob:ucb_acquisition}) become $n(t+n)$ and $n(t+n+1)$-dimensional optimization problems, respectively. The convexity of the kernelized problems allows for scalable solutions, although the computational cost scales as $\mathcal{O}(n(t+n))$. Notably, this cost is still more efficient than the multi-task GP, which scales as $\mathcal{O}(n^3 t^3)$ and involves non-convex optimization \citep{bonilla2007multi}. See Appendix~\ref{sec:eff_ucb} for details on the kernelized problems.

\subsection{Proof of Lemma~\ref{lemma:representor}}\label{proof:representor}
\begin{lemma}[\textbf{Kernelized formulation}]\label{lemma:representor}
    MAP and (\ref{prob:ucb_acquisition}) can be recasted into convex optimisation:
    \noindent
    \small
    \begin{subequations}
    \begin{center}
    \vspace{-2em}
    \begin{tabular}{p{5.2cm}p{8.0cm}}
    \begin{equation}
    \begin{aligned}
    &\text{(Reformulated MLE)}\\
    &\max_{
    \substack{
    U_t \in \mathbb{R}^{n t}\\
    V_t \in \mathbb{R}^{n t}\\
    \tilde{A} \in \mathbb{R}^{n^2}
    }} 
    \mathcal{L}_t^\mathrm{MAP}(U_t, \tilde{A}, V_t \mid D_t)\\
    &\text{s.t.} \quad U_t^\top K^{-1}_{\mathcal{Q}_t^u} U_t \leq L_v^2,\\
    &\quad \quad V_t^\top K^{-1}_{\mathcal{Q}_t^v} V_t \leq L_v^2,\\
    &\quad \quad 1 - \delta_A \geq A_{ij} \geq \delta_A, \,\, \forall i,j,\\
    &\quad \quad \sum_{j} A_{ij} = 1, \,\, \forall j
    \label{prob:mle_rep}
    \end{aligned}
    \end{equation} &
    \begin{equation}
    \begin{aligned}
    &\text{(Reformulated acquisition function)}\\
    &\max_{\substack{
    U_t \in \mathbb{R}^{n t}, \,\, \boldsymbol{z}\in\mathbb{R}^n\\
    V_t \in \mathbb{R}^{n t}, \tilde{A} \in  \mathbb{R}^{n^2}
    }} \mathcal{A}[\boldsymbol{z}] - \mathcal{A}[\boldsymbol{z}_t]\\
    &\text{s.t.}\quad 
    \left[\begin{array}{l}
    U_t \\ \boldsymbol{z} \end{array}\right]^{\top}
    K_{\mathcal{Q}_t^u,x}^{-1}
    \left[\begin{array}{l}
    U_t \\ \boldsymbol{z}
    \end{array}\right]   
    \leq L_v^2, \\
    &\quad \quad V_t^\top K^{-1}_{\mathcal{Q}_t^v} V_t \leq L_v^2,\\
    &\quad \quad\mathcal{L}_t^\mathrm{MAP}(U_t, \tilde{A}, V_t \mid D_t) \geq \mathcal{L}_t^\mathrm{MAP}(\hat{u}_t, \hat{A}_t, \hat{v}_t) - \beta_t(|\mathcal{Q}_t^{uv}|),\\
    &\quad \quad\ell_t(U_t \mid D_{\mathcal{Q}^u_t}) \geq \ell_t(\hat{U}_t \mid D_{\mathcal{Q}^u_t}) - \beta_t^u(|\mathcal{Q}_t^{u}|),\\
    &\quad \quad \ell_t(V_t \mid D_{\mathcal{Q}^v_t}) \geq \ell_t(\hat{V}_t \mid D_{\mathcal{Q}^v_t}) - \beta_t^v(|\mathcal{Q}_t^{v}|),\\
    &\quad \quad 1 - \delta_A \geq A_{ij} \geq \delta_A, \,\, \forall i,j,\\
    &\quad \quad \sum_{j} A_{ij} = 1, \,\, \forall j
    \label{prob:acquisition_rep}
    \end{aligned}
    \end{equation}
    \end{tabular}
    \end{center}
    \vspace{-2em}
    \end{subequations}
    \normalsize
\end{lemma}
\begin{proof}
We begin by the convexity, then we show the equivalence to MAP, and the acquisition function maximisation problem~(\ref{prob:ucb_acquisition}).

\subsubsection{Proof of convexity.}
\paragraph{Reformulated MLE}
The function $\psi_\tau(y,y^\prime) := \log(e^y+e^{y^\prime}) - p_\tau y - (1 - p_\tau) y^\prime$, $p_\tau \in \{0, 1\}$ is a convex function because $\nabla \psi_\tau(y_\tau, y_\tau^\prime) = 0$, thus the Hessian is nonnegative. Then, when assume $z_\tau = \tilde{v}(x_\tau), \eta_\tau = \tilde{u}(x_\tau)$, our negative log likelihood function $-\mathcal{L}_t(z_\tau, z_\tau^\prime, \eta_\tau, \eta_\tau^\prime)$ is also a convex function with $\nabla \mathcal{L}_t(z_\tau, z_\tau^\prime, \eta_\tau, \eta_\tau^\prime) = 0$.

For graph convolution part, the function $\psi_\tau^\prime(A) := \log(e^{Au}+e^{A u^\prime}) - p_\tau A u - (1 - p_\tau) A u^\prime$ is also convex with respect to A because its Hessian is nonnegative:
\begin{align}
    \frac{\partial^2}{\partial A^2} \psi_\tau^\prime(A)
    = (u - u^\prime)^2 \frac{e^{Au}e^{A u^\prime}}{(e^{Au} + e^{A u^\prime})^2} 
    \geq 0.
\end{align}
Similarly, this function is convex with respect to $u$ and $u^\prime$. We introduce the function $P(u, u^\prime) = \nicefrac{e^{Au}}{e^{Au} + e^{Au^\prime}}$
\begin{align}
    \frac{\partial^2}{\partial u^2} \psi_\tau^\prime(u)
    &= A^2 P(u) (1 - P(u)),\\
    \frac{\partial^2}{\partial {u^\prime}^2} \psi_\tau^\prime(u)
    &= A^2 P(u^\prime) (1 - P(u^\prime)),\\
    \frac{\partial^2}{\partial u \partial u^\prime} \psi_\tau^\prime(u, u^\prime)
    &= - A^2 P(u, u^\prime) (1 - P(u, u^\prime)),
\end{align}
Therefore, Hessian matrix $H$ is
\begin{align}
    H &= 
    A^2 P(u,u^\prime) (1 - P(u,u^\prime))
    \begin{bmatrix}
    1 & -1 \\
    -1 & 1
    \end{bmatrix}
\end{align}
and the Hessian matrix is positive semi-definite because all eigenvalues are non-negative. Thus, this is also convex.

\paragraph{Reformulated acquisition function}
The GSF is the convex combination of utilities independent of both $\tilde{A}$ and $\boldsymbol{z}$. Thus, the aggregate operation is simply reduced to the linear combination of $\boldsymbol{z}$. Under the convex constraint of optimistic MLE, the linear combination of convex functions with nonzero weights is also a convex function. And the weight function of GSF is nonzero by definition.

\subsubsection{Proof of MLE reformulation.}
The joint likelihood $\mathcal{L}_t$ in Eq.~(\ref{eq:joint_likelihood}) only depends on the values $(\hat{u}(x_\tau,:), \hat{u}(x_\tau^\prime,:), \hat{v}(x_\tau,:), \hat{v}(x_\tau^\prime,:)) = (u_\tau, u_\tau^\prime, v_\tau, v_\tau^\prime)$, where $u_\tau = (u_\tau^{(i)})_{i \in [n]}$.
As such, we only need to optimise over $(u_\tau, u_\tau^\prime, v_\tau, v_\tau^\prime)$ subject to that they are functions in $\mathcal{H}_{k_i}$ with norm less or equal to $L_v$. 
Furthermore, Algorithm~\ref{alg:SBO} sets $x^\prime_\tau = x_{\tau-1}$, thereby $u_\tau^\prime = u_{\tau-1}$, $v_\tau^\prime = v_{\tau-1}$. Hence, we can further reduce the optimisation variables to $(u_\tau, v_\tau)$. 

Here, we only assume $\tilde{u}(\cdot, i) \in \mathcal{H}_{k_i}$, then the norm bound constraints are only subject to $\tilde{u}$. Note that our kernel is vector-valued, so we use the following notation to describe:
\begin{align}
    U_t^\top K_{\mathcal{Q}_t^u}^{-1} U_t := \left( (U_t^{(i)})^\top (K^{(i)}_{\mathcal{Q}_t^u})^{-1} U_t^{(i)} \right)_{i=1}^n
\end{align}
where $K^{(i)}_{\mathcal{Q}_t^u} := (k_v(x_{\tau_1}, x_{\tau_2})_{\tau1, \tau_2 \in \mathcal{Q}_t^u}$. The same applies to $V_t$ and corresponding kernel. As such, the constraint in Prob.~(\ref{prob:mle_rep}) consists of $n$ kernel bound constraints. Each constraint is direct application of representor theorem \citep{scholkopf2001generalized}.

\subsubsection{Proof of acquisition function reformulation.}
Prob~(\ref{prob:ucb_acquisition}) can be formally written as
\begin{equation}
\begin{aligned}
    \max_{\tilde{u}} \quad &\mathcal{A}[\tilde{u}(x, :)] - \mathcal{A}[\tilde{u}(x_t, :)],\\
    \text{s.t. }\quad 
    &\tilde{u}, \tilde{A}, \tilde{v} \in \mathcal{B}^{u,A,v},\\
    &\mathcal{L}_t(\tilde{u}, \tilde{A}, \tilde{v}) \geq \mathcal{L}_t(\hat{u}, \hat{A}, \hat{v}) - \beta_t^{u,A,v}.
\end{aligned}\label{prob:acquisition_infin}
\end{equation}

This has an infinite-dimensional function variable, thereby being intractable. Similar to the MLE reformulation, we can recast to finite, tractable optimization problem. 

\paragraph{Simplest setting.}
First, we will start with the simplest case where $|\mathcal{Q}^u_t| = T$, as such
\begin{equation}
\begin{aligned}
    \max_{\tilde{u}} \quad &\mathcal{A}[\tilde{u}(x, :)] - \mathcal{A}[\tilde{u}(x_t, :)],\\
    \text{s.t. }\quad 
    &\tilde{u} \in \mathcal{B}^{u},\\
    &\mathcal{L}_t(\tilde{u}) \geq \mathcal{L}_t(\hat{u}) - \beta_t^{u}.
\end{aligned}\label{prob:acquisition_infin2}
\end{equation}

And we show the equivalence to the following kernelized formulation.
\begin{equation}
\begin{aligned}
&\max_{
Z_{0:t} \in \mathbb{R}^{n (t + 1)}, \,\, \boldsymbol{z}\in\mathbb{R}^n
} \mathcal{A}[\boldsymbol{z}] - \mathcal{A}[\boldsymbol{z}_t]\\
&\text{s.t.}\quad 
\left[\begin{array}{l}
Z_{0:t} \\ \boldsymbol{z} \end{array}\right]^{\top}
K_{\mathcal{Q}_t^u,x}^{-1}
\left[\begin{array}{l}
Z_{0:t} \\ \boldsymbol{z}
\end{array}\right]   
\leq L_u^2, \\
&\quad \quad Z_{0:t}^\top K^{-1}_{0:t} Z_{0:t} \leq L_v^2,\\
&\quad \quad \ell_t(Z_{0:t}) \geq \mathcal{L}_t(\hat{u}_t) - \beta^u_t.
\label{prob:acquisition_rep2}
\end{aligned}
\end{equation}

Let $\tilde{u}$ be any feasible solution of the above innter optimisaton problem, and $\tilde{z} = \tilde{u}(x,:)$ and $\tilde{Z}_{0:t} = (\tilde{u}(x_\tau,:))_{\tau=0}^t$ be the corresponding utility value. Consider the minimum-norm interpolation problem,
\begin{equation}
\begin{aligned}
    \min_{s \in \mathcal{B}^v} \quad 
    &\lVert s \rVert^2\\
    \text{s.t. }\quad 
    &s(x_\tau, :) = \tilde{z}_\tau, \forall \tau \in \{0\} \cup [t],\\
    &s(x) = \tilde{z}.
\end{aligned}\label{prob:min_interpolate}
\end{equation}
By representer theorem, this problem admits an optimal solution with the form $\alpha^\top k_{0:t, x}$, where $k_{0:t, x} := \{ k(w, \cdot) \}_{w \in \{x_0,\cdots,x_t, x\}}$. Thus, Prob.~(\ref{prob:min_interpolate}) can be reduced to
\begin{equation}
\begin{aligned}
    \min_{\alpha \in \mathbb{R}^{t+2}} \quad 
    &\alpha^\top K_{0:t, x} \alpha\\
    \text{s.t. }\quad 
    &K_{0:t, x} \alpha = \left[\begin{array}{l}
\tilde{Z}_{0:t} \\ \tilde{z} \end{array}\right].
\end{aligned}\label{prob:min_interpolate_rep}
\end{equation}
Then the optimal solution of this Prob.~(\ref{prob:min_interpolate_rep}) is
\begin{equation}
\begin{aligned}
    \alpha^\top K_{0:t, x} \alpha
    &= (K_{0:t, x} \alpha)^\top K_{0:t, x}^{-1} K_{0:t, x} \alpha,\\
    &= 
    \left[\begin{array}{l}
\tilde{Z}_{0:t} \\ \tilde{z} \end{array}\right]^T K_{0:t, x}^{-1} \left[\begin{array}{l}
\tilde{Z}_{0:t} \\ \tilde{z} \end{array}\right].
\end{aligned}
\end{equation}
Since $\tilde{u}$ is an interpolant by construction of $(\tilde{Z}_{0:t}, \tilde{z})$. We have 
\begin{equation}
\begin{aligned}
    \left[\begin{array}{l}
\tilde{Z}_{0:t} \\ \tilde{z} \end{array}\right]^T K_{0:t, x}^{-1} \left[\begin{array}{l}
\tilde{Z}_{0:t} \\ \tilde{z} \end{array}\right] \leq \lVert u \rVert^2 \leq L_u^2,
\end{aligned}
\end{equation}
yielding the first constraint. As the LL function only depends on $(\tilde{Z}_{0:t})$, it holds that
\begin{equation}
    \mathcal{L}(\tilde{Z}_{0:t}, \mid D_{\mathcal{Q}^u_t}) = \mathcal{L}_t(\tilde{v}) \geq \mathcal{L}_t(\hat{u}_t) - \beta_t^u,
\end{equation}
and the objective satisfy
\begin{equation}
    \mathcal{A}[\boldsymbol{z}] - \mathcal{A}[\boldsymbol{z}_t]
    = \mathcal{A}[\tilde{u}(x,:)] - \mathcal{A}[\tilde{u}(x_t,:)].
\end{equation}
Therefore, a set $(\tilde{Z}_{0:t}, \tilde{z})$ is a feasible solution for Prob.~(\ref{prob:acquisition_rep2}), with the same objective as $\tilde{v}$ for the the infinite dimensional Prob.~(\ref{prob:acquisition_infin2}).

Next, we show that for any feasible solution for Prob.~(\ref{prob:acquisition_rep2}), we can find a corresponding feasible solution of Prob.~(\ref{prob:acquisition_infin2}) with the same objective value. 
Let $(Z_{0:t}, z)$ be a feasible solution of Prob.~(\ref{prob:acquisition_rep2}). We construct 
\begin{equation}
\begin{aligned}
    \tilde{u}_z = \left[\begin{array}{l}
\tilde{Z}_{0:t} \\ \tilde{z} \end{array}\right]^T 
K_{0:t, x}^{-1} k_{0:t, x}(\cdot),
\end{aligned}
\end{equation}
Hence,
\begin{equation}
\begin{aligned}
    \lVert \tilde{u}_z \rVert^2 = \left[\begin{array}{l}
\tilde{Z}_{0:t} \\ \tilde{z} \end{array}\right]^T 
K_{0:t, x}^{-1} \left[\begin{array}{l}
\tilde{Z}_{0:t} \\ \tilde{z} \end{array}\right] \leq L_u^2,
\end{aligned}
\end{equation}
and it can be checked that $\tilde{u}_z(x_\tau) = \boldsymbol{z}_\tau, \forall \tau \in \{0\} \cup [t]$ and $\tilde{u}_z(x) = \boldsymbol{z}$. So $\mathcal{L}_t(\tilde{u}_z) = \mathcal{L}(\tilde{Z}_{0:t} \mid D_t) \geq \mathcal{L}(\hat{v}) - \beta_t^u$. 
And the objectives satisfy $\mathcal{A}[\tilde{u}_z(x)] - \mathcal{A}[\tilde{u}_z(x_t)] = \mathcal{A}[\boldsymbol{z}] - \mathcal{A}[\boldsymbol{z}_t]$. So it is proved that for any feasible solution of Prob.~(\ref{prob:acquisition_rep2}), we can find a corresponding feasible solution of Prob.~(\ref{prob:acquisition_infin2}) with the same objective value.

\paragraph{Original setting}
The $V_t$ constraint is the same with the previous MLE reformulation. Also, the optimistic MLE bound needs to modify to $\mathcal{L}_t(\hat{u}, \hat{A}, \hat{v})$ as it involves $A$ and $v$ estimate. Then, we can show the equivalence of this Prob.~(\ref{prob:acquisition_infin}) with the constraint in Prob.~(\ref{prob:acquisition_rep}).
\end{proof}

\subsection{Predictive confidence bound}
Using the same idea, we can obtain the predictive confidence bounds. Here, $x$ is given as the prediction point, then the upper confidence bound $\overline{u}(x,i)$ and lower confidence bound $\underline{u}(x,i)$ become
\begin{equation}
\begin{aligned}
    \overline{u}(x,i) := &\max_{\substack{
    U_t \in \mathbb{R}^{t+1} \,\, V_t \in \mathbb{R}^{t}, \boldsymbol{z}\in\mathbb{R}\\ \tilde{A} \in \mathbb{R}^{n^2},
    }} \boldsymbol{z} \\
    &\text{s.t.}\quad 
    \left[\begin{array}{l}
    U_t \\ \boldsymbol{z} \end{array}\right]^{\top}
    K_{\mathcal{Q}^u_t,x}^{-1}
    \left[\begin{array}{l}
    U_t \\ \boldsymbol{z}
    \end{array}\right]   
    \leq L_v^2, \\
    &\quad \quad V_t^\top K^{-1}_{\mathcal{Q}_t^v} V_t \leq L_v^2,\\
    &\quad \quad\mathcal{L}_t^\mathrm{MAP}(U_t, \tilde{A}, V_t \mid D_t) \geq \mathcal{L}_t^\mathrm{MAP}(\hat{u}_t, \hat{A}_t, \hat{v}_t) - \beta_t(|\mathcal{Q}_t^{uv}|),\\
    &\quad \quad\ell_t(U_t \mid D_{\mathcal{Q}^u_t}) \geq \ell_t(\hat{U}_t \mid D_{\mathcal{Q}^u_t}) - \beta_t^u(|\mathcal{Q}_t^{u}|),\\
    &\quad \quad \ell_t(V_t \mid D_{\mathcal{Q}^v_t}) \geq \ell_t(\hat{V}_t \mid D_{\mathcal{Q}^v_t}) - \beta_t^v(|\mathcal{Q}_t^{v}|),\\
    &\quad \quad 1 - \delta_A \geq A_{ij} \geq \delta_A, \,\, \forall i,j,\\
    &\quad \quad \sum_{j} A_{ij} = 1, \,\, \forall j
\end{aligned}\label{prob:max_ucb_rep}
\end{equation}

\begin{equation}
\begin{aligned}
    \underline{u}(x,i) := &\min_{\substack{
    U_t \in \mathbb{R}^{t+1} \,\, V_t \in \mathbb{R}^{t}, \boldsymbol{z}\in\mathbb{R}\\ \tilde{A} \in \mathbb{R}^{n^2},
    }} \boldsymbol{z} \\
    &\text{s.t.}\quad 
    \left[\begin{array}{l}
    U_t \\ \boldsymbol{z} \end{array}\right]^{\top}
    K_{\mathcal{Q}^u_t,x}^{-1}
    \left[\begin{array}{l}
    U_t \\ \boldsymbol{z}
    \end{array}\right]   
    \leq L_v^2, \\
    &\quad \quad V_t^\top K^{-1}_{\mathcal{Q}_t^v} V_t \leq L_v^2,\\
    &\quad \quad\mathcal{L}_t^\mathrm{MAP}(U_t, \tilde{A}, V_t \mid D_t) \geq \mathcal{L}_t^\mathrm{MAP}(\hat{u}_t, \hat{A}_t, \hat{v}_t) - \beta_t(|\mathcal{Q}_t^{uv}|),\\
    &\quad \quad\ell_t(U_t \mid D_{\mathcal{Q}^u_t}) \geq \ell_t(\hat{U}_t \mid D_{\mathcal{Q}^u_t}) - \beta_t^u(|\mathcal{Q}_t^{u}|),\\
    &\quad \quad \ell_t(V_t \mid D_{\mathcal{Q}^v_t}) \geq \ell_t(\hat{V}_t \mid D_{\mathcal{Q}^v_t}) - \beta_t^v(|\mathcal{Q}_t^{v}|),\\
    &\quad \quad 1 - \delta_A \geq A_{ij} \geq \delta_A, \,\, \forall i,j,\\
    &\quad \quad \sum_{j} A_{ij} = 1, \,\, \forall j
\end{aligned}\label{prob:max_lcb_rep}
\end{equation}

\subsection{Projection weight function}
We decompose the projection weight function to the following:
\begin{align}
    w^u_t(x_t, x_t^\prime) := \lVert \overline{\delta}^u_t(x_t, x_t^\prime, :) - \underline{\delta}^u_t(x_t, x_t^\prime, :) \rVert,
\end{align}
where 
\begin{equation}
\begin{aligned}
    \overline{\delta}^u_t (x_t,x_t^\prime, i) = &\max_{\substack{
    U_t \in \mathbb{R}^{t+1} \,\, V_t \in \mathbb{R}^{t}, \boldsymbol{z}\in\mathbb{R}\\ \tilde{A} \in \mathbb{R}^{n^2},
    }} \boldsymbol{z} - \max_{t \in \mathcal{Q}^u_t} U_t \\
    &\text{s.t.}\quad 
    \left[\begin{array}{l}
    U_t \\ \boldsymbol{z} \end{array}\right]^{\top}
    K_{\mathcal{Q}^u_t,x}^{-1}
    \left[\begin{array}{l}
    U_t \\ \boldsymbol{z}
    \end{array}\right]   
    \leq L_v^2, \\
    &\quad \quad V_t^\top K^{-1}_{\mathcal{Q}_t^v} V_t \leq L_v^2,\\
    &\quad \quad\mathcal{L}_t^\mathrm{MAP}(U_t, \tilde{A}, V_t \mid D_t) \geq \mathcal{L}_t^\mathrm{MAP}(\hat{u}_t, \hat{A}_t, \hat{v}_t) - \beta_t(|\mathcal{Q}_t^{uv}|),\\
    &\quad \quad\ell_t(U_t \mid D_{\mathcal{Q}^u_t}) \geq \ell_t(\hat{U}_t \mid D_{\mathcal{Q}^u_t}) - \beta_t^u(|\mathcal{Q}_t^{u}|),\\
    &\quad \quad \ell_t(V_t \mid D_{\mathcal{Q}^v_t}) \geq \ell_t(\hat{V}_t \mid D_{\mathcal{Q}^v_t}) - \beta_t^v(|\mathcal{Q}_t^{v}|),\\
    &\quad \quad 1 - \delta_A \geq A_{ij} \geq \delta_A, \,\, \forall i,j,\\
    &\quad \quad \sum_{j} A_{ij} = 1, \,\, \forall j
\end{aligned}
\end{equation}
\begin{equation}
\begin{aligned}
    \underline{\delta}^u_t (x_t,x_t^\prime, i) = &\min_{\substack{
    U_t \in \mathbb{R}^{t+1} \,\, V_t \in \mathbb{R}^{t}, \boldsymbol{z}\in\mathbb{R}\\ \tilde{A} \in \mathbb{R}^{n^2},
    }} \boldsymbol{z} - \max_{t \in \mathcal{Q}^u_t} U_t \\
    &\text{s.t.}\quad 
    \left[\begin{array}{l}
    U_t \\ \boldsymbol{z} \end{array}\right]^{\top}
    K_{\mathcal{Q}^u_t,x}^{-1}
    \left[\begin{array}{l}
    U_t \\ \boldsymbol{z}
    \end{array}\right]   
    \leq L_v^2, \\
    &\quad \quad V_t^\top K^{-1}_{\mathcal{Q}_t^v} V_t \leq L_v^2,\\
    &\quad \quad\mathcal{L}_t^\mathrm{MAP}(U_t, \tilde{A}, V_t \mid D_t) \geq \mathcal{L}_t^\mathrm{MAP}(\hat{u}_t, \hat{A}_t, \hat{v}_t) - \beta_t(|\mathcal{Q}_t^{uv}|),\\
    &\quad \quad\ell_t(U_t \mid D_{\mathcal{Q}^u_t}) \geq \ell_t(\hat{U}_t \mid D_{\mathcal{Q}^u_t}) - \beta_t^u(|\mathcal{Q}_t^{u}|),\\
    &\quad \quad \ell_t(V_t \mid D_{\mathcal{Q}^v_t}) \geq \ell_t(\hat{V}_t \mid D_{\mathcal{Q}^v_t}) - \beta_t^v(|\mathcal{Q}_t^{v}|),\\
    &\quad \quad 1 - \delta_A \geq A_{ij} \geq \delta_A, \,\, \forall i,j,\\
    &\quad \quad \sum_{j} A_{ij} = 1, \,\, \forall j
\end{aligned}
\end{equation}

\subsection{Proof of Theorem~\ref{thm:regret}}\label{proof:regret}
We begin by introducing the known proofs.

\subsubsection{Known results}
We introduce useful theorems from literature.
\begin{theorem}[Theorem 3.6 in \citet{xu2024principled}]\label{thm:duel_error}
    For any estimate $\tilde{v}_{t+1} \in \mathcal{B}^v_{t+1}$ measurable with respect to the filtration $\mathcal{F}_t$, we have, with probability at least $1 - \delta, \forall t \geq 1, (x,x^\prime) \in \mathcal{X} \times \mathcal{X}$,
    \begin{align}
        \lvert (\tilde{v}_{t+1}(x) - \tilde{v}_{t+1}(x^\prime)) - (v(x) - v(x^\prime)) \rvert
        \leq
        2 \left( 2 L_v + \lambda^{-\nicefrac{1}{2}} \sqrt{\beta(\epsilon, \nicefrac{\delta}{2}, |\mathcal{Q}_t^v|)} \right)
        \sigma^{vv^\prime}_{t+1}(x, x^\prime). 
    \end{align}
    where
    \begin{align}
        \beta(\epsilon, \nicefrac{\delta}{2}, |\mathcal{Q}_t^v|)
        &= \mathcal{O} \left(
        \sqrt{|\mathcal{Q}^v_t| \log \frac{|\mathcal{Q}^v_t| \mathcal{N}(\mathcal{B}^v, \epsilon, \lVert \cdot \rVert_\infty)}{\delta}} + \epsilon |\mathcal{Q}^v_t| + \epsilon^2 |\mathcal{Q}^v_t|
        \right),\\
        \left( \sigma^{vv^\prime}_{t+1}(x, x^\prime) \right)^2 &= k^{vv^\prime}(\omega, \omega) - k^{vv^\prime}(\omega_{1:t}, \omega)^\top \left( K^{vv^\prime}_{t-1} + \lambda I \right) k^{vv^\prime}(\omega_{1:t}, \omega),\\
        k^{vv^\prime}((x, x^\prime), (y,y^\prime)) &:= k(x,y) + k(x^\prime, y^\prime),\\
        \omega &:= (x, x^\prime),\\
        \omega_{1:t-1} &:= \left( (x_\tau, x^\prime_\tau) \right)_{\tau=1}^{t-1},\\
        K^{vv^\prime}_{t-1} &:= \left( k^{vv^\prime}((x_{\tau_1}, x^\prime_{\tau_1}), (x_{\tau_2}, x^\prime_{\tau_2}))
        \right)_{\tau_1 \in [t-1], \tau_2 \in [t-1]},
    \end{align}
    and $\lambda$ is a positive regularization constant.
\end{theorem}

\subsubsection{Supporting results}
We introduce the supporting lemmas for the main proof.
\begin{theorem}\label{thm:cumulative_error}
    With probability at least $1 - \delta$, for $\tilde{v}_t\in\mathcal{B}_t^v$ that is measurable with respect to the filtration $\mathcal{F}_t$,
    \begin{align}
        \sum_{t\in\mathcal{Q}_t^v}\lvert \tilde{v}_{t}(x_t) - \tilde{v}_{t}(x_t^\prime) - (v(x_t) - v(x_t^\prime)) \rvert
        \leq\sum_{t\in\mathcal{Q}_t^v} w_t^v(x_t, x_t^\prime) 
        = \mathcal{O}\left( \sqrt{\beta_T \gamma^{vv^\prime}_T |\mathcal{Q}_t^v|} \right)
    \end{align}
    where
    \begin{align}
        \beta_T &:= \beta(\nicefrac{1}{T}, \delta, |\mathcal{Q}_t^v|)\\
        &= \mathcal{O} \left(
        \sqrt{|\mathcal{Q}^v_t| \log \frac{|\mathcal{Q}^v_t| \mathcal{N}(\mathcal{B}^v, \nicefrac{1}{T}, \lVert \cdot \rVert_\infty)}{\delta}} \right),\\
        \gamma^{vv^\prime}_T &:= \max_{\Omega \subset \mathcal{X} \times \mathcal{X}; |\Omega| = |\mathcal{Q}_t^v|} \frac{1}{2} \log \left\lvert
        I + \lambda^{-1} K^{vv^\prime}_{\Omega}
        \right\rvert,\\
        K^{vv^\prime}_{\Omega} &:= 
        \left( k^{vv^\prime}((x, x^\prime), (y, y^\prime))
        \right)_{(x, x^\prime), (y, y^\prime) \in \Omega}.
    \end{align}
\end{theorem}
\begin{proof}
The first inequality follows by the definition. 
\begin{align}
&\sum_{t\in\mathcal{Q}_T^v}w_t^v(x_t, x_t^\prime)\\
=&\sum_{t\in\mathcal{Q}_T^v}\sup_{\tilde{v}\in\mathcal{B}_t^v}\lvert (\tilde{v}(x_t) - \tilde{v}(x_t^\prime)) - (v(x_t) - v(x_t^\prime)) \rvert\\
\leq&\sum_{t\in\mathcal{Q}_T^v}
        2 \left( 2 L_v + \lambda^{-\nicefrac{1}{2}} \sqrt{\beta(\epsilon, \nicefrac{\delta}{2}, |\mathcal{Q}_t^v|)} \right)
        \sigma^{vv^\prime}_{t+1}(x, x^\prime)\tag{Thm.~\ref{thm:duel_error}}\\
\leq&\left( 2 L_v + \lambda^{-\nicefrac{1}{2}} \sqrt{\beta(\epsilon, \nicefrac{\delta}{2}, |\mathcal{Q}_T^v|)} \right) \sum_{t\in\mathcal{Q}_T^v}\sigma^{vv^\prime}_{t+1}(x, x^\prime) \tag{Monotonicity of $\beta$ in $t$} \\
\leq&\mathcal{O}\left( \sqrt{\beta_T \gamma^{vv^\prime}_T |\mathcal{Q}_T^v|} \right), \tag{Lem. 4 in \cite{kern_bandits}} 
\end{align} 
\end{proof}

We further introduce useful definitions to keep simple notations.
\begin{definition}[\textbf{Sorting basis swap}]\label{def:sorting}
    Let $\phi(\textbf{a})$ be a sorting function that reorders the $n$-dimensional input vector $\textbf{a}$ to be in an ascending order, and $\phi_\textbf{b}(\textbf{a})$ be a sorting function that reorders the $n$-dimensional input vector $\textbf{a}$ based on the ascending order of the vector $\textbf{b}$. Thus, $\phi(\textbf{a}) = \phi_\textbf{a}(\textbf{a})$, but $\phi_\textbf{a}(\textbf{a}) \neq \phi_\textbf{b}(\textbf{a})$ when $\textbf{a} \neq \textbf{b}$.
\end{definition}

\begin{definition}[\textbf{Weight sorting}]\label{def:weight_sort}
    Let $\phi_\textbf{a}^{-1}(\textbf{w})$ be a sorting function that reorders the $n$-dimensional weight vector $\textbf{w}$ based on the ascending order of the vector $\textbf{a}$. Thus, $\textbf{w}^\top \phi_\textbf{b}(\textbf{a}) = \phi_\textbf{b}^{-1}(\textbf{w})^\top \textbf{a}$.
\end{definition}

\begin{lemma}[\textbf{Hardy-Littlewood-Polya inequality}]\label{lemma:hlp}
    Using the notation in Definition~\ref{def:sorting}, \ref{def:weight_sort}, $\forall \textbf{u}_1, \textbf{u}_2 \in \mathcal{B}^u, \textbf{u}_1 \neq \textbf{u}_2, \mathcal{A}(\textbf{u}) = \textbf{w}^\top \phi(\textbf{u})$,
    \begin{align}
        \textbf{w}^\top \phi_{\textbf{u}_1}(\textbf{u}_1) \leq \textbf{w}^\top \phi_{\textbf{u}_2}(\textbf{u}_1) \leq
        \textbf{w}^\top \phi_{-\textbf{u}_1}(\textbf{u}_1).
    \end{align}
    Or equivalently,
    \begin{align}
        \phi_{\textbf{u}_1}^{-1}(\textbf{w})^\top \textbf{u}_1 \leq \phi_{\textbf{u}_2}^{-1}(\textbf{w})^\top \textbf{u}_1 \leq \phi_{-\textbf{u}_1}^{-1}(\textbf{w})^\top \textbf{u}_1.
    \end{align}
\end{lemma}

\begin{assump}[\textbf{Optimality of acquisition function}]\label{assump:optimality}
    Let $\hat{u} = \argmax_{\tilde{u} \in \mathcal{B}^u_{t}} \mathcal{A}[\tilde{u}(x_t, :)] - \mathcal{A}[\tilde{u}(x_{t-1}, :)]$ be the estimated upper bound of social utility difference. We have
    \begin{align}
        \hat{u}(x_t, :) - u(x_t, :) &> \textbf{0},\\
        u(x_{t-1}, :) - \hat{u}(x_{t-1}, :) &> \textbf{0},
    \end{align}
    with probability at least $1 - \delta$.
\end{assump}
Assumption~\ref{assump:optimality} is justifiable as GSF is a monotonic function (see Proposition~\ref{proposition:gsf}), i.e., if $\mathcal{A}[\textbf{u}_1] \geq \mathcal{A}[\textbf{u}_2]$, then $\textbf{u}_1 \succeq \textbf{u}_2$ holds. The acquisition function maximizes the difference $\mathcal{A}[\tilde{u}(x_t, :)] - \mathcal{A}[\tilde{u}(x_{t-1}, :)]$, this is effectively viewed as $\max_{\tilde{u} \in \mathcal{B}^u_{t}} \mathcal{A}[\tilde{u}(x_t, :)] - \min_{\tilde{u} \in \mathcal{B}^u_{t}} \mathcal{A}[\tilde{u}(x_{t-1}, :)]$. By Lemma~\ref{lemma:conf_set}, $\max_{\tilde{u} \in \mathcal{B}^u_{t}} \mathcal{A}[\tilde{u}(x_t, :)] \geq u(x_t, :)$ and $\min_{\tilde{u} \in \mathcal{B}^u_{t}} \mathcal{A}[\tilde{u}(x_{t-1}, :)] \leq u(x_{t-1}, :)$ hold with probability at least $1 - \delta$. Thus, when we view $\max_{\tilde{u} \in \mathcal{B}^u_{t}} \mathcal{A}[\tilde{u}(x_t, :)] \approx \mathcal{A}[\hat{u}(x_t, :)]$ and $\min_{\tilde{u} \in \mathcal{B}^u_{t}} \mathcal{A}[\tilde{u}(x_{t-1}, :)] \approx \mathcal{A}[\hat{u}(x_{t-1}, :)]$, then monotonicity offers $\hat{u}(x_t, :) \succ u(x_t, :)$ and $u(x_{t-1}, :) \succ \hat{u}(x_{t-1}, :)$. Therefore, assumption can be held.

\begin{assump}[\textbf{Lipchitz continuity}]\label{assump:lipchitz}
    Under Assumption~\ref{assump:optimality}, let $\textbf{u}_1 = \hat{u}(x_t, :) - u(x_t,:), \textbf{u}_2 = u(x_{t-1}, :) - \hat{u}(x_{t-1}, :)$ be utility vectors that are strictly positive $\textbf{u}_1, \textbf{u}_2 > \textbf{0}$ and non-identical $\lVert \textbf{u}_1 - \textbf{u}_2 \rVert \geq \delta_u > 0$ is strictly positive, $\textbf{w}_1 = \phi^{-1}_{\textbf{u}_1}(\textbf{w}), \textbf{w}_2 = \phi^{-1}_{\textbf{u}_2}(\textbf{w})$ be the weight vectors that reorders $\textbf{w} > \textbf{0}$ based on corresponding sorting basis. Then we assume the following Lipchitz continuity
    \begin{align}
        \lVert \textbf{w}_1 - \textbf{w}_2 \lVert 
        \leq C_{L_1} \lVert \textbf{u}_1 - \textbf{u}_2 \rVert,\\
        \lVert \textbf{u}_1 + \textbf{u}_2 \lVert \leq C_{L_2} \lVert \textbf{u}_1 - \textbf{u}_2 \rVert,
    \end{align}
    where $C_{L_1}, C_{L_2}$ are a Lipchitz constant.
\end{assump}
Lipchitz continuity assumption for sorting function can be seen in many literature (e.g., \citet{anil2019sorting, nguyen2021lipschitz}). If $\textbf{u}_1 = \textbf{u}_2$, $\textbf{w}_1 = \textbf{w}_2$ hold due to the sorting function, thus we can view this inequality translates the alignment of weight space into utility space. As the GSF is a piecewise linear function, and it is Lipchitz continuous. Here, we also emphasize that $\lVert \textbf{u}_1 - \textbf{u}_2 \rVert > \delta_u$ because $\textbf{u}_1$ is conditioned on $x_t$ but $\textbf{u}_2$ is conditioned on $x_{t-1}$. As shown in Theorem~\ref{thm:cumulative_error}, the error bound of pointwise utility estimation is submodular, i.e., monotonically decreasing function. Thus, the right hand side is always positive.

\begin{lemma}[\textbf{weighted difference}]\label{lemma:owa}
    Under Assumption~\ref{assump:lipchitz}, we have
    \begin{align}
        \lvert \phi^{-1}_{\textbf{u}_1}(\textbf{w})^\top \textbf{u}_1 - \phi^{-1}_{\textbf{u}_2}(\textbf{w})^\top \textbf{u}_2 \rvert \leq L^\prime_w \lVert \textbf{u}_1 - \textbf{u}_2 \rVert,
    \end{align}
    where $L^\prime_w = \lVert \textbf{w} \rVert + L_u C_{L_1}$.
\end{lemma}
\begin{proof}
    Let us denote $\phi^{-1}_{\textbf{u}_1}(\textbf{w}) = \textbf{w}_1$, $\phi^{-1}_{\textbf{u}_2}(\textbf{w}) = \textbf{w}_2$. We have
    \begin{align}
        &\lvert \phi^{-1}_{\textbf{u}_1}(\textbf{w})^\top \textbf{u}_1 - \phi^{-1}_{\textbf{u}_2}(\textbf{w})^\top \textbf{u}_2 \rvert \\
        =&
        \lvert \textbf{w}_1^\top \textbf{u}_1 - \textbf{w}_1^\top \textbf{u}_2 + \textbf{w}_1^\top \textbf{u}_2 - \textbf{w}_2^\top \textbf{u}_2 \rvert \tag{Adding and substract the same term} \\
        \leq&
        \lvert \textbf{w}_1^\top (\textbf{u}_1 - \textbf{u}_2) \rvert + \lvert (\textbf{w}_1^\top  - \textbf{w}_2)^\top \textbf{u}_2 \rvert \tag{Triangle inequality} \\
        \leq& 
        \lVert \textbf{w} \rVert \lVert \textbf{u}_1 - \textbf{u}_2 \rVert + \lVert \textbf{w}_1 - \textbf{w}_2 \rVert \lVert \textbf{u}_2 \rVert, \tag{Cauchy-Schwarz inequality}\\
        \leq&
        \lVert \textbf{w} \rVert \lVert \textbf{u}_1 - \textbf{u}_2 \rVert + L_u C_{L_1}\lVert \textbf{u}_1 - \textbf{u}_2 \rVert, \tag{Assumption~\ref{assump:lipchitz} and Assumption~\ref{assump:bounded_norm}}\\
        =&
        (\lVert \textbf{w} \rVert + L_u C_{L_1} ) \lVert \textbf{u}_1 - \textbf{u}_2 \rVert. \notag
    \end{align}
\end{proof}
\begin{lemma}[\textbf{Inverse weighted difference}]\label{lemma:iwd}
    Under Assumption~\ref{assump:lipchitz}, we have
    \begin{align}
        \lvert \phi^{-1}_{-\textbf{u}_1}(\textbf{w})^\top \textbf{u}_1 - \phi^{-1}_{\textbf{u}_2}(\textbf{w})^\top \textbf{u}_2 \rvert \leq L_w \lVert \textbf{u}_1 - \textbf{u}_2 \rVert,
    \end{align}
    where $L_w = \lVert \textbf{w} \rVert + L_u C_L, \quad C_L = C_{L_1} C_{L_2}$.
\end{lemma}
\begin{proof}
    Let us denote $\phi^{-1}_{-\textbf{u}_1}(\textbf{w}) = \textbf{w}_1^\prime$, $\phi^{-1}_{\textbf{u}_2}(\textbf{w}) = \textbf{w}_2$. We have
    \begin{align}
        &\lvert \phi^{-1}_{-\textbf{u}_1}(\textbf{w})^\top \textbf{u}_1 - \phi^{-1}_{\textbf{u}_2}(\textbf{w})^\top \textbf{u}_2 \rvert \\
        =&
        \lvert \textbf{w}_1^{\prime, \top} \textbf{u}_1 - \textbf{w}_1^{\prime, \top} \textbf{u}_2 + \textbf{w}_1^{\prime, \top} \textbf{u}_2 - \textbf{w}_2^\top \textbf{u}_2 \rvert \tag{Adding and substract the same term} \\
        \leq&
        \lvert \textbf{w}_1^{\prime, \top} (\textbf{u}_1 - \textbf{u}_2) \rvert + \lvert (\textbf{w}_1^{\prime, \top}  - \textbf{w}_2)^\top \textbf{u}_2 \rvert \tag{Triangle inequality} \\
        \leq& 
        \lVert \textbf{w} \rVert \lVert \textbf{u}_1 - \textbf{u}_2 \rVert + \lVert \textbf{w}_1^{\prime, \top} - \textbf{w}_2 \rVert \lVert \textbf{u}_2 \rVert, \tag{Cauchy-Schwarz inequality}\\
        \leq&
        \lVert \textbf{w} \rVert \lVert \textbf{u}_1 - \textbf{u}_2 \rVert + L_u C_{L_1}\lVert \textbf{u}_1 + \textbf{u}_2 \rVert, \tag{Assumption~\ref{assump:lipchitz} and Assumption~\ref{assump:bounded_norm}}\\
        \leq&
        \lVert \textbf{w} \rVert \lVert \textbf{u}_1 - \textbf{u}_2 \rVert + L_u C_{L_1} C_{L_2} \lVert \textbf{u}_1 - \textbf{u}_2 \rVert, \tag{Assumption~\ref{assump:lipchitz}}\\
        =&
        (\lVert \textbf{w} \rVert + L_u C_L ) \lVert \textbf{u}_1 - \textbf{u}_2 \rVert. \notag
    \end{align}
\end{proof}

\begin{lemma}[\textbf{Instantaneous regret}]\label{lemma:inst_regret}
    Under asumptions~\ref{assump:optimality},\ref{assump:lipchitz}, $\forall u \in \mathcal{B}^u, x,x^\prime \in \mathcal{X}, w_i \in \textbf{w}$,
    \begin{equation} 
    \begin{aligned}
    &\mathcal{A}[\tilde{u}_t(x^*, :)]-\mathcal{A}[\tilde{u}_t(x_t, :)] \\
    \leq & L_\mathcal{A} \lvert \tilde{u}_{t}(x_t) - \tilde{u}_{t}(x_t^\prime) - (u(x_t) - u(x_t^\prime)) \rvert \\
    \leq &
    L_\mathcal{A} w^u_t(x_t, x_{t-1})
    \end{aligned}
    \end{equation}
    where $L_\mathcal{A} := \sqrt{n} (\lVert \textbf{w} \rVert + 2 L_u C_L)$.
\end{lemma}
\begin{proof}
    By the acquisition function maximization, we have
    \begin{align}
     &\mathcal{A}[u(x^\star, :)]-\mathcal{A}[u(x_t, :)] \notag\\
    =&\mathcal{A}[u(x^\star, :)]-\mathcal{A}[u(x_{t-1}, :)]+\mathcal{A}[u(x_{t-1}, :)]-\mathcal{A}[u(x_t, :)] \tag{cancel out}\\
     \leq&\mathcal{A}[\tilde{u}_t(x_{t}, :)]-\mathcal{A}[\tilde{u}_t(x_{t-1}, :)]-(\mathcal{A}[u(x_t, :)]-\mathcal{A}[u(x_{t-1}, :)]), \tag{Optimality of Line~\ref{alg_line:query} in Alg.~\ref{alg:SBO}}\\
     \leq& \textbf{w}^\top \left( \phi_{\textbf{u}_1}[\textbf{u}_1]-\phi_{\textbf{u}_2}[\textbf{u}_2]-(\phi_{\textbf{u}_3}[\textbf{u}_3]-\phi_{\textbf{u}_4}[\textbf{u}_4]) \right), \tag{GSF definition}
    \end{align}
    For simplicity, we denote $\textbf{u}_1 = \tilde{u}_t(x_{t}, :), \textbf{u}_2 = \tilde{u}_t(x_{t-1}, :), \textbf{u}_3 = u(x_t, :), \textbf{u}_4 = u(x_{t-1}, :)$, and use Definition~\ref{def:sorting} for sorting function $\phi$ notation. 
    By simple rearrangement, we get
    \begin{align}
        &\textbf{w}^\top \left( \phi_{\textbf{u}_1}[\textbf{u}_1]-\phi_{\textbf{u}_2}[\textbf{u}_2]-(\phi_{\textbf{u}_3}[\textbf{u}_3]-\phi_{\textbf{u}_4}[\textbf{u}_4]) \right) \notag \\
        =& 
        \textbf{w}^\top \left(\phi_{\textbf{u}_1}[\textbf{u}_1]-\phi_{\textbf{u}_3}[\textbf{u}_3]-(\phi_{\textbf{u}_2}[\textbf{u}_2]-\phi_{\textbf{u}_4}[\textbf{u}_4]) \right), \tag{Rearrange the terms}\\
        \leq&
        \textbf{w}^\top \left(\phi_{\textbf{u}_3}[\textbf{u}_1]-\phi_{\textbf{u}_3}[\textbf{u}_3]-(\phi_{\textbf{u}_2}[\textbf{u}_2]-\phi_{\textbf{u}_2}[\textbf{u}_4]) \right). \tag{Lemma~\ref{lemma:hlp}}
    \end{align}
    Here, we slightly abuse notation $\textbf{w}^\top \phi_{\textbf{u}_3}[\textbf{u}_1] = \phi_{\textbf{u}_3}^{-1}(\textbf{w})^\top \textbf{u}_1$. This means that we rearrange the weight $\textbf{w}$ instead of utility vector $\textbf{u}_1$. Then we have 
    \begin{align}
        &\textbf{w}^\top \left(\phi_{\textbf{u}_3}[\textbf{u}_1]-\phi_{\textbf{u}_3}[\textbf{u}_3]-(\phi_{\textbf{u}_4}[\textbf{u}_2]-\phi_{\textbf{u}_4}[\textbf{u}_4]) \right)\\
        =&
        \phi_{\textbf{u}_3}^{-1}(\textbf{w})^\top [ \underbrace{\textbf{u}_1 - \textbf{u}_3}_\text{positive by Assumption~\ref{assump:optimality}} ] - \phi_{\textbf{u}_2}^{-1}(\textbf{w})^\top [\underbrace{\textbf{u}_2 - \textbf{u}_4}_\text{positive by Assumption~\ref{assump:optimality}}], \tag{weight rearrangement}\\
        \leq &
        \phi_{\textbf{u}_3 - \textbf{u}_1}^{-1}(\textbf{w})^\top [\textbf{u}_1 - \textbf{u}_3] - \phi_{\textbf{u}_2 - \textbf{u}_4}^{-1}(\textbf{w})^\top [\textbf{u}_2 - \textbf{u}_4], \tag{Lemma~\ref{lemma:hlp}}\\
        \leq &
        (\lVert \textbf{w} \rVert + 2 L_u C_L) \lVert \textbf{u}_1 - \textbf{u}_3 - (\textbf{u}_2 - \textbf{u}_4) \rVert, \tag{Lemma~\ref{lemma:iwd} and $\lVert \textbf{u}_2 - \textbf{u}_4 \rVert \leq 2 L_v$}\\
        \leq &
        \sqrt{n} (\lVert \textbf{w} \rVert + 2 L_u C_L) \lvert \tilde{u}_{t}(x_t) - \tilde{u}_{t}(x_t^\prime) - (u(x_t) - u(x_t^\prime)) \rvert, \tag{Cauchy-Schwarz for individual utility}\\
        \leq &
        L_\mathcal{A} w^u_t(x_t, x_{t-1}), \tag{Definition of $w^u_t(x_t, x_{t-1})$ (supremum)}
    \end{align}
\end{proof}

\subsection{Main proof}

\subsubsection{Case-I: Given and invertible matrix $A$}
\paragraph{Confidence set.}
The ground-truth matrix $A$ is known and invertible. In this case, we do not need to learn $A$, so we can restrict the confidence set, 
\begin{equation}
\begin{aligned}
&\mathcal{B}_t^{v}=\{\tilde{v} \mid \ell_t(\tilde{v} \mid  D_{\mathcal{Q}_t^v})\geq \ell_t^\mathrm{MLE}-\beta^{v}_t \}\\
&\mathcal{B}_t^{u}=\{\tilde{u} \mid \ell_t(\tilde{u} \mid  D_{\mathcal{Q}_t^u})\geq \ell_t^\mathrm{MLE}-\beta^{u}_t \}\\
&\mathcal{B}_t^{v, A, u}=\{(\tilde{v}, \tilde{A}, \tilde{u}) \mid \ell_t(\tilde{A}, \tilde{u}, \tilde{v} \mid D_{\mathcal{Q}_t^u}, D_{\mathcal{Q}_t^v})\geq \ell_t^\mathrm{MLE}-\beta^{v, A, u}_t, \tilde{v}=\tilde{A}\tilde{u}, \tilde{A}=A, \tilde{v} \in \mathcal{B}_t^{v}, \tilde{u} \in \mathcal{B}_t^{u}\}.\\
\end{aligned}
\end{equation}

\paragraph{Instantaneous regret.}
\begin{align}
 &\mathcal{A}[u(x^\star, :)]-\mathcal{A}[u(x_t), :] \notag\\
 \leq& L_{\mathcal{A}} \lvert \tilde{u}_t(x_t, :)-\tilde{u}_t(x_{t-1}, :) - \left( {u}(x_t, :)-u(x_{t-1}, :) \right) \rvert, \tag{Lemma~\ref{lemma:inst_regret}} \\
 \leq& L_{\mathcal{A}} \lVert A^{-1} \rVert \lvert \tilde{v}_t(x_t, :)- \tilde{v}_t(x_{t-1}, :) - \left( {v}(x_t, :)-v(x_{t-1}, :) \right) \rvert. \tag{Cauchy-Schwarz}
\end{align}

\paragraph{Cumulative regret.}
By using Theorem~\ref{thm:cumulative_error}, our cumulative regret is:
\begin{align}
 R_T
 =&\sum_{t \in [T]} \mathcal{A}[u(x^\star, :)]-\mathcal{A}[u(x_t), :] \notag\\
 \leq&
 \sum_{t \in [T]} L_{\mathcal{A}} \lVert A^{-1} \rVert \lvert \tilde{v}_t(x_t, :)- \tilde{v}_t(x_{t-1}, :) - \left( {v}(x_t, :)-v(x_{t-1}, :) \right) \rvert, \notag\\
 \leq& 
 \mathcal{O}\left(n L_{\mathcal{A}}  \sqrt{\beta_T \gamma^{vv^\prime}_T T} \right) \tag{Theorem~\ref{thm:cumulative_error} and Lemma~\ref{lemma:graph}}
\end{align}

\paragraph{Cumulative queries.}
Obviously, we do not even need to query the ground truth $u$. Thus,
\begin{align}
    |\mathcal{Q}^u_T| &= 0.
\end{align}

\subsubsection{Case-II: Unknown but identifiable $A$}
\paragraph{Confidence set.}
Altough the matrix $A$ can be non-invertibible, the linear relationship can constrain the confidence set, as such,
\begin{equation}
\begin{aligned}
&\mathcal{B}_t^{v}=\{\tilde{v} \mid \ell_t(\tilde{v} \mid  D_{\mathcal{Q}_t^v})\geq \ell_t^\mathrm{MLE}-\beta^{v}_t \}\\
&\mathcal{B}_t^{u}=\{\tilde{u} \mid \ell_t(\tilde{u} \mid  D_{\mathcal{Q}_t^u})\geq \ell_t^\mathrm{MLE}-\beta^{u}_t \}\\
&\mathcal{B}_t^{v, A, u}=\{(\tilde{v}, \tilde{A}, \tilde{u}) \mid \ell_t(\tilde{A}, \tilde{u}, \tilde{v} \mid D_{\mathcal{Q}_t^u}, D_{\mathcal{Q}_t^v})\geq \ell_t^\mathrm{MLE}-\beta^{v, A, u}_t, \tilde{v}=\tilde{A}\tilde{u}, \tilde{v} \in \mathcal{B}_t^{v}, \tilde{u} \in \mathcal{B}_t^{u}\}.\\
\end{aligned}
\end{equation}

\paragraph{Instantaneous regret.}
The result is exactly the same with Lemma~\ref{lemma:inst_regret}.

\paragraph{Cumulative regret.}
Consider the following stopping criterion,
\begin{align*}
    w_t^u(x_t, x_{t-1}) \geq \max \left\{ \frac{1}{t^q}, w_t^v(x_t, x_{t-1}) \right\},
\end{align*}

Then, we have
\begin{align}
 &\sum_{t\in[T]}(\mathcal{A}[u(x^\star, :)]-\mathcal{A}[u(x_t, :)])\\
\leq&
L_\mathcal{A} \sum_{\tau \in [T]} w_t^u(x_\tau, x_{\tau-1}), \tag{Lemma~\ref{lemma:inst_regret}}\\
\leq&
L_\mathcal{A} \sum_{\tau \in \mathcal{Q}^u_T} w_\tau^u(x_\tau, x_{\tau-1}) + 
L_\mathcal{A} \sum_{\tau \in [T] \backslash \mathcal{Q}^u_T} \max \left\{ \frac{1}{t^q}, w_\tau^v(x_\tau, x_{\tau-1}) \right\}, \tag{stopping criterion}\\
\leq&
L_\mathcal{A} \sum_{\tau \in \mathcal{Q}^u_T} w_\tau^u(x_\tau, x_{\tau-1}) +  L_\mathcal{A} \sum_{\tau \in [T] \backslash \mathcal{Q}^u_T} w_\tau^v(x_\tau, x_{\tau-1}) + 
L_\mathcal{A} \sum_{\tau \in [T] \backslash \mathcal{Q}^u_T} \frac{1}{\tau^q}, 
\\
\leq&
\mathcal{O}\left(
L_\mathcal{A} \sqrt{\beta_T^u \gamma^{uu^\prime}_T |\mathcal{Q}^u_T|} + L_\mathcal{A} \sqrt{\beta_T^v \gamma^{vv^\prime}_T (T - |\mathcal{Q}^u_T|)} + L_\mathcal{A} (T - |\mathcal{Q}^u_T|)^{1-q}
\right) \tag{Theorem~\ref{thm:cumulative_error}}
\end{align}
This is the tightest bound. For visibility and interpretability, we simplify
\begin{align}
 R_T
 \leq&
 \mathcal{O}\left(
 L_\mathcal{A} T^{1-q} +
 L_\mathcal{A} \sqrt{\beta_T^u \gamma^{uu^\prime}_T T} 
 + 
L_\mathcal{A} \sqrt{\beta_T^v \gamma^{vv^\prime}_T T} 
\right), \tag{$|\mathcal{Q}^u_T| < T$}\\
 \leq&
 \mathcal{O}\left(
 L_\mathcal{A} T^{1-q} +
 L_\mathcal{A} \sqrt{(\beta_T^u \gamma^{uu^\prime}_T + \beta_T^v \gamma^{vv^\prime}_T) T} 
\right), \tag{factor out}
\end{align}

\paragraph{Cumulative queries.}
\begin{align}
|\mathcal{Q}_T^u|
=&\sum_{t\in\mathcal{Q}_T^u}1,\notag\\
\leq&T^q\sum_{t\in\mathcal{Q}_T^u}\frac{1}{{t^q}}, \label{eq:harmonic}\\
\leq&T^q\sum_{t\in\mathcal{Q}_T^u} w_t^u(x_t, x_{t-1}), \tag{stopping criterion} \\
=&
\mathcal{O}\left(
T^q \sqrt{\beta_T^u \gamma^{uu^\prime}_T |\mathcal{Q}_T^u|} 
\right) \tag{Theorem~\ref{thm:cumulative_error}}
\end{align}

Here, by setting $\epsilon=\frac{1}{T}$, we have
\begin{align}
\beta^u_T=\mathcal{O}\left(\sqrt{
|\mathcal{Q}_T^u| 
\log\frac{T \mathcal{N}(\mathcal{B}^u, \nicefrac{1}{T},\lVert \cdot \rVert_\infty)
}{\delta}}\right).
\end{align}

Hence,
\begin{align}
|\mathcal{Q}_T^u|
\leq &
\mathcal{O}\left(
T^q |\mathcal{Q}_T^u|^{3/4} \sqrt{\gamma^{uu^\prime}_T} 
\left( \log\frac{T \mathcal{N}(\mathcal{B}^u, \nicefrac{1}{T},\lVert \cdot \rVert_\infty)
}{\delta} \right)^{1/4}
\right), \label{eq:3_4}\\
|\mathcal{Q}_T^u|^{1/4} 
\leq & \
\mathcal{O}\left(
T^q \sqrt{\gamma^{uu^\prime}_T} 
\left( \log\frac{T \mathcal{N}(\mathcal{B}^u, \nicefrac{1}{T},\lVert \cdot \rVert_\infty)
}{\delta} \right)^{1/4}
\right),\\
|\mathcal{Q}_T^u| 
\leq & 
\mathcal{O}\left(
T^{4q} (\gamma^{uu^\prime}_T)^2 
\log\frac{T \mathcal{N}(\mathcal{B}^u, \nicefrac{1}{T},\lVert \cdot \rVert_\infty)
}{\delta}
\right),
\end{align}

Here, we consider the optimal $q$.
For upper bound, at least we want $|\mathcal{Q}_T^u| \leq T$, otherwise we have to query $u$ every iteration. Based on this, we have
\begin{align}
    \mathcal{O}\left(
    T^{4q} L_k
    \right) &\leq T,\\
    \mathcal{O}\left(
    T^{4q}
    \right) &\leq T,\\
    4q \leq 1,\\
    q \leq \frac{1}{4},
\end{align}
That is to say, by picking $q \leq \frac{1}{4}$, we can get a sublinear regret bound for the cumulative queries of $u$.

\subsubsection{Case-III: Unidentifiable $A$}
\paragraph{Confidence set.}
The matrix $A$ and public $u$ information is not useful anymore. Thus, we just only query private $u$. 
\begin{equation}
\begin{aligned}
&\mathcal{B}_t^{u}=\{\tilde{u} \mid \ell_t(\tilde{u} \mid  D_{\mathcal{Q}_t^u})\geq \ell_t^\mathrm{MLE}-\beta^{u}_t \}\\
\end{aligned}
\end{equation}

\paragraph{Instantaneous regret.}
\begin{align}
 &\mathcal{A}[u(x^\star, :)]-\mathcal{A}[u(x_t), :] \notag\\
 \leq& L_{\mathcal{A}} \lvert \tilde{u}_t(x_t, :)-\tilde{u}_t(x_{t-1}, :) - \left( {u}(x_t, :)-u(x_{t-1}, :) \right) \rvert, \tag{Lemma~\ref{lemma:inst_regret}}
\end{align}

\paragraph{Cumulative regret.}
By using Theorem~\ref{thm:cumulative_error}, our cumulative regret is:
\begin{align}
 R_T
 =&\sum_{t \in [T]} \mathcal{A}[u(x^\star, :)]-\mathcal{A}[u(x_t), :] \notag\\
 \leq&
 \sum_{t \in [T]} L_{\mathcal{A}} \lvert \tilde{u}_t(x_t, :)- \tilde{u}_t(x_{t-1}, :) - \left( {u}(x_t, :)-u(x_{t-1}, :) \right) \rvert, \notag\\
 \leq& 
 \mathcal{O}\left(L_{\mathcal{A}}  \sqrt{\beta_T \gamma^{uu^\prime}_T T} \right) \tag{Theorem~\ref{thm:cumulative_error} and Lemma~\ref{lemma:graph}}
\end{align}

\paragraph{Cumulative queries.}
Obviously, we end up querying the ground truth $u$ all the time. Thus,
\begin{align}
    |\mathcal{Q}^u_T| &= T.
\end{align}

\subsection{Proof of the kernel-specific bounds in Table~\ref{tab:kern_spec_bounds}}
To focus on kernel specific term only, we reduce the constants
\begin{align}
    R_T &\leq \mathcal{O}\left(
    T^{1 - \frac{q}{4}} + \sqrt{\beta_{T} \gamma^{vv^\prime}_{T} T}
    \right),\\
    |\mathcal{Q}^u_t| &= T^{q} \left( \gamma_T^{vv^\prime} \right)^2 \log\mathcal{N}(\mathcal{B}^v, T^{-1},\|\cdot\|_\infty),
\end{align}
Recall
$$
    \beta_T=\mathcal{O}\left(\sqrt{T 
    \log\frac{T \mathcal{N}(\mathcal{B}^v, \nicefrac{1}{T},\lVert \cdot \rVert_\infty)}{\delta}}\right).
$$
For kernel specific bound, we have, 

\paragraph{Linear kernel}
$$
\log\mathcal{N}(\mathcal{B}^v, T^{-1},\|\cdot\|_\infty) = \mathcal{O} \left(\log\frac{1}{\epsilon} \right)=\mathcal{O}\left(\log T\right).
$$
The corresponding $k^{vv^\prime}((x,x^\prime),(y,y^\prime)) = x^\top y + {x^\prime}^\top y^\prime = \langle(x, x^\prime), (y, y^\prime) \rangle$, which is also linear. Thus, by Theorem. 5 in~\cite{srinivas2012information},
$$
\gamma_T^{vv^\prime}=\mathcal{O}(\log T). 
$$
Hence, 
\begin{align}
    R_T &\leq \mathcal{O}\left(
    T^{1-\frac{q}{4}} + T^{3/4} (\log T)^{3/4}
    \right),\\
    |\mathcal{Q}^u_t| &\leq \mathcal{O}\left( T^{q} (\log T)^{3}\right).
\end{align}

\paragraph{Squared exponential kernel}
$$
\log\mathcal{N}(\mathcal{B}^v, T^{-1},\|\cdot\|_\infty)=\mathcal{O}\left((\log\frac{1}{\epsilon})^{d+1}\right)=\mathcal{O}\left((\log T)^{d+1}\right).
$$~(Example 4,~\cite{zhou2002covering}). By Thm. 4 in~\cite{kandasamy2015high}, we have,
$$
\gamma_T^{vv^\prime}=\mathcal{O}((\log T)^{d+1}). 
$$
Hence, 
\begin{align}
    R_T &\leq \mathcal{O}\left(
    T^{1-\frac{q}{4}} + T^{3/4} (\log T)^{3/4 (d+1)}
    \right),\\
    |\mathcal{Q}^u|_t &\leq \mathcal{O}\left( T^q (\log T)^{3(d+1)}\right).
\end{align}

\paragraph{Mátern kernel}
$$
\log\mathcal{N}(\mathcal{B}^v, T^{-1},\|\cdot\|_\infty)=\mathcal{O}\left((\frac{1}{\epsilon})^{\nicefrac{d}{\nu}}\log\frac{1}{\epsilon}\right)=\mathcal{O}\left(T^{\nicefrac{d}{\nu}}\log T\right).
$$~(by Thm. 5.1 and Thm. 5.3 in \cite{xu2024lower}). By Thm. 4 in~\cite{kandasamy2015high}, we have,
$$
\gamma_T^{vv^\prime}=\mathcal{O}\left(T^{\frac{d(d+1)}{2\nu+d(d+1)}}\log T\right). 
$$
where $\nu > \frac{d ( d+3+\sqrt{d^2+14d+17} )}{4}$.

Hence,
\begin{align}
    R_T &\leq 
    \mathcal{O}\left(
    T^{1-\frac{q}{4}} + T^{\frac{d}{4\nu} + \frac{d(d+1)}{4\nu + 2d(d+1)}} (\log T)^{3/4}
    \right),\\
    |\mathcal{Q}^u|_t &\leq \mathcal{O}\left( T^{q + \frac{d}{v}+ \frac{2d(d+1)}{2\nu + d(d+1)}} (\log T)^{3}\right).
\end{align}

\subsection{Proof of Theorem~\ref{thm:asymptotic_conv}}\label{proof:asymptotic_conv}
We first introduce the supporting results, then we prove Theorem~\ref{proof:asymptotic_conv}.

\subsubsection{Supporting results}
\begin{lemma}[\textbf{Strongly convex MAP estimation}]\label{lemma:convex_map}
    The log posterior defined in Eq.~(\ref{eq:map}) is strongly convex with respect to $A$.
\end{lemma}
\begin{proof}
    By Eq.~(\ref{eq:map}), the (unnormalised) negative log posterior can be written as:
    \begin{align}
        \mathcal{L}^\text{MAP}_t := -\mathcal{L}_t(\hat{u}, \hat{A}, \hat{v}) - \log p(A) - \log p(u) - \log p(v) \label{prob:map}
    \end{align}
    Here, we assume the priors for $u$ and $v$ are the same uniform distribution $\mathcal{U}(u; -L_v, L_v)$, where the range is the same due to the range preservation Lemma~\ref{lemma:graph}. Then, its log prior becomes $-\log p(u) = -\log p(v) = \log(2L_v)$, and these are constant, thereby negligible in terms of the optimisation. Similarly, the normalising constant (also known as Bayesian evidence, marginal likelihood) is also constant, thereby negligible.
    
    Original log likelihood function was convex yet not strongly convex with respect to $A$ because the Hessian matrix is positive semi-definite instead of positive definite. By adding the negative log prior term as regularliser, Eq.~(\ref{prob:map}) becomes strongly convex. 
    
    As the sum of strongly convex functions is strongly convex, we will show the row-wise $A_i$ is strongly convex. 
    First, we unpack the Eq.~(\ref{prob:map}) for graph $A$ related parts. By symmetric structure for $u$ and $v$, we only extract for $u$ at $t=1$ step for brevity,
    \begin{equation}
    \begin{aligned}
        \mathcal{L}^\text{MAP}
        =&
        \log \left( e^{A_{i}^\top u(x,:)} + e^{A_{i}^\top u(x^\prime,:)}  \right) - \textbf{1}^u A_i^\top u(x,:) - (1 - \textbf{1}^u) A_i^\top u(x,:)\\
        &- \sum_{i \in V} (\kappa_i - 1) \log A_{i} + \xi \sum_{i,j \in V} A_{ij},
    \end{aligned}
    \end{equation}
    Then, we introduce the function $P = \nicefrac{e^{A_{i}^\top u(x,:)}}{e^{A_{i}^\top u(x,:)} + e^{A_{i}^\top u(x^\prime,:)}}$. Then, the Hessian matrix $H$ is,
    \begin{align}
        H_{jk}
        =&
        \frac{\partial^2 \mathcal{L}^\text{MAP}}{\partial A_{ij} \partial A_{ik}},\\
        =&
        P(1-P)(u(x,j) - u(x^\prime,j))(u(x,k) - u(x^\prime,k)) + \delta_{jk} \frac{\kappa_i}{A^2_i} + 2 \xi
    \end{align}
    where $\delta_{jk}$ is the Kronecker delta. 
    The first term is a positive semi-definite matrix as show in Lemma~\ref{lemma:representor}. The second term is a diagonal matrix, which is positive definite because $\alpha_i - 1 > 0$ and $A_i > 0$. The sum of a positive definite matrix and positive semi-definite matrix is positive definite. Therefore, Hessian is positive definite.
    
    The smallest eigenvalue of Hessian $\lambda_\text{min}(H)$ is at least as large as the smallest eigenvalue of the positive definite matrix, namely the second term, as such:
    \begin{align}
        \lambda_\text{min}(H) \geq \frac{\kappa_i - 1}{\delta_A^2} + 2 \xi > 0.
    \end{align}
    As we confirmed, all elements are positive, thereby the minimum eigenvalues of Hessian is strictly positive.
    Therefore, our MAP loss function is stringly convex with respect to $A$.
\end{proof}

\subsubsection{Main proof}
\begin{proof}
We first prove the graph estimation error, then we show the utility estimation error convergence. 

\subsubsection{Graph identification error}
Firstly, we define the minimum excess risk (MER; \citet{xu2022minimum}):
\begin{align}
    \text{MER} := \inf_{\tilde{A}} \mathbb{E}_{D_{\mathcal{Q}_t^v}}[\mathcal{L}_\text{MAP}(\tilde{A})] - \mathcal{L}_\text{MAP}(A),
\end{align}
Here, $\mathcal{L}_\text{MAP}$ is `omniscient' loss function when true hyperparamter $A$ is given. Even if $A$ estimation is perfect, our log likelihood estimate should have some error according to the randomness of the data generating process. Thus, the second term express the fundemental limit of Bayesian learning, and we can interpret this as aleatoric uncertainty. The first term, on the other hand, represents the empirical estimate from the observed data, i.e., 
$\mathbb{E}_{D_{\mathcal{Q}_t^u}}[\mathcal{L}_\text{MAP}(\tilde{A})] = \mathcal{L}^\text{MAP}_t(A)$. 
Thus, MER represents reducible risk by gathering more data, thus we can interpret MER as epistemic uncertainty. 

\citet{xu2022minimum} theoretically analyses the connection between MER and the information-theoretic quantity, particularly conditional mutual information, then derived 
the following asymptotic convergence rate,
\begin{align}
    \text{MER} \leq \mathcal{O}\left( \frac{n^2}{2|\mathcal{Q}_t^u|}\right), \label{eq:mer_bound}
\end{align}
for the logistic regression cases with binary feedback in Theorem 5, which is the same with our case.

By Definition~\ref{defn:convex} of strong convexity,
\begin{equation}
    \mathcal{L}_\text{MAP}(\hat{A}) \geq \mathcal{L}_\text{MAP}(A) + \nabla \mathcal{L}_\text{MAP}(A)^\top (\hat{A} - A) + \frac{m}{2} \lVert \hat{A} - A \rVert^2
\end{equation}
Since $A$ is the ground truth matrix, the optimality assures its gradient is zero. Thus, the inequality can simplify
\begin{align}
    \mathcal{L}_\text{MAP}(\hat{A}) &\geq \mathcal{L}_\text{MAP}(A) + \frac{m}{2} \lVert \hat{A} - A \rVert^2,\\
    \mathcal{L}_\text{MAP}(\hat{A}) - \mathcal{L}_\text{MAP}(A) &\geq \frac{m}{2} \lVert \hat{A} - A \rVert^2,
\end{align}
This left-hand side is exactly the MER since $\mathcal{L}_\text{MAP}(A)$ is irreducible risk as $A$ is a ground-truth parameter, and $\hat{A} := \argmin_{\tilde{A} \in \mathcal{A}^{u,v,B}} \mathcal{L}^\text{MAP}_t(\tilde{A} \mid D_{|\mathcal{Q}_t^u|})$ is reducible risk. Then, we have
\begin{align}
    \text{MER} \geq \frac{m}{2} \lVert \hat{A} - A \rVert^2, \label{eq:error_bound}
\end{align}

By Eq.~(\ref{eq:mer_bound}), we have
\begin{align}
    \lVert \hat{A} - A \rVert^2 \leq& \mathcal{O}\left( \frac{m n^2}{2|\mathcal{Q}_t^u|}\right), \tag{Eqs.~(\ref{eq:mer_bound}) and (\ref{eq:error_bound})}\\
    \lVert \hat{A} - A \rVert \leq& \mathcal{O}\left( \frac{n \sqrt{m}}{\sqrt{2 |\mathcal{Q}_t^u|}}\right) \tag{square root},
\end{align}

Next, we analyse the upper bound of strong convexity constant $m$. 
Recall Lemma~\ref{lemma:convex_map} that the strong convexity constant is only dependent on the prior term, thereby
\begin{align}
    m 
    &\leq \max_{i}\left( \frac{\kappa_i - 1}{A_{ij}^2} \right) + 2\xi,\\
    &\leq \frac{\overline{\kappa} - 1}{\delta_A^2} + 2 \xi,
\end{align}
where $\overline{\kappa} = \max_{i,j \in V} \kappa_{i,j}$. As $\overline{\kappa}, \xi$ are user-defined parameters, we set 
\begin{align}
    \frac{\overline{\kappa} - 1}{\delta_A^2} + 2 \xi &= \frac{1}{n^2}, \label{eq:hypers_constraint}\\
    \overline{\kappa} &= 1 + \frac{\delta_A^2}{n^2} - 2 \xi \delta_A^2,
\end{align}
where $\kappa_i > 1$, $\xi < \frac{1}{2n^2}$ for the positivity. Then $\sqrt{m} \leq \frac{1}{n^2}$, leading to
\begin{align}
    \lVert \hat{A} - A \rVert &\leq \mathcal{O}\left( \frac{1}{\sqrt{2 |\mathcal{Q}_t^u|}}\right),
\end{align}
The equal constraint Eq.~(\ref{eq:hypers_constraint}) shows that $\kappa_i$ almost equal to 1 for all $i,j$, we need `flat' prior for graph $A$ estimate.

\subsubsection{Pointwise utility estimation error}
By definition of $w_t^u(x_t, x_t^\prime)$, which is the supremum of the pointwise error, we can upper bound
\begin{align}
    \lvert \tilde{u}_t(x_t) - \tilde{u}_t(x_t^\prime) -(u(x_t) - u(x_t^\prime)) \rvert 
    &\leq 
    w_t^u(x_t, x_t^\prime),
\end{align}
Here, we introduce the following notation:
\begin{align}
    W(T) := \sum_{\tau \in \mathcal{Q}^u_t} w(\tau) := w_t^u(x_t, x_{t-1}),
\end{align}
By Theorem~\ref{thm:duel_error}, $w(t)$ is submodular with respect to $t$, and $W(T)$ is monotonic with respect to $T$ because it is cumulative sum of positive values. Then by Theorem~\ref{thm:cumulative_error}, we have
\begin{align}
    W(T) 
    &\leq \mathcal{O}\left(
    \sqrt{\beta_T^u \gamma^{uu^\prime}_T |\mathcal{Q}_T^u|} 
    \right),\\
    &\leq \mathcal{O}\left(
    |\mathcal{Q}_T^u|^{3/4} \left(\gamma^{uu^\prime}_T \right)^{1/2} 
    \left( |\mathcal{Q}^u_t| \log\frac{T \mathcal{N}(\mathcal{B}^v, \nicefrac{1}{T},\lVert \cdot \rVert_\infty)}{\delta}\right)^{1/4}
    \right),\\
    &=
    \mathcal{O}\left(
    |\mathcal{Q}_T^u|^{3/4} L_{k,\mathcal{Q}^u_t}^{1/4}
    \right),
\end{align}
where $L_{k,\mathcal{Q}^u_t} := \left( \gamma^{uu^\prime}_T \right)^{2} |\mathcal{Q}^u_t| \log\frac{T \mathcal{N}(\mathcal{B}^v, \nicefrac{1}{T},\lVert \cdot \rVert_\infty)}{\delta}$ is the kernel-dependent term.

By submodularity and monotonicity, we have
\begin{align}
    w(t) \leq \frac{W(t)}{t}
\end{align}
for large $t$, i.e., $t \gg 1$, because
\begin{align}
    w(t) &\geq w(T), \tag{submodularity, $t \leq T$}\\
    \sum_{\tau=1}^T w(\tau) &\geq \sum_{\tau=1}^T w(T), \tag{submodular inequality holds $\tau \leq T$ for all $\tau$}\\
    W(T) &\geq T w(T), \tag{$W(T)$ definition}\\
    w(T) &\leq \frac{W(T)}{T}.
\end{align}
Here, we assume the running horizons $Q, T \gg 1$. Then we have
\begin{align}
    w(|\mathcal{Q}_T^u|) \leq \mathcal{O} \left(\frac{|\mathcal{Q}_T^u|^{3/4} L^{1/4}_{k,Q}}{Q}\right) = \mathcal{O} \left(|\mathcal{Q}_T^u|^{-1/4} L^{1/4}_{k,\mathcal{Q}^u_t}\right).
\end{align}
\end{proof}

\section{Extensions}
\subsection{Gaussian Process model approach}\label{app:gp}
\begin{assump}[\textbf{Direct feedback}]\label{assump:direct}
    At step $t$, if query point $x_t$ is evaluated, we get a noisy evaluation of $u^{(i)}$, $u^{(i)}_t = u^{(i)}(x_t) + \xi_t$ , where $\xi_t$ is i.i.d. $\sigma^{(i)}$-sub-Gaussian noise with fixed $\sigma^{(i)} > 0$.
\end{assump}

\paragraph{Modelling.}
Considering the data-generation process under Defn.~\ref{def:bandwagon}, we employ zero-mean multi-task GP (MTGP) regression model \citep{bonilla2007multi}. For simplicity, define $u: \cX \times E \to \RR$ as our utility function taking location query $x$ and agent index as arguments, i.e. $u(x, i)$ is the utility for query $x$ and agent $i$. We place a prior over $u$ as $u\sim\cG\cP(0, k_X \otimes k_E)$\footnote{$\otimes$ denotes Kronecker product.}, which is distributed as,
\begin{align*}
    u(x, :) \sim \cN(0, k_{X}(x,x) \times K_E),
\end{align*}
where $K_E$ is the kernel across agents and $k_X$ is the kernel across options $x$. Under the bandwagon model, with a graph $G$ (and its adjacency matrix $A$), we have,
\begin{align}\label{eq:graph_convolution}
    v(x, :) = A u(x, :) \sim \cN(0, k_X(x,x)\times AK_EA^\top),
\end{align}
due to the linearity of the graph convolution, we naturally get $v$ being itself an induced MTGP
\begin{align}
    v \sim \cG\cP(0, k_X \times A k_E A^\top), \text{ such that } \operatorname{Cov}(v(x, i), v(x', j)) = k_X(x,x') \times (AK_EA^\top)_{ij}
\end{align}\label{eq:mtgp}
This holds true for the inverse case, where $B := (A + \lambda I)^{-1}$, resulting in $u(x,:) \approx B v(x, :)$. Here, $\lambda$ is a regularization term and we typically set a fixed small positive value (e.g., 1e-4). 

\paragraph{Estimate social graph.}
Interestingly, Eq.~(\ref{eq:mtgp}) tells us that the graph adjacent matrix $A$ is merely the kernel hyperaparameter for $v$. Thus, similarly to optimise other kernel hyperparameters, we can estimate the graph $A$ (or equivalently, $B$) through maximum likelihood estimation (MLE) of log marginal likelihood (LML), with a slight modification:
\begin{align}\label{eq:mll}
    \log \mathbb{P}(U_{\mathcal{Q}^u_t} \mid D_{\mathcal{Q}^v_t}, X_{\mathcal{Q}^u_t}, B)
    := \frac{1}{n}\sum_{i=1}^n \log \mathcal{N} \left( U^{(i)}_{\mathcal{Q}^u_t} ; m^{u^{(i)}}_{\mathcal{Q}^v_t}(X_{\mathcal{Q}^u_t}), C^{u^{(i)}}_{\mathcal{Q}^v_t}(X_{\mathcal{Q}^u_t}, X_{\mathcal{Q}^u_t}) \right),
\end{align}
where $D_{\mathcal{Q}^u_t} := (X_{\mathcal{Q}^u_t}, U_{\mathcal{Q}^u_t})$ is the $u$ observations, and $m^{u^{(i)}}_{\mathcal{Q}^v_t}$ and $C^{u^{(i)}}_{\mathcal{Q}^v_t}$ represent the predictive mean and covariance of $u(x,i)$ using the GP conditioned on $D_{\mathcal{Q}^v_t} := (X_{\mathcal{Q}^v_t}, V_{\mathcal{Q}^v_t})$ through Eq.~(\ref{eq:mtgp}).
\begin{align*}
    m^{u^{(i)}}_{\mathcal{Q}^v_t}(x) &= \left[k_X(x, X_{\mathcal{Q}^v_t}) \otimes k_E^{\prime (i)} \right]^\top \Sigma^{-1} (B V_{\mathcal{Q}^v_t}),\\
    C^{u^{(i)}}_{\mathcal{Q}^v_t}(x,x^\prime) &= k_X(x,x^\prime) \times k_E^{\prime (i)} - \left[k_X(x, X_{\mathcal{Q}^v_t}) \otimes k_E^{\prime (i)} \right]^\top \Sigma^{-1} \left[k_X(x, X_{\mathcal{Q}^v_t}) \otimes k_E^{\prime (i)} \right] ,
\end{align*}
where $\Sigma := k_X(X_{\mathcal{Q}^v_t}, X_{\mathcal{Q}^v_t}) \otimes k_E^{\prime (i)} + D_{\sigma^2} \otimes I $, $k_E^{\prime (i)} := \big( B K_E B^\top \big)^{(i)}$ is the $i$-th column of the matrix $B K_E B^\top$, $D_{\sigma^2}$ is the diagonal matrix whose $(i,i)$-th element is the $v^{(i)}$ noise variance. 

Note that the GP is conditioned on $v$, not $u$. $D_{\mathcal{Q}^u_t}$ is used as `test dataset' to estimate $B$. This offers $|\mathcal{Q}^v_t| \neq |\mathcal{Q}^u_t|$, where typical MTGP requires $X_{\mathcal{Q}^v_t} = X_{\mathcal{Q}^u_t}$. Thus, this formulation allows us to separate the predictive contribution on $\mathcal{Q}^v_t$ and $\mathcal{Q}^u_t$, allowing \emph{decoupled} query for $u$ and $v$.

\subsubsection{Inference from Direct Utility Feedback}
To estimate the posterior predictive distribution of $u$ including the uncertainty of $B$ estimate requires extensive MCMC approximation. For the computational efficiency and closed-form propagation, we adopt the Laplace approximation:
\begin{align*}
    \mathbb{P}(B \mid D_{\mathcal{Q}^u_t}, D_{\mathcal{Q}^v_t}) \approx &\mathcal{N}(B; \hat{B}_{\mathcal{Q}_t}, \Lambda^{-1}_{\mathcal{Q}_t}),\\
    \hat{B}_{\mathcal{Q}_t} = \arg\max_{B} \log \mathbb{P}(U_{\mathcal{Q}^u_t} \mid D_{\mathcal{Q}^v_t}, X_{\mathcal{Q}^u_t}, B), \quad
    &\Lambda_{\mathcal{Q}_t} = - \nabla_B \nabla_B \log \mathbb{P}(U_{\mathcal{Q}^u_t} \mid D_{\mathcal{Q}^v_t}, X_{\mathcal{Q}^u_t}, B) \mid_{B = \hat{B}}.
\end{align*}
The Hessian for covariance can be conveniently estimated via auto-differentiation. Now all of our variables are Gaussian, offering the closed-form uncertainty propagation:
\begin{corollary}[\textbf{Uncertainty propagation}]\label{corollary:unc_prop}
    Given $B \sim \mathcal{N}(B; \hat{B}_{\mathcal{Q}_t}, \Lambda^{-1}_{\mathcal{Q}_t})$ and $v(x,i) \mid D_{\mathcal{Q}^v_t} \sim \mathcal{GP}(m^{u^{(i)}}_{\mathcal{Q}^v_t}, C^{u^{(i)}}_{\mathcal{Q}^v_t})$, the posterior predictive distribution of $u$ becomes the closed-form:
    \begin{align*}
    \mathbb{P}(u(x,i) \mid D_{\mathcal{Q}^u_t}, D_{\mathcal{Q}^v_t}) &\approx \mathcal{N} \left( u(x,i); 
    \mu^{u^{(i)}}_{\mathcal{Q}_t}(x), \sigma^{u^{(i)}}_{\mathcal{Q}_t}(x) \right)\\
    \mu^{u^{(i)}}_{\mathcal{Q}_t}(x) = \hat{B}_{Q_t}(i,:) {m^{v}_{Q^v_t}}(x), \quad
    \left(\sigma^{u^{(i)}}_{\mathcal{Q}_t}(x)\right)^2 &= \sum_{j=1}^n \left(\prod_{j=1}^n \bar{b}^2_{ij} \bar{v}^{(j)^2}(x) - \prod_{j=1}^n \hat{b}^2_{ij} m^{v^{(j)^2}}_{Q^v_t}(x) \right).
\end{align*}
\end{corollary}
We obtain the closed-form UCB $\bar{u}(x,i) := m^{u^{(i)}}_{\mathcal{Q}_t}(x) + \beta_t^{1/2} \sigma^{u^{(i)}}_{\mathcal{Q}_t}(x)$ for the acqusition function, and $\Psi(B) := \mathcal{A}^\prime[\text{diag}(\Lambda^{-1}_{\mathcal{Q}_t})]$ for stopping criterion.
\begin{proof}
    Given graph convolution operation in Eq.~\ref{eq:graph_convolution}, we can decompose the $i$-th truthful utility $u(x,i)$ as such:
    \begin{align}
        u(x, i) = B(i,:) v(x, :) = \sum_{j=1}^n b_{ij} v(x, j), \label{eq:decomp_graph}
    \end{align}
    where $b_{ij}$ is the element of $B$ at $i$-th row and $j$-th column. 
    Here, $b_{ij}$ is independent of $v$, thus this linear operation is the product of two independent random variables. The expectation and variance of such case is known \citep{goodman1960exact}, as such:
    \begin{align}
        \mu^{u^{(i)}}_{\mathcal{Q}_t}
        &=\mathbb{E} \left[\sum_{j=1}^n b_{ij} v(x, j) \right], \tag{Eq.~\ref{eq:decomp_graph}}\\
        &=\sum_{j=1}^n \mathbb{E}[b_{ij} v(x, j)], \tag{linearity of expectation}\\
        &=\sum_{j=1}^n \mathbb{E}[b_{ij}] \mathbb{E}[v(x, j)], \tag{independence}\\
        &=\sum_{j=1}^n \hat{b}_{ij} m^{v^{(j)}}_{Q^v_t}(x), \tag{predictive mean of $b_{ij}$ and $u$}\\
        &=\hat{B}_{Q_t}(i,:) {m^{v}_{Q^v_t}}(x). \tag{inner product}
    \end{align}
    \begin{align}
        \sigma^{u^{(i)^2}}_{\mathcal{Q}_t}(x)
        &=\mathbb{V} \left[\sum_{j=1}^n b_{ij} v(x, j) \right], \tag{Eq.~\ref{eq:decomp_graph}}\\
        &= \sum_{j=1}^n \mathbb{V} \left[b_{ij} v(x, j) \right], \tag{independence}\\
        &= \sum_{j=1}^n \left( \prod_{j=1}^n (\mathbb{V}[b_{ij}] + \mathbb{E}[b_{ij}]^2)(\mathbb{V}[v(x, j)] + \mathbb{E}[v(x, j)]^2) - \prod_{j=1}^n \mathbb{E}[b_{ij}]^2 \mathbb{E}[v(x, j)]^2 \right) \tag{independence}\\
        &= \sum_{j=1}^n \left( \prod_{j=1}^n (\sigma_{b_{ij}}^2 + \hat{b}_{ij}^2)(C^{v^{(j)}}_{Q^v_t}(x) + {m^{v^{(j)}}_{Q^v_t}}^2(x)) - \prod_{j=1}^n \left( \hat{b}_{ij} {m^{v^{(j)}}_{Q^v_t}}(x) \right)^2 \right) \tag{posterior mean}\\
        &= \sum_{j=1}^n \left(\prod_{j=1}^n \bar{b}^2_{ij} \bar{v}^{(j)^2}(x) - \prod_{j=1}^n \hat{b}^2_{ij} m^{v^{(j)^2}}_{Q^v_t}(x) \right), \tag{notation change}
    \end{align}
    where $\bar{b}^2_{ij} := \sigma_{b_{ij}}^2 + \hat{b}_{ij}^2$ 
    and ${\bar{v}^{(j)^2}} (x) := C^{v^{(j)}}_{Q^v_t}(x) + {m^{v^{(j)}}_{Q^v_t}}^2(x)$.
\end{proof}

\subsection{Future directions}\label{app:positive}
\subsubsection{Positive social influence}
When consider the our objective is now changed to:
\begin{align}
    x^\star = \argmax_{x \in \mathcal{X}} \mathcal{A}[v(x,:)].
\end{align}
Note that now we consider $v(x,:)$ as the truthful utilities, and more costly to query than $u(x,:)$. We can think individual utility $u(x,:)$ can be based on misunderstanding, but the meeting can mitigate this confusion, then we get truthful $v(x,:)$ thanks to the social interaction. This can be seen in real-world, highlighting our OpenReview discussion is exactly the same, where each review is based on individual utility, yet the discussion can mitigate this misunderstanding thus the utility after discussion $v(x,:)$ can be regarded as truthful. Then, our problem becomes easier than before; we can cheaply observe $u(x,:)$, and $v(x,:) = A u(x,:)$, meaning that we do not need to consider the invertibility issues. Moreover, importantly, the impossibility theorem also can be mitigated. That means, of course $\mathcal{A}[u(x,:)] \neq \mathcal{A}[v(x,:)]$ is still valid, yet we can say the following Pareto Front containment Theorem:
\begin{theorem}[\textbf{Pareto Front Containment}]\label{thm:pareto-front-containment}
    For any social graph G the Pareto front corresponding to non-truthful utilities is a subset of the original Pareto front of truthful utilities. 
\end{theorem}
\begin{proof}
    Before we define the Pareto fronts of truthful and non-truthful utilities, we define two preferences $\succeq_u$ and $\succeq_v$ as 
    \begin{align}
      x \succeq_u x'\quad \text{iff} \quad u(x,i) \geq u(x',i) \quad \forall i\\
      x \succeq_v x'\quad \text{iff} \quad v(x,i) \geq v(x',i) \quad \forall i
    \label{eq:def-background-pref}
    \end{align}

    Consider the Pareto front of truthful utilities as $P_u:=\{x|  \forall x'\in \mathcal{X} x'\nsucc_{u}x\}$ and analogously for non-truthful utilities as $P_v:=\{x| \forall\text{ } x'\in \mathcal{X} x'\nsucc_{v}x\}$. We want to show that $P_v\subseteq P_u$. Before we show that, we prove that the following holds for any unknown social influence graph $G$
    \begin{align*}
        \forall\text{ }x,x'\in\mathcal{X} x\succeq_u x' \implies x\succeq_v x'
    \end{align*} 
    Given the transition matrix of graph $G$ as $A$, we know that $a_{ij}\geq 0$ and $\sum_ja_{ij}=1$. Consider 
    \begin{align*}
       x &\succeq_u x'\\
       \quad \forall i\text{ } u(x,i) &\geq u(x',i) \quad (\text{Definition }~\ref{eq:def-background-pref})\\
       \sum_{i}a_{ij}u(x,i) &\geq \sum_{i}a_{ij}u(x',i) \quad (a_{ij}\geq 0)\\
       A_j^Tu(x,:)&\geq A_j^Tu(x',:) \quad (A_i:=\text{ }j^{\text{th}}\text{ row of }A)\\
        v(x,j) &\geq v(x',j) \quad \forall j\\
        x &\succeq_v x'
    \end{align*}
    Since, the Pareto set is defined as $P_v=\{x| \forall\text{ } x'\in \mathcal{X} x'\nsucc_{v}x\}$, we can rewrite the condition of $x$ not being strictly dominated by $x'$ as either $x$ weakly domainates $x'$ or is incomparable 
    \begin{align*}
        x'\nsucc_{v}x &= (x \succeq_v x') \lor [(x\nsucceq_v x')\land (x'\nsucceq_v x)] \\
        &= [(x \succeq_v x') \lor (x\nsucceq_v x')] \land [(x \succeq_v x') \lor (x'\nsucceq_v x)]\\
        &= (x \succeq_v x') \lor (x'\nsucceq_v x)
    \end{align*}
    Then $P_v=\{x| \forall\text{ } x'\in \mathcal{X} (x \succeq_v x') \lor (x'\nsucceq_v x)\}$ and similarly, $P_u=\{x| \forall\text{ } x'\in \mathcal{X} (x \succeq_u x') \lor (x'\nsucceq_u x)\}$. 
    To show that $P_v\subseteq P_u$, consider $x\in P_v$, then $x$ needs to satisfy $(x \succeq_v x') \land (x'\nsucceq_v x)$ $\forall x'\in\mathcal{X}$. We divide the condition into two cases, 

    \par{\textbf{Case 1:} $\forall x'\in\mathcal{X}$ $x'\nsucceq_v x$} 

    Since $x \succeq_u x'\implies x \succeq_v x'$, considering the contrapositive we have $x'\nsucceq_v x \implies x \nsucceq_u x'$. Then $x\in P_u$. 

    \par{\textbf{Case 2:} $\forall x'\in\mathcal{X}$ $x \succeq_v x'$}
    
    This case is rarer as it implies that $P_v=\{x\}$. We can re-write the case as, $\forall x'\in \mathcal{X}$
    \begin{align*}
        \forall j \quad v(x,j)&\geq v(x',j)\quad (\text{Definition }~\ref{eq:def-background-pref})\\
        \forall j \quad A_j^Tu(x,:)&\geq A_j^Tu(x',:)
    \end{align*}
    By fixing any $j$ we can say that $x=\argmax_{x\in\mathcal{X}}A_j^Tu(x,:)$. Therefore $x\in P_u$. 
\end{proof}

Therefore, we can use private votes to explore the Pareto front $\mathcal{X}^\star$, then the consensus is contained $x^\star \in \mathcal{X}^\star$.
This can naturally lead to the combination of multi-objective BO, where we use private vote $u$ to estimate the Pareto Front $\mathcal{X}^\star$, then we search the consensus $x^\star$ within the estimated Pareto Front set. 

Let $\text{vol}(\mathcal{X})$ be the volume of the domain. Then, we have $\text{vol}(\mathcal{X}^\star) \leq \text{vol}(\mathcal{X})$ since $\mathcal{X}^\star \subseteq \mathcal{X}$. As shown in \citet{kandasamy2016gaussian}, the maximum information gain (MIG) depends on the domain volume. Let $\Psi_t(\mathcal{X})$ be the MIG at $t$-th iteration over the domain $\mathcal{X}$. For instance, with the squared exponential kernel, MIG can be expressed as $\Psi_t(\mathcal{X}) \propto \text{vol}(\mathcal{X}) \log(t)^{d+1}$. This means that the regret of our algorithm should improve by a factor of $\Psi_t(\mathcal{X}^g) / \Psi_t(\mathcal{X}) = \text{vol}(\mathcal{X}^g) / \text{vol}(\mathcal{X})$. Thus, in this case, we can provably better regret convergence bound when individual votes are available. This is actually what our OpenReview system does.

\subsubsection{Other extensions}
We can extend our approaches to batch cases by using common techniques in BO communities \citep{azimi2010batch, gonzalez2016batch, adachi2022fast, adachi2023bayesian, adachi2024quadrature, balandat2020botorch}. Extending to heterogeneous and dynamic settings is also an important direction for future research, where the probabilistic choice function approach~\citep{benavoli2023learning} offers a promising pathway. Integrating preference explanation technique~\citep{hu2022explaining} to SBO could also improve the interpretability of the whole framework.

\section{Hyperparameters}\label{app:hypers}
We summarized the comprehensive list of hyperparameters used in this work and their settings in Table~\ref{tab:hypers}. Most of these are standard in typical GP-UCB approaches. The newly introduced hyperparameters are primarily tunable in a data-driven manner, and we provided a sensitivity analysis in the experiment section for those that are not.

\subsection{Update kernel hyperparameters}\label{app:hypers_update}
By Assumption~\ref{assump:bounded_norm}, there exists a large enough constant $L_v$ that upper bounds the norm of the ground-truth latent black-box utility function $u,v$. However, a tight estimate of this upper bound may be unknown to us in practice, while the execution of our algorithm explicitly relies on knowing a bound $L_v$ (in Prob.~(\ref{prob:ucb_acquisition}), $L_v$ is a key parameter). 

So it is necessary to estimate the norm bound $L_v$ using the online data. Suppose our guess is $\hat{L}$. It is possible that $\hat{L}$ is even smaller than the ground-truth function norm $\|v\|$. To detect this underestimate, we observe that, with the correct setting of $L_v$ such that $L_v \geq \|v\|$, we have that by Lemma~\ref{lemma:conf_set} and the definition of MAP estimate,
$$
\mathcal{L}_t^\mathrm{MAP}(\hat{u}_{t \mid \hat{L}}, \hat{A}_{t \mid \hat{L}}, \hat{v}_{t \mid \hat{L}}) \geq \mathcal{L}_t^\mathrm{MAP}(u, A, v) \geq \mathcal{L}_t^\mathrm{MAP}(\hat{u}_{t \mid \hat{L}}, \hat{A}_{t \mid \hat{L}}, \hat{v}_{t \mid \hat{L}})-\beta_{t \mid \hat{L}},
$$
where $\hat{u}_{t \mid \hat{L}}$ is the MAP estimate function with function norm bound $\hat{L}$ and  $\beta_t$ is the corresponding parameter as defined in Lemma~\ref{lemma:conf_set} with norm bound $\hat{L}$. We also have $2\hat{L}$ is a valid upper bound on $\|v\|$ and thus,     
$$
\mathcal{L}_t^\mathrm{MAP}(\hat{u}_{t \mid 2\hat{L}}, \hat{A}_{t \mid 2\hat{L}}, \hat{v}_{t \mid 2\hat{L}}) \geq \mathcal{L}_t^\mathrm{MAP}(u, A, v) \geq \mathcal{L}_t^\mathrm{MAP}(\hat{u}_{t \mid 2\hat{L}}, \hat{A}_{t \mid 2\hat{L}}, \hat{v}_{t \mid 2\hat{L}})-\beta_{t \mid 2\hat{L}},
$$
Therefore, 
$$
\mathcal{L}_t^\mathrm{MAP}(\hat{u}_{t \mid \hat{L}}, \hat{A}_{t \mid \hat{L}}, \hat{v}_{t \mid \hat{L}}) \geq \mathcal{L}_t^\mathrm{MAP}(u, A, v) \geq \mathcal{L}_t^\mathrm{MAP}(\hat{u}_{t \mid 2\hat{L}}, \hat{A}_{t \mid 2\hat{L}}, \hat{v}_{t \mid 2\hat{L}})-\beta_{t \mid 2\hat{L}},
$$
That is to say, $\mathcal{L}_t^\mathrm{MAP}(\hat{u}_{t \mid \hat{L}}, \hat{A}_{t \mid \hat{L}}, \hat{v}_{t \mid \hat{L}})$ needs to be greater than or equal to $\mathcal{L}_t^\mathrm{MAP}(\hat{u}_{t \mid 2\hat{L}}, \hat{A}_{t \mid 2\hat{L}}, \hat{v}_{t \mid 2\hat{L}})-\beta_{t \mid 2\hat{L}}$ when $\hat{L}$ is a valid upper bound on $\|v\|$. 

Therefore, we can use the heuristic: every time we find that
$$
\mathcal{L}_t^\mathrm{MAP}(\hat{u}_{t \mid \hat{L}}, \hat{A}_{t \mid \hat{L}}, \hat{v}_{t \mid \hat{L}}) <\mathcal{L}_t^\mathrm{MAP}(\hat{u}_{t \mid 2\hat{L}}, \hat{A}_{t \mid 2\hat{L}}, \hat{v}_{t \mid 2\hat{L}}) -\beta_{t \mid 2\hat{L}},
$$
we double the upper bound guess $\hat{L}$.

\paragraph{Optimize the kernel hyperparamters}
Unlike the GP, our likelihood model does not have the analytical form of marginal likelihood. Thus, we adopt the leave-one-out cross-validation (LOO-CV) as the optimization loss (See Section 5.3 in \citet{williams2006gaussian}). That is, we leave one out from the observed dataset and compute the negative log posterior of the left one dataset, and averaging all samples. We optimize the kernel hyperparameters by minimizing this LOO-CV. 

\section{Experiments}\label{app:exp}
An RBF kernel is used by default unless otherwise specified, and for each optimization iteration, the inputs are rescaled to the unit cude $[0, 1]^d$. The initial dataset consist of $5$ randomly sampled pairs $(x_t, x_t^\prime)$ from the domain $\cX$ with labels generated according to Assumption~\ref{assump:pairwise}. A $x_t$ is determined by $\argmax_{x \in \cX}{\alpha(x, x_{t-1})}$ in each iteration then votes are queried sequentially. All experiments were repeated $10$ times under different seeds and initial datasets. Hyperparameters such as kernel lengthscales were tuned online at each iteration (see Appendix~\ref{app:hypers} for details). Optimization problems are solved using the interior-point nonlinear optimizer IPOPT \citep{wachter2006implementation}, interfaced via the symbolic framework CasADi \citep{andersson2019casadi}. Models are implemented in GPyTorch \citep{gardner2018gpytorch}, and experiments are conducted using a laptop PC\footnote{MacBook Pro 2019, 2.4 GHz 8-Core Intel Core i9, 64 GB 2667 MHz DDR4}. Computational time is discussed in Appendix~\ref{app:compute_time}. 

\subsection{Toy example 1}
The instrinsic utilities of the influencer ($u^{(1)}$) and the follower ($u^{(2)}$) are defined as follows:
\begin{align}
    u^{(1)} &:= 0.3\mathcal{N}(x; 0.35,0.05) + 1.2\mathcal{N}(x; 0.45,0.18) + 0.8\mathcal{N}(x; 0.75,0.1),\\
    u^{(2)} &:= 0.5\mathcal{N}(x; 0.25, 0.1) + 0.8\mathcal{N}(x; 0.65, 0.15) + 0.4\mathcal{N}(x; 0.85, 0.05),
\end{align}
The social-influence graph is defined as $A := \begin{pmatrix} 0.9 & 0.1  \\ 0.6 & 0.4\\ \end{pmatrix}$, and the aggregation function is utilitarian $\mathcal{A} := \sum_{i=1}^2 \frac{1}{2} u^{(i)}$.

\subsection{Real-world tasks}
\textbf{1. Thermal comfort:} 
Three office workers (influencer, follower, altruist) wishes to optimize the thermal condition (set temperature and air speed) under different garments and activity conditions, leading to varying thermal insulation and metabolic generation. The facilitator sets aggregation rule as egalitarian because they considers uncomfortable thermal conditions deteriorates productivity and health. We calculate each utility function under the given conditions using a simulator \citep{tartarini2020pythermalcomfort}, following the ASHRAE industrial standard.

\textbf{2. TeamOpt:} 
Four office workers (a leader, a follower, and two indecisive members) hold a meeting to select team members for the next project, which includes eight workers in total, including themselves. The leader prioritizes skill set diversity to enhance productivity, while the others focus on inter-member compatibility: one considers average compatibility, another seeks to maximize the worst-case scenario, and the last prefers a balanced team with minimal variance in compatibility. The facilitator applies an egalitarian rule, ensuring that the worst-off member is at least satisfied with the team. Each potential team forms a graph with an adjacency matrix representing skills and compatibility, and summary statistics define each utility function. The inter-graph kernel was computed using the graph diffusion kernel \citep{zhi2023gaussian}.

\textbf{3. TripAdvisor:} 
Three colleagues (two influencers and a follower) are deciding on a hotel for their upcoming group retreat, using the TripAdvisor website. One prefers a luxurious hotel with the highest ranking, another seeks a budget-friendly option with reasonable reviews, and the third prioritizes the hotel with the highest overall review score. The facilitator (group leader) chooses $\rho=0.5$ to balance both egalitarian and utilitarian aspects. We used the TripAdvisor New Zealand Hotel dataset \citep{tripadviser}.

\textbf{4. EnergyTrading:} 
Three firms (a large corporation, a niche startup, and a joint venture) form a strategic collaboration to enhance their energy trading business. Their profits rely on a machine learning model that predicts day-ahead market prices, which in turn depends on a demand dataset. Due to the scarcity of such data, they jointly invest in a market research project. However, since the demand data is spatiotemporal, they must decide on the optimal location for data collection. Each firm's utility function is based on the information gain for the selected location, and their internal datasets lead to heterogeneous utilities. The facilitator (project leader) adopts $\rho=0.5$ to balance egalitarian and utilitarian considerations. We used the UK-wide energy demand dataset \citep{meterdataset, grunewald2019specific}.

\subsubsection{Thermal comfort} 
We compute the utility function using the PMV (predicted mean vote), which is the estimation from the large collection of human preference dataset that predicts the values from $[-3, 3]$, where $-3$ means very cold, and $3$ means very hot, and $0$ means comfortable at the current condition. We take the minus absolute PMV, with 0 is the maximum for each agent's utility. The input values are two dimensional continuous values $x \in \mathcal{R}^2$, where the first dimension denotes the temperature $[15, 35]$ with degrees Celcius, and the second dimension is the air velocity $[0.3, 1.5]$.

The agents has different conditions of the garments and activity conditions, resulting in the relative air velocity difference. The conditions are as follows:

\textbf{Activity}
\begin{compactenum}
    \item[(agent 1)] Seated, heavy limb movement
    \item[(agent 2)] House cleaning
    \item[(agent 3)] Writing
\end{compactenum}

\textbf{Garments}
\begin{compactenum}
    \item[(agent 1)] 'Executive chair', 'Thick trousers', 'Long-sleeve long gown', "Boots", 'Ankle socks'
    \item[(agent 2)] "Thin trousers", "T-shirt", "Shoes or sandals"
    \item[(agent 3)] 'Standard office chair', 'Long sleeve shirt (thin)', 'Long-sleeve dress shirt', 'Slippers'
\end{compactenum}
Then PyThermalComfort \citep{tartarini2020pythermalcomfort} computes the corresponding metabolic generation and thermal insulation based on ASHRAE industrial standard. The social-influence graph is defined as $A := \begin{pmatrix} 0.8 & 0.1 & 0.1  \\ 0.6 & 0.1 & 0.3 \\ 0.4 & 0.3 & 0.3 \\ \end{pmatrix}$, and the aggregation function is egalitarian $\rho=0.1$.

\subsubsection{TeamOpt} 
We have modified the setting from \citet{wan2023bayesian, adachi2024adaptive}. Each team is represented by graphs with 8 members from 11 candidates. Such teams are positioned on the node of the supergraph, of which edge is the similarity between teams defined as the Jaccord index. 
There are two additional information on nodes; skills and inter-member compatibility. Both are represented as continuous values, generated from row-wise Dirichlet distribution, resulting in $n \times n$ square matrices. There are four agents who has the different utility function.

\textbf{Utility functions}
\begin{compactenum}
    \item[(agent 1)] Skill set diversity: this is measured by the entropy of the skill set matrix, assuming the optimal team is when each member is specialised in one skill, and the whole skill distribution is close to uniform. 
    \item[(agent 2)] mean compatibility: averaging the compatibility matrix.
    \item[(agent 3)] minimax compatibility: take minimum element of the compatibility matrix.
    \item[(agent 4)] flat compatibility: take minimum variance of the compatibility matrix.
\end{compactenum}
The social-influence graph is defined as $A := \begin{pmatrix} 0.7 & 0.1 & 0.1 & 0.1 \\ 0.4 & 0.4 & 0.1 & 0.1 \\ 0.3 & 0.2 & 0.3 & 0.2 \\  0.3 & 0.2 & 0.2 & 0.3 \\ \end{pmatrix}$, and the aggregation function is egalitarian $\rho=0.1$.

\subsubsection{TripAdviser:} 
We used the TripAdvisor New Zealand Hotel dataset \citep{tripadviser} which consists of three dimensional data, price, number of review, and review rank. We denote $p(x)$ as price, $r(x)$ as number of review, and $R(x)$ as the review rank.
There are three agents who has the different utility function.

\textbf{Utility functions}
\begin{compactenum}
    \item[(agent 1)] luxury: 0.5 p(x) + 0.5 R(x)
    \item[(agent 2)] budget: -0.5 p(x) + r(x) + 0.5 r(x)
    \item[(agent 3)] review: R(x)
\end{compactenum}
The social-influence graph is defined as $A := \begin{pmatrix} 0.5 & 0.3 & 0.2 \\ 0.1 & 0.8 & 0.1 \\ 0.2 & 0.1 & 0.7\\ \end{pmatrix}$, and the aggregation function is egalitarian $\rho=0.5$.

\subsubsection{EnergyTrading:} 
We used METER dataset, UK-wide energy demand dataset \citep{meterdataset, grunewald2019specific}. Due to the privacy reason, we cannot identify which smart meter id corresponds to the geography in the UK. Therefore, we place the all meter dataset into the hypothetical two-dimensional space, which placed based on the similarity between the time series data. Intuitively, the energy demand should have some geographical relationship, for instance Scotland is cooler than England, thereby the heating energy is more necessary. As such estimated two dimensional space locations are further transformed into continuous space, by interpolating by GP model. We refer to this GP as oracle GP. We also assume three firms use GPs as their demand prediction model. We refer to these individual GPs as internal models. As we use GP, the maximum information gain can be reasonably approximated by the predictive variance \citep{srinivas2010gaussian}. Then, we allocate different number of datasets to each internal models, resulting in different information gain functions. We use this predictive variance of each internal model as the utility functions. A large corporation is assumed to have the largest and diverse data, and a niche startup has handful of data but is very niche where the large corporation does not have. A joint venture is the mixture of niche data. These data distribution creates the different maximizers of utility functions.

The social-influence graph is defined as $A := \begin{pmatrix} 0.8 & 0.1 & 0.1 \\ 0.6 & 0.1 & 0.3 \\ 0.4 & 0.3 & 0.3\\ \end{pmatrix}$, and the aggregation function is egalitarian $\rho=0.5$.

\subsection{Gaussian process based model}\label{app:gp_exp}
Now we tested our GP-based algorithm with popular baselines: (a) GP-UCB \citep{srinivas2010gaussian}: This baseline solves Prob.~(\ref{objective: SBO}) as a single-objective BO, conditioning a GP on aggregated votes, $y_t := \mathcal{A}[u(x_t,:)]$. We compare scenarios where the GP is conditioned on truthful utilities $(u)$ versus non-truthful ones $(v)$. (b) Multi-task (MT) GP-UCB \citep{kandasamy2016gaussian}: This baseline models utilities using a simple MTGP without incorporating the graph structure, employing our acquisition function from Prob.~(\ref{prob:ucb_acquisition}) for querying. We compare cases where the GP is conditioned only on truthful utilities $(u)$ versus both truthful and non-truthful utilities $(u, v)$. If the non-truthful utilities $(v)$ provide useful low-fidelity information, the convergence rate for $(u, v)$ should improve relative to $(u)$ alone; otherwise, it might remain the same or deteriorate.

\paragraph{Social-influence graph.} 
\begin{figure}
  \centering
  \includegraphics[width=0.6\hsize]{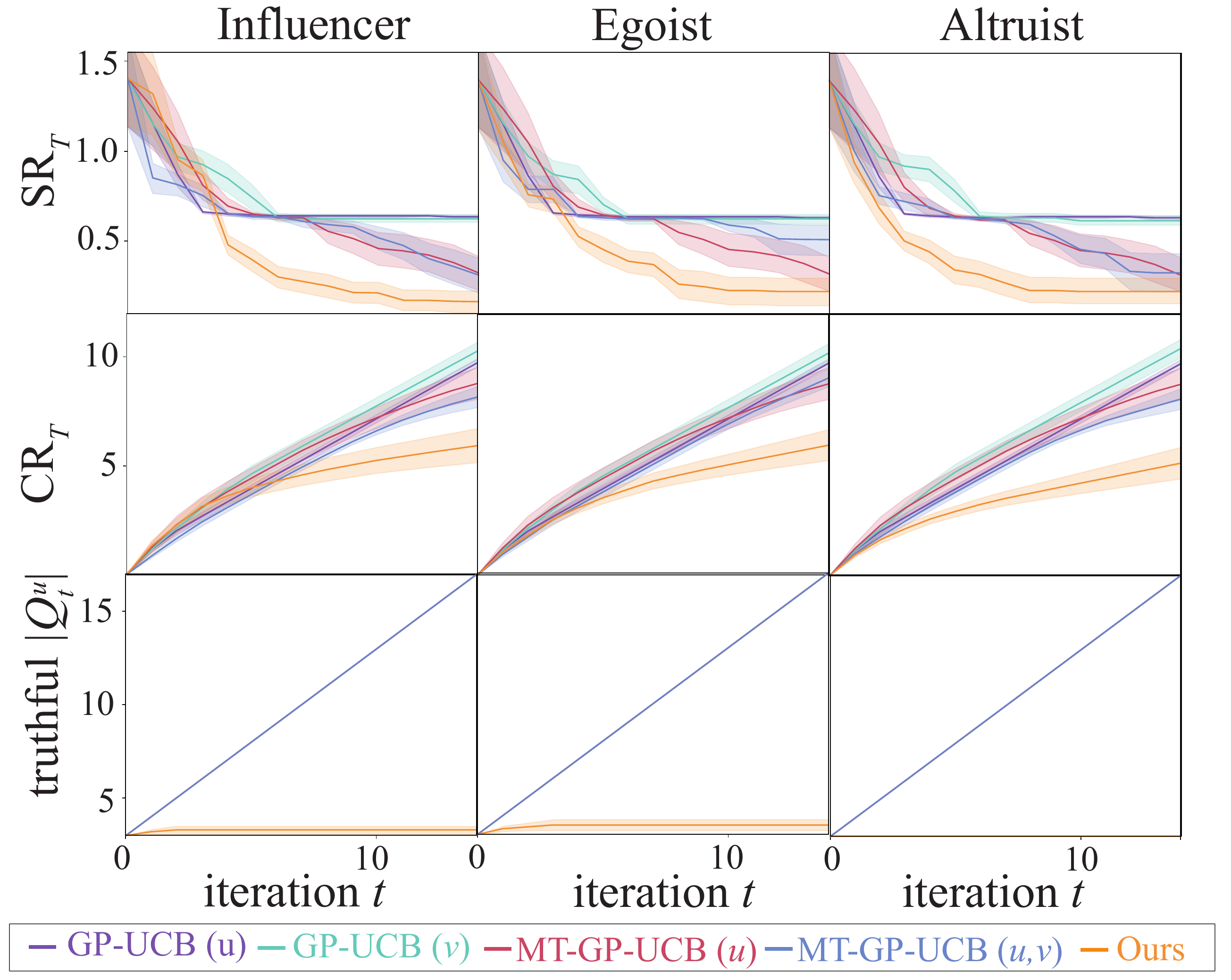}
  \caption{Simple regret, cumulative regret, and cumulative queries on different social influence graph.}
  \label{fig:gp_social_influence}
  \vspace{-1em}
\end{figure}
To examine the effect of the social-influence graph $A$, we varied $A$ using the same funcitons and graph in Figure~\ref{fig:social_influence}. Figure~\ref{fig:gp_social_influence} highlights the robustness and efficacy of our algorithm. Our approach consistently outperforms the baselines in both simple and cumulative regret. Notably, the cumulative number of truthful queries $|\mathcal{Q}^u_T|$ grows logarithmically, requiring only a few queries overall, while the baselines require a linear growth in $|\mathcal{Q}^u_T|$. This demonstrates our algorithm’s sample efficiency for expensive $u$ queries. A closer look reveals that GP-UCB often gets stuck in local maxima. This is a known limitation of GP-UCB under model misspecification \citep{berkenkamp2019no}, which necessitates additional exploration through an increased $\beta$ parameter \citep{bogunovic2021misspecified}, requiring more iterations for $\beta$ to grow. In contrast, our convolutional kernel GP captures the correlations in corrupted $v$, providing better extrapolation and avoiding misspecification issues compared to the vanilla GP model. Contrastingly, a naïve combination of MT-GP-UCB $(u,v)$ fails to accelerate in most cases because standard correlation learning in MTGP is not the convolution learning. It often performs worse than using only truthful data $(u)$, as \citet{mikkola2023multi} explains that incorporating unreliable information can actually decelerate convergence.

\paragraph{Aggregate function.} 
\begin{figure}
  \centering
  \includegraphics[width=0.6\hsize]{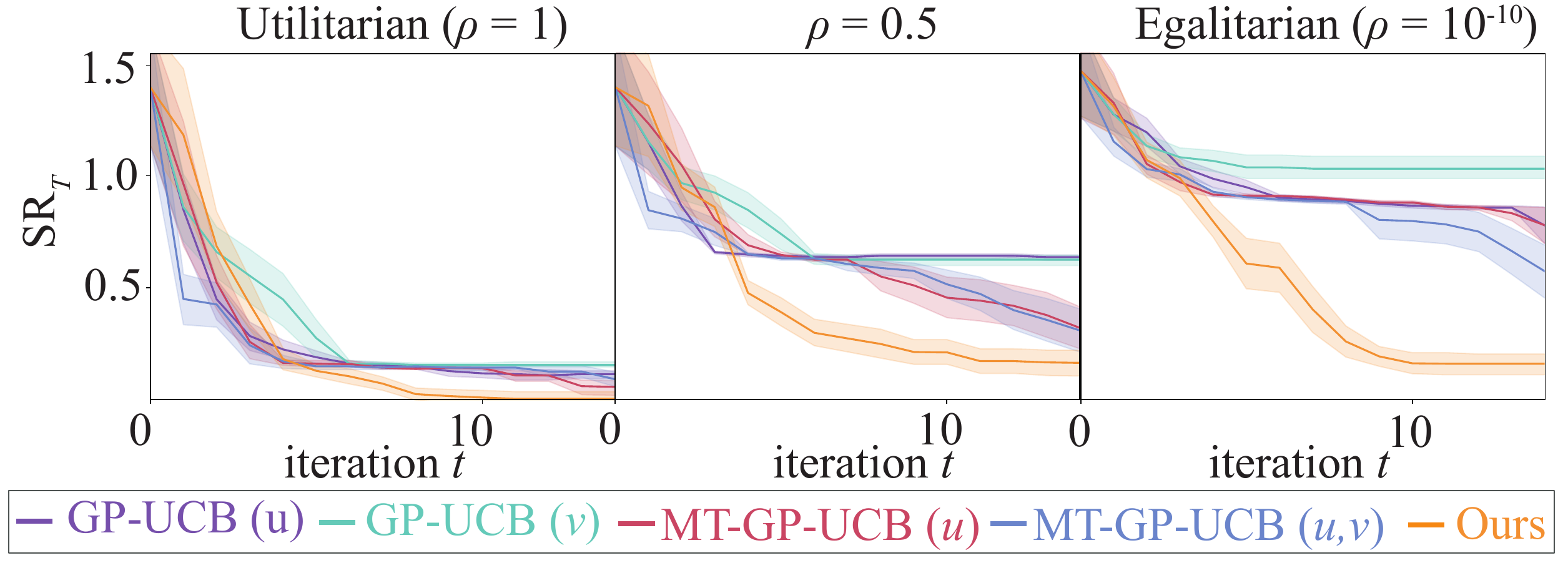}
  \caption{Simple regret, cumulative regret, and cumulative queries on different aggregation function.}
  \label{fig:gp_aggregate}
  \vspace{-1em}
\end{figure}
We further tested the effect of the aggregation function by varying $\rho \in [1, 0.5, 10^{-10}]$ in Eq.~(\ref{eq:gini}), keeping the influencer-follower matrix $A$ fixed. Smaller $\rho$ values lead to more pronounced differences in our model. This makes sense, as $\rho$ controls the degree to which minority preferences, such as those of the follower in this case, are prioritized, thereby accentuating model misspecification issues as $\rho$ decreases.

\subsection{Computation time}\label{app:compute_time}
\begin{figure}[htb!]
  \centering
  \includegraphics[width=0.9\hsize]{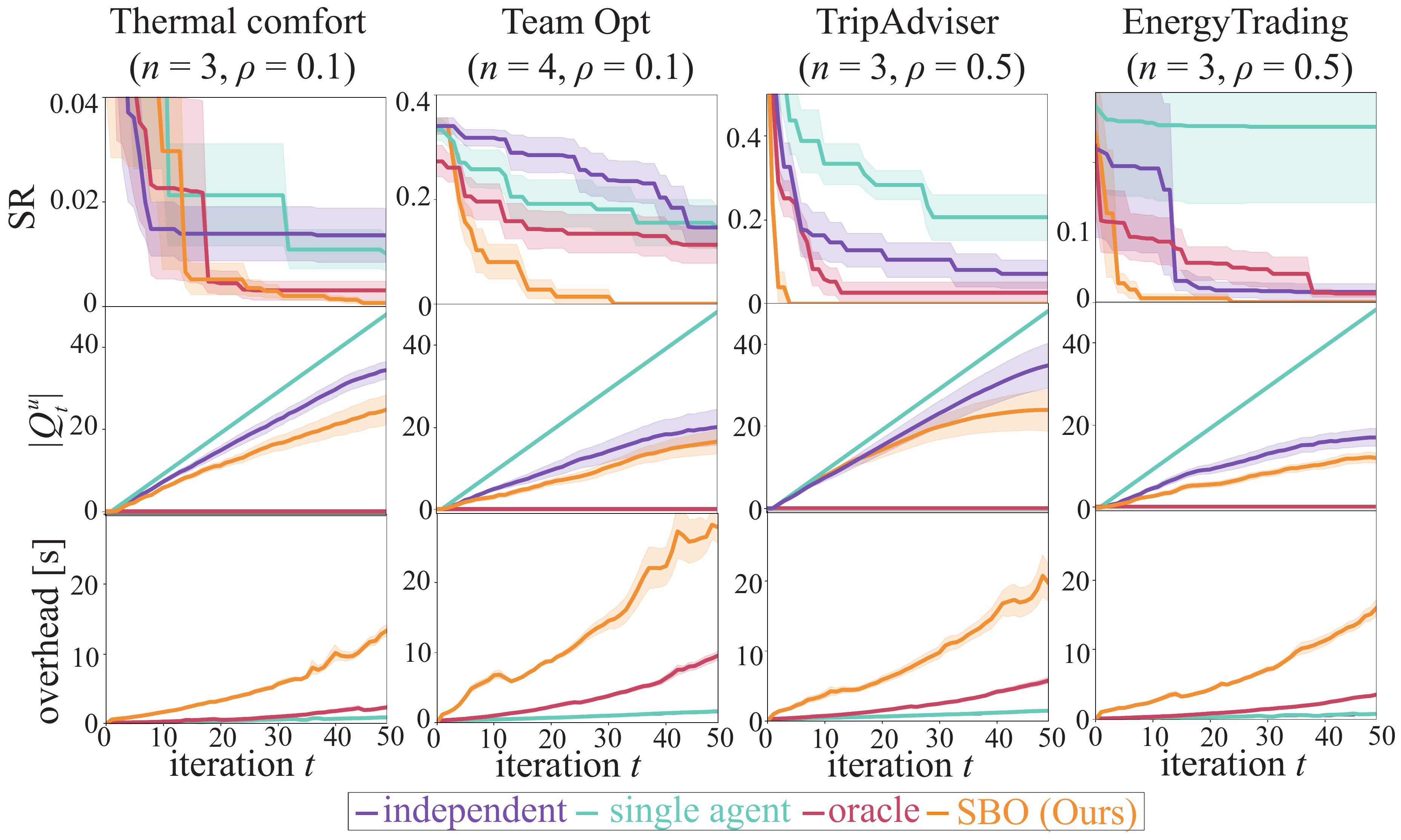}
  \caption{Computation time on real-world tasks.}
  \label{fig:overhead}
  \vspace{-1em}
\end{figure}
We report the computation time on Figure~\ref{fig:overhead}. As we can see, although our algorithm is the slowest, each query only takes within 30 seconds at iteration $t=50$. The complexity is $\mathcal{O}(n(t + n))$, and the computation time scales linearly as shown in the figure. Taking tens of seconds in multi-agentic scenario is common as it is inherently expensive computation. 

\end{document}